\documentclass[english]{article}
\usepackage[T1]{fontenc}
\usepackage[latin9]{luainputenc}
\usepackage{geometry}
\usepackage{amstext}
\usepackage{amssymb}
\usepackage{feyn}
\usepackage{tikz}
\usepackage{graphicx}
\usepackage{array}
\usepackage{booktabs}

\usepackage{amsmath}
\usepackage{amsfonts}
\usepackage{mathrsfs}

\usepackage[normalem]{ulem}
\usepackage{listings,braket}
\usepackage{caption}
\usepackage{subcaption}
\usepackage{float}
\usepackage{color,soul}
\definecolor{darkgreen}{rgb}{0,0.35,0}
\definecolor{Rood}{rgb}{1, 0, 0}

\makeatletter
\usepackage[english]{babel}
\usepackage{feyn}

\makeatother

\begin{document}
	
	\title{\bf Gauge-invariant scalar and vector operators in the $SU\left(2\right)\times U\left(1\right)$
		Higgs model: Ward identities and renormalization}

	{\author{ \textbf{Giovani~Peruzzo$^1$}\thanks{gperuzzofisica@gmail.com}\\\\\
			\textit{{\small $^1$Instituto de F\'{i}sica, Universidade Federal Fluminense,
			}}\\
			\textit{{\small Campus da Praia Vermelha, Av. Litor\^{a}nea s/n, }}\\
			\textit{{\small 24210-346, Niter\'{o}i, RJ, Brasil}}\\
		}
		
		\date{}
		
		\maketitle
		
		\begin{abstract}
			In this work, we investigate the renormalization of the gauge-invariant scalar and vector composite operators proposed in \cite{Dudal:2023jsu} to describe the $SU(2)\times U(1)$ Higgs model from a gauge-invariant perspective. To establish the relationship between the counterterms, we use the Algebraic Renormalization approach \cite{Piguet:1995er}. Therefore, we also derive the set of Ward identities of the model after introducing these composite operators. Using these Ward identities, we demonstrate important exact relations between the correlation functions of elementary fields and composite fields. 
			
		\end{abstract}

\section{Introduction}

In the previous works \cite{Dudal:2019pyg,Capri:2020ppe,Dudal:2020uwb,Dudal:2021pvw,Dudal:2021dec}, we developed a gauge-invariant setup for the $U(1)$ and $SU(2)$ Higgs models \cite{Higgs:1964pj,Englert:1964et}. This was obtained by considering correlation functions of local composite gauge-invariant operators. Such operators were carefully chosen according to their quantum numbers, and the asymptotic states that might be associated with them. For each of the known particles in the spectrum, we considered a gauge invariant operator with the same quantum numbers and analyzed its correlation functions. In \cite{Capri:2020ppe,Dudal:2020uwb,Dudal:2019aew}, we also explicitly showed and compared the advantages of working with correlation functions of gauge invariant fields over correlation functions of elementary field. Most recently \cite{Dudal:2023jsu}, we started to extend this approach to the $SU(2)\times U(1)$ due its importance for the Standard Model physics, and, in this paper, we discuss some of the technicalities necessary to use these gauge-invariant operators, such as the renormalization.  \par

When we quantize a gauge field theory in the continuous space-time, we have the freedom to choose the gauge fixing. Therefore, physical observables should not be gauge dependent quantities. This simple requirement would be easily fulfilled if the Faddeev-Popov method \cite{Faddeev:1967fc} worked in all cases. At least formally, by using this method, correlation functions of gauge invariant operators are gauge-independent \cite{Weinberg:1996kr}. This is the technical motivation for developing a gauge-invariant setup to the $SU(2)\times U(1)$ Higgs model. In general, however, correlation functions of elementary fields are gauge-dependent mathematical objects. Therefore, extracting physical information from them is a non-trivial task. Frequently, we rely on intuition and the set of Nielsen identities \cite{Nielsen:1975fs,Piguet:1984js} to accomplish this. However, gauge dependent correlation functions are not always well-behaved; for example, they can contain non-physical thresholds. Recently \cite{Dudal:2020uwb,Dudal:2019aew}, we presented examples within the context of the $SU(2)$ and $U(1)$ Higgs models where, for example, the positivity of the K\"all\'en-Lehmann spectral representation is violated. In the same work, we showed that none of these problems occur in correlation functions of gauge-invariant operators. \par   

Unfortunately, for non-Abelian theories, the Faddeev-Popov method breaks down in the non-perturbative regime due to the existence of Gribov-Singer copies \cite{Gribov:1977wm,Singer:1978dk}. The Faddeev-Popov method also does not perform well with non-linear gauges, such as the Maximal Abelian Gauge (MAG)\cite{tHooft:1981bkw}. In this specific example, the theory is renormalizable at the quantum level only after we change the original gauge by adding quartic interactions to the ghost fields. Here, we do not address these particular issues regarding the Faddeev-Popov method and leave these problems for future work. On purpose, we hope that a gauge-invariant or a BRST-invariant description will facilitate comparisons between different methods. The first immediate application can be found in the RGZ (Refined Gribov-Zwanziger) theory \cite{Dudal:2008sp}, which recently received a BRST-invariant formulation \cite{Capri:2016aqq}. \par

In addition to the gauge independence, correlation functions of gauge-invariant operators possess other desirable properties. The one that caught our attention the most is the positivity of the K\"all\'en-Lehmann representation. We explicitly verified this in the $U(1)$ and $SU(2)$ Higgs models. Later, \cite{Maas:2020kda} also verified the same in the presence of a fermionic field. In fact, positivity is not a property that can be taken for granted. As pointed out by Kugo and Ojima \cite{Kugo:1979gm}, in a BRST invariant theory it is necessary to demonstrate that the physical space defined by the cohomology of the BRST-charge $Q_{BRST}$ has a positive semi-definite inner product. Under certain assumptions, such as \emph{asymptotic completeness}, \cite{Kugo:1979gm} argued that, at least perturbatively, the Higgs models have this properties, which unsures also the unitarity of the theory. The mechanism behind it is the so-called \emph{quartet mechanism}. Positivity violation of the K\"all\'en-Lehmann representation is a sign that something strange is occurring. In a gauge field theory, we can only reach this conclusion after ruling out the effects caused by a lack of gauge invariance. According to what we just mentioned, if one does find violation for correlation functions of gauge-invariant operators, then something has drastically changed in the theory. In some case, such situation is used to explain the fact that some particles (e.g quarks and gluons) do not show up in the physical spectrum. Indeed, we can mention some works regarding the gluon propagator \cite{Cucchieri:2007md,Cucchieri:2007rg,Aguilar:2008xm,Bowman:2007du}, especially in lattice studies, that indicates an association between positivity violation and gluons confinement. \par

In the Standard Model context, the confinement of quarks and gluons is attribute to the strong interaction, described by the $SU(3)$ color interaction (QCD). The $SU(2)\times U(1)$ sector, responsable for the electroweak interaction and where the Higgs enters, does not exhibit this property. For several reasons, the electroweak force cannot form bound states as the strong interaction does. Therefore, it seems that a possible confinement phase in the $SU(2)\times U(1)$ Higgs model has more theoretical than physical value in the current status of our universe. As indicated in the work of Fradkin and Shenker \cite{Fradkin:1978dv}, which uses lattice techniques, indeed, one can find confining phases in Higgs models. In special, in the fundamental $SU(N)$ Higgs model the confining phase and the Higgs phase are smoothly connected. For more recent references on this subject, see \cite{Greensite:2021fyi,Maas:2017wzi}.\par

The feature just mentioned about the fundamental Higgs model- that confining and Higgs phases are smoothly connected- can produce interesting insights when analyzed from the point of view of gauge-invariant operators \cite{Frohlich:1980gj,Frohlich:1981yi}. Due to the Higgs mechanism, we have a very special situation in which the scalar field $\varphi$ is expanded around one of the minima of the effective potential, say $\vartheta$. Therefore, the first contribution to a local composite operators built with $\varphi$ comes from this constant part $\vartheta$. It turns out that many correlation functions of elementary fields are equivalent to correlation functions of gauge-invariant composite operators, at least at the tree-level. Such a situation is not seen in other models, such as the QCD or the QED. In both cases, to achieve that, we need to work with non-local operators or introduce auxiliary fields, see \cite{Stueckelberg:1938hvi,Stueckelberg:1938zz,Ruegg:2003ps,Capri:2016aqq}. As elementary fiels are usually associated with elementary particles and composite operators with bound states, this particular situation we found in the Higgs model may cause some confusion. The best way to address this is perhaps to assert that the particle is the same in both cases as there is no physical means of distinguishing between them. Indeed, in the $SU(2)$ and $U(1)$ fundamental Higgs models we were able to show that by using Ward identities \cite{Dudal:2021dec}. \par 

We mentioned some results from the lattice. In the recent times, we have seen an increasing number of studies on the Higgs model in lattice gauge theory; see \cite{Maas:2017wzi} for a review on the subject. Since lattice simulations have become one of the most important resources for extracting non-perturbative information, it is crucial to find ways to compare results in the continuum with those from the lattice. In lattice simulations, it is not necessary to fix the gauge, and due to the Elitzur theorem \cite{Elitzur:1975im}, the correlation functions of gauge variant operators vanish. Therefore, gauge-invariant operators are also crucial in lattice gauge theories. This provides us with further motivation to develop a gauge-invariant framework.  \par

In \cite{Dudal:2023jsu}, we propose a novel set of gauge-invariant operators for the $SU(2)\times U(1)$ Higgs model by utilizing the aforementioned features of the Higgs mechanism and the Stueckelberg field method \cite{Capri:2016aqq}. In this paper, we will present the renormalization of the $SU(2)\times U(1)$ Higgs model in the presence of these composite operators in the Landau gauge. We determine the counterterms and renormalization factors by utilizing general renormalization theory and the set of Ward identities (symmetries) that exist in this model. In other words, we follow the Algebraic Renormalization approach \cite{Piguet:1995er}. Therefore, we are not committed \emph{a priori} to any specific regularization or renormalization scheme. \par

In general, when we introduce a composite operator coupled with a source in the starting action, we do not expect to obtain a new Ward identity. However, we show that in the $SU(2)\times U(1)$ Higgs model with these specific operators, there are three additional Ward identities, two of which are non-integrated equations. These Ward identities not only help control the ultraviolet divergences of the correlation functions of these operators, but also relate some of their renormalization factors to those of the fundamental fields. We can mention the renormalization factors associated with the scalar operator $O(x)$, which are completely determined by the renormalization factor of the quartic coupling $\lambda$ and the counterterm necessary to renormalize the tadpoles. \par

One of these new Ward identities involves the equation of motion for the Abelian gauge field $X_{\mu}(x)$. This identity exists  only because one of the vector gauge-invariant operators $O_{\mu}(x)$ is precisely the $U(1)$ current. Using this Ward identity, we demonstrate several results, such as the fact that the anomalous dimension of $O_{\mu}(x)$ is zero, which is consistent with the fact that this is a current. We also show that the transverse part of $\langle O_{\mu}(x)O_{\nu}(y) \rangle$ is given by $\langle X_{\mu}(x)X_{\nu}(y) \rangle$, and that the longitudinal part is constant, implying that no propagating mode can be associated with it. Since we use functional identities to demonstrate this, these results should also be non-perturbative. \par

The paper is organized as follows. In Section \ref{sec:Review-of-the},
we present a brief review of the $SU\left(2\right)\times U\left(1\right)$
Higgs model and the Higgs mechanism. In Section \ref{sec:Gauge-invariant-operators},
we introduce the gauge-invariant operators and the localizing
procedure. In Section \ref{sec:BRST-symmetry-and}, we discuss the
BRST symmetry and the gauge fixing term in the Landau gauge. In Section
\ref{sec:Composite-operators-and}, we give the proper treatment to
the composite operators, by introducing each of them in the starting
action together with a source. In Section \ref{sec:Symmetries-and-Ward},
we present the Ward identities of the model in the presence of these
composite operators. In Section \ref{sec:All-orders-renormalization},
counterterms and renormalization factor are determined. In Section
\ref{sec:Exploring-some-of}, we derive a few important relation between
correlation functions helped by the new Ward identities. In Section
\ref{sec:Conclusions}, we present our conclusions and perspective
to this work. 

\section{Brief review of the $SU\left(2\right)\times U\left(1\right)$ Higgs model\protect\label{sec:Review-of-the}}

\subsection{The Higgs action}

We will consider the Euclidian $SU\left(2\right)\times U\left(1\right)$ Higgs model
for a charged scalar field $\varphi\left(x\right)$ in the fundamental
representation. This model is defined by the following action:
\begin{eqnarray}
	S_{\textrm{Higgs}} & = & \int d^{4}x\left[\frac{1}{4}W_{\mu\nu}^{a}W_{\mu\nu}^{a}+\frac{1}{4}X_{\mu\nu}X_{\mu\nu}+\left(D_{\mu}\varphi\right)^{\dagger}D_{\mu}\varphi+\lambda\left(\varphi^{\dagger}\varphi-\frac{\vartheta^{2}}{2}\right)^{2}\right],\label{eq:higgs_action}
\end{eqnarray}
where 
\begin{eqnarray}
	D_{\mu}\varphi & = & \partial_{\mu}\varphi-ig\frac{\tau^{a}}{2}W_{\mu}^a\varphi-\frac{1}{2}ig'X_{\mu}\varphi,\nonumber \\
	W_{\mu\nu}^{a} & = & \partial_{\mu}W_{\nu}^{a}-\partial_{\nu}W_{\mu}^{a}+g\varepsilon^{abc}W_{\mu}^{b}W_{\nu}^{c},\nonumber \\
	X_{\mu\nu} & = & \partial_{\mu}X_{\nu}-\partial_{\nu}X_{\mu}.
\end{eqnarray}
Notice that the generators of the $SU(2)$ group are constructed in terms of the Pauli matrices $\left\{ \tau^{1},\tau^{2},\tau^{3}\right\}$. Therefore, we have the usual communtation and anticommutation relations:
\begin{eqnarray}
	\left\{ \tau^{a},\tau^{b}\right\}  & = & 2\delta^{ab},\nonumber \\
	\left[\tau^{a},\tau^{b}\right] & = & 2i\varepsilon^{abc}\tau^{c},
\end{eqnarray}
where $\varepsilon^{abc}$ ($\varepsilon^{123}=1$) is the 3D Levi-Civita
symbol. $W_{\mu}^{a}$ and $X_{\mu}$ are the $SU\left(2\right)$
and $U\left(1\right)$ gauge fields, respectively. Also note that we
have three different massless couplings: the gauge couplings
$\left(g,g'\right)$ and the quartic coupling $\lambda$. $\vartheta$
is a massive parameter related to the vacuum expectation value (VEV)
of the scalar field. 

The Higgs action (\ref{eq:higgs_action}) is invariant under the following
local gauge transformations:
\begin{description}
	\item [{\emph{SU(2)}}] \emph{transformations}
\end{description}
\begin{eqnarray}
	\varphi^{'} & = & e^{-ig\frac{\tau^{a}}{2}\omega^{a}}\varphi,\nonumber \\
	\frac{\tau^{a}}{2}W_{\mu}^{'a} & = & e^{-ig\frac{\tau^{b}}{2}\omega^{b}}\frac{\tau^{a}}{2}W_{\mu}^{a}e^{ig\frac{\tau^{c}}{2}\omega^{c}}-\frac{\tau^{a}}{2}\partial_{\mu}\omega^{a},\nonumber \\
	X_{\mu}^{'} & = & X_{\mu},\label{eq:su2_gauge_tranf}
\end{eqnarray}

\begin{description}
	\item [{\emph{U(1)}}] \emph{transformations}
\end{description}
\begin{eqnarray}
	\varphi^{'} & = & e^{-\frac{1}{2}ig'\omega}\varphi,\nonumber \\
	W^{'}_{\mu} & = & W_{\mu},\nonumber \\
	X_{\mu}^{'} & = & X_{\mu}-\partial_{\mu}\omega.\label{eq:u1_gauge_transf}
\end{eqnarray}
To quantize this theory we break these symmetries by choosing a gauge fixing, but we will
get back to this point later. Now, let us discuss the vacuum structure
of the theory.

\subsection{Higgs Mechanism}

The vacuum of the theory is determined by the minimum of the effective
potential $V\left(\varphi\right)$. In this case, the tree level potential, also called the Higgs potential, is given by
\begin{eqnarray}
	V\left(\varphi\right) & = & \lambda\left(\varphi^{\dagger}\varphi-\frac{\vartheta^{2}}{2}\right)^{2}.\label{eq:Higgs_potential}
\end{eqnarray}
Notice that $V\left(\varphi\right)$ has minima when $\varphi^{\dagger}\varphi=\frac{\vartheta^{2}}{2}$,
which implies that $\left\langle \varphi\right\rangle \neq0$ (VEV of
$\varphi$). Consequently, the vacuum is not invariant by the full $SU\left(2\right)\times U\left(1\right)$
group, \emph{i.e., }the symmetry is spontaneously broken. Let us choose
the minimum
\begin{eqnarray}
	\varphi_{0} & = & \left(\begin{array}{c}
		0\\
		\frac{\vartheta}{\sqrt{2}}
	\end{array}\right).\label{eq:potencial}
\end{eqnarray}
and then quantize the theory around this minimum. Therefore, all  the original generators of the $SU\left(2\right)\times U\left(1\right)$
group are broken, \emph{i.e.},
\begin{gather}
	\frac{\tau^{a}}{2}\varphi_{0}\neq0,\qquad I\varphi_{0}\neq0.
\end{gather}
However, we can change to a basis with
\begin{eqnarray}
	Q & = & \frac{\tau^{3}}{2}+\frac{I}{2}
\end{eqnarray}
as one of the generators. $Q$ is the electric charge that generates
the residual Abelian electromagnetic symmetry, denoted by $U_{EM}\left(1\right)$.
Notice that 
\begin{eqnarray}
	Q\varphi_{0} & = & 0.
\end{eqnarray}
Therefore, we have the well-known breaking pattern $SU\left(2\right)\times U\left(1\right)\sim U_{EM}\left(1\right)$.
Thus, according to the theory \cite{Higgs:1964pj,Englert:1964et,Guralnik:1964eu}, we should have three massive vector bosons
and one massless vector boson, followed by three Goldstone modes.

By following the usual quantization approach, we rewrite the scalar
field as
\begin{eqnarray}
	\varphi\left(x\right) & = & \varphi_{0}+\frac{1}{\sqrt{2}}\left(\begin{array}{c}
		i\rho^{1}\left(x\right)+\rho^{2}\left(x\right)\\
		H\left(x\right)-i\rho^{3}\left(x\right)
	\end{array}\right)\nonumber \\
	& = & \frac{1}{\sqrt{2}}\left[\left(\vartheta+H\left(x\right)\right)I+i\rho^{a}\left(x\right)\tau^{a}\right]\left(\begin{array}{c}
		0\\
		1
	\end{array}\right),\label{eq:phi_rewitten}
\end{eqnarray}
where $H\left(x\right)$ and $\rho^{a}\left(x\right)$, the Higgs
field and the would be Goldstone fields, respectively, will be the
dynamical fields. To properly explore the residual $U_{EM}\left(1\right)$ symmetry,
we introduce a 2D notation, where the upper Greek indices run from
1 to 2, and $\varepsilon^{\alpha\beta}$ ($\varepsilon^{12}=1$) is
the 2D Levi-Civita symbol. In reference to the Pauli matrices, we
designate components 1 and 2 as the off-diagonal components, and component
3 as the diagonal component. Therefore, (\ref{eq:higgs_action})
can be rewritten as
\begin{eqnarray}
	S_{\textrm{Higgs}} & = & \int d^{4}x\left\{ \frac{1}{4}W_{\mu\nu}^{\alpha}W_{\mu\nu}^{\alpha}+\frac{1}{4}W_{\mu\nu}^{3}W_{\mu\nu}^{3}+\frac{1}{4}X_{\mu\nu}X_{\mu\nu}\right.\nonumber \\
	&  & +\frac{1}{2}\left(\partial_{\mu}H\right)^{2}+\frac{1}{2}\left(\partial_{\mu}\rho^{\alpha}\right)^{2}+\frac{1}{2}\left(\partial_{\mu}\rho^{3}\right)^{2}\nonumber \\
	&  & +\frac{1}{2}gW_{\mu}^{\alpha}\left[\left(\partial_{\mu}H\right)\rho^{\alpha}-\left(\partial_{\mu}\rho^{\alpha}\right)\left(\vartheta+H\right)-\varepsilon^{\alpha\beta}\left(\partial_{\mu}\rho^{\beta}\right)\rho^{3}+\varepsilon^{\alpha\beta}\left(\partial_{\mu}\rho^{3}\right)\rho^{\beta}\right]\nonumber \\
	&  & +\frac{1}{2}gW_{\mu}^{3}\left[\left(\partial_{\mu}H\right)\rho^{3}-\left(\partial_{\mu}\rho^{3}\right)\left(\vartheta+H\right)-\varepsilon^{\alpha\beta}\left(\partial_{\mu}\rho^{\alpha}\right)\rho^{\beta}\right]\nonumber \\
	&  & +\frac{1}{8}g^{2}W_{\mu}^{\alpha}W_{\mu}^{\alpha}\left[\left(\vartheta+H\right)^{2}+\rho^{\beta}\rho^{\beta}+\rho^{3}\rho^{3}\right]\nonumber \\
	&  & +\frac{1}{8}g^{2}W_{\mu}^{3}W_{\mu}^{3}\left[\left(\vartheta+H\right)^{2}+\rho^{\alpha}\rho^{\alpha}+\rho^{3}\rho^{3}\right]\nonumber \\
	&  & +\frac{1}{4}gg'W_{\mu}^{3}X_{\mu}\left(\vartheta+H\right)^{2}-\frac{1}{2}gg'\left(\vartheta+H\right)\varepsilon^{\alpha\beta}W_{\mu}^{\alpha}\rho^{\beta}X_{\mu}\nonumber \\
	&  & -\frac{1}{4}gg'W_{\mu}^{3}X_{\mu}\rho^{\alpha}\rho^{\alpha}-\frac{1}{4}gg'W_{\mu}^{3}X_{\mu}\rho^{3}\rho^{3}+\frac{1}{2}gg'W_{\mu}^{\alpha}X_{\mu}\rho^{3}\rho^{\alpha}+\frac{1}{2}gg'W_{\mu}^{3}X_{\mu}\rho^{3}\rho^{3}\nonumber \\
	&  & +\frac{1}{2}g'X_{\mu}\left[-\left(\partial_{\mu}H\right)\rho^{3}-\left(\partial_{\mu}\rho^{3}\right)\left(\vartheta+H\right)+\varepsilon^{\alpha\beta}\left(\partial_{\mu}\rho^{\alpha}\right)\rho^{\beta}\right]\nonumber \\
	&  & +\frac{1}{8}g'{}^{2}X_{\mu}X_{\mu}\left[\left(\vartheta+H\right)^{2}+\rho^{\alpha}\rho^{\alpha}+\rho^{3}\rho^{3}\right]\nonumber \\
	&  & \left.+\frac{\lambda}{4}\left[\left(\vartheta+H\right)^{2}+\rho^{\alpha}\rho^{\alpha}+\rho^{3}\rho^{3}\right]^{2}\right\} ,\label{eq:Higgs_action_rewritten}
\end{eqnarray}

To have an idea of the propagating modes of the theory after the Spontaneous
Symmetry Breaking (SSB), let us analyze the quadratic part of (\ref{eq:Higgs_action_rewritten}),
namely, 
\begin{eqnarray}
	\left.S_{\textrm{Higgs}}\right|_{quad} & = & \int d^{4}x\left\{ \frac{1}{4}\left(\partial_{\mu}W_{\nu}^{\alpha}-\partial_{\nu}W_{\mu}^{\alpha}\right)^{2}+\frac{1}{8}\vartheta^{2}g^{2}W_{\mu}^{\alpha}W_{\mu}^{\alpha}\right.\nonumber \\
	&  & +\frac{1}{4}\left(\partial_{\mu}W_{\nu}^{3}-\partial_{\nu}W_{\mu}^{3}\right)^{2}+\frac{1}{4}\left(\partial_{\mu}X_{\nu}-\partial_{\nu}X_{\mu}\right)^{2}\nonumber \\
	&  & +\frac{1}{2}\left(\partial_{\mu}H\right)^{2}+\lambda\vartheta^{2}H^{2}+\frac{1}{2}\left(\partial_{\mu}\rho^{\alpha}\right)^{2}+\frac{1}{2}\left(\partial_{\mu}\rho^{3}\right)^{2}\nonumber \\
	&  & -\frac{1}{2}\vartheta gW_{\mu}^{\alpha}\left(\partial_{\mu}\rho^{\alpha}\right)-\frac{1}{2}\vartheta gW_{\mu}^{3}\left(\partial_{\mu}\rho^{3}\right)-\frac{1}{2}\vartheta g'X_{\mu}\left(\partial_{\mu}\rho^{3}\right)\nonumber \\
	&  & \left.+\frac{1}{8}\vartheta^{2}g^{2}W_{\mu}^{3}W_{\mu}^{3}+\frac{1}{4}\vartheta^{2}gg'W_{\mu}^{3}X_{\mu}+\frac{1}{8}\vartheta^{2}g'{}^{2}X_{\mu}X_{\mu}\right\} .\label{eq:Higgs_action_rewritten-1}
\end{eqnarray}
We do not need to worry about the mixing terms 
\begin{gather*}
	-\frac{1}{2}\vartheta gW_{\mu}^{\alpha}\left(\partial_{\mu}\rho^{\alpha}\right)-\frac{1}{2}\vartheta gW_{\mu}^{3}\left(\partial_{\mu}\rho^{3}\right)-\frac{1}{2}\vartheta g'X_{\mu}\left(\partial_{\mu}\rho^{3}\right),
\end{gather*}
as we can eliminate them by choosing a convenient gauge fixing. Immediately,
we conclude that the gauge fields $W_{\mu}^{1}$ and $W_{\mu}^{2}$
have equal masses,
\begin{equation}
	m_{W} = \frac{\vartheta g}{2}.
\end{equation}
The same thing happens with the Higgs field, which acquires the mass
\begin{equation}
	m_{H} = \sqrt{2\lambda}\vartheta.
\end{equation}
The massive term in the last line of (\ref{eq:Higgs_action_rewritten-1})
can be rewritten as 
\begin{equation}
	\frac{1}{8}\vartheta^{2}g^{2}W_{\mu}^{3}W_{\mu}^{3}+\frac{1}{4}\vartheta^{2}gg'W_{\mu}^{3}X_{\mu}+\frac{1}{8}\vartheta^{2}g'{}^{2}X_{\mu}X_{\mu} = \frac{1}{8}\vartheta^{2}\left(gW_{\mu}^{3}+g'X_{\mu}\right)^{2}.\label{eq:massive_term}
\end{equation}
So, if we change the fields $W_{\mu}^{3}$ and $X_{\mu}$ to
\begin{gather}
	Z_{\mu} = \cos\theta_{W}W_{\mu}^{3}+\sin\theta_{W}X_{\mu},\qquad
	A_{\mu} = -\sin\theta_{W}W_{\mu}^{3}+\cos\theta_{W}X_{\mu},
\end{gather}
where the Weinberg angle \cite{Weinberg:1967tq} $\theta_{W}$ is defined by
\begin{gather}
	\cos\theta_{W} = \frac{g}{\sqrt{g^{2}+g'{}^{2}}}, \qquad
	\sin\theta_{W} = \frac{g'}{\sqrt{g^{2}+g'{}^{2}}},
\end{gather}
then (\ref{eq:massive_term}) takes the form
\begin{gather*}
	\frac{1}{8}\vartheta^{2}g^{2}W_{\mu}^{3}W_{\mu}^{3}+\frac{1}{4}\vartheta^{2}gg'W_{\mu}^{3}X_{\mu}+\frac{1}{8}\vartheta^{2}g'{}^{2}X_{\mu}X_{\mu}=\frac{1}{8}\frac{\vartheta^{2}g^{2}}{\cos^{2}\theta_{W}}Z_{\mu}^{2}.
\end{gather*}
Therefore, $Z_{\mu}$ is a massive field,
\begin{eqnarray}
	m_{Z} & = & \frac{\vartheta g}{2\cos\theta_{W}},
\end{eqnarray}
whereas $A_{\mu}$, the photon field, is massless. 

\subsection{Residual $U_{EM}\left(1\right)$ symmetry}

Examining the Higgs action after SSB, we see that there is no component
$W_{\mu}^{1}$ or $W_{\mu}^{2}$ left alone. Likewise, for components
$\rho^{1}$ and $\rho^{2}$. They are always in global  $SO\left(2\right)\simeq U\left(1\right)$ invariant combinations. Therefore,
we conclude that the whole action and the vacuum are invariant with respect the global
$U_{EM}\left(1\right)$ transformations
\begin{gather}
	W'{}_{\mu}^{\alpha}=R^{\alpha\beta}W_{\mu}^{\beta},\quad\rho'{}^{\alpha}=R^{\alpha\beta}\rho^{\beta},\quad Z'_{\mu}=Z_{\mu},\quad A'_{\mu}=A_{\mu},\quad H'=H,\quad\rho'{}^{3}=\rho^{3},\label{eq:SO1_tranf}
\end{gather}
\emph{i.e.}
\begin{gather}
	\varphi'=e^{i\omega Q}\varphi,\quad\varphi'{}^{\dagger}=\varphi^{\dagger}e^{-i\omega Q},
\end{gather}
where $R^{\alpha\beta}$ are the components of a 2D rotation matrix
\begin{eqnarray}
	R & = & \left(\begin{array}{cc}
		\cos\omega & \sin\omega\\
		-\sin\omega & \cos\omega
	\end{array}\right).
\end{eqnarray}
It is common to express everything in terms of the complex field 
\begin{gather}
	W_{\mu}^{+}=\frac{W_{\mu}^{1}+iW_{\mu}^{2}}{\sqrt{2}},\quad W_{\mu}^{-}=\frac{W_{\mu}^{1}-iW_{\mu}^{2}}{\sqrt{2}},\quad\rho^{+}=\frac{\rho^{1}+i\rho^{2}}{\sqrt{2}},\quad\rho^{-}=\frac{\rho^{1}-i\rho^{2}}{\sqrt{2}},
\end{gather}
thus (\ref{eq:SO1_tranf}) takes the form
\begin{gather}
	W'{}_{\mu}^{+}=e^{-i\omega}W_{\mu}^{+},\quad W'{}_{\mu}^{-}=e^{i\omega}W_{\mu}^{-},\quad\rho'{}^{+}=e^{-i\omega}\rho^{+},\quad\rho'{}^{-}=e^{i\omega}\rho^{-},\nonumber \\
	Z'_{\mu}=Z_{\mu},\quad A'_{\mu}=A_{\mu},\quad H'=H,\quad\rho'{}^{3}=\rho^{3}.\label{eq:U(1)EM}
\end{gather}
By looking at (\ref{eq:U(1)EM}), we see that $W'{}_{\mu}^{+}$ and
$W'{}_{\mu}^{-}$ are complex electric charged fields. They are both
associated with the massive charged $W^{+}$ and $W^{-}$ bosons.

\section{Gauge-invariant operators\protect\label{sec:Gauge-invariant-operators}}

\subsection{Universal operators and the 't Hooft operators}

In the previous work \cite{Dudal:2023jsu}, we constructed a set of very interesting
gauge-invariant operators. This was achieved by employing the $U\left(1\right)$
gauge invariant scalar operator
\begin{eqnarray}
	\Phi\left(x\right) & = & e^{-\frac{ig'}{2}\int d^{4}y\left(\frac{1}{\partial^{2}}\right)_{x,y}\left(\partial_{\mu}X_{\mu}\right)\left(y\right)}\varphi\left(x\right).
\end{eqnarray}
Then, in addition to the usual gauge-invariant operators
\begin{eqnarray}
	O\left(\varphi\right) & = & \varphi^{\dagger}\varphi-\frac{\vartheta^{2}}{2},\nonumber \\
	O_{\mu}\left(\varphi\right) & = & \frac{1}{2i}\left(\varphi^{\dagger}D_{\mu}\varphi-\left(D_{\mu}\varphi\right)^{\dagger}\varphi\right),\label{eq:traditional_gauge_inv_op}
\end{eqnarray}
one can build the 't Hooft \cite{tHooft:1980xss,Dudal:2021dec} type of operators
\begin{eqnarray}
	O_{\mu}^{3}\left(\Phi\right) & = & \frac{1}{2i}\left(\Phi^{\dagger}\mathscr{D}_{\mu}\Phi-\left(\mathscr{D}_{\mu}\Phi\right)^{\dagger}\Phi\right),\nonumber \\
	O_{\mu}^{+}\left(\Phi\right) & = & -i\Phi^{T}\tau^{2}\mathscr{D}_{\mu}\Phi=\Phi^{T}\left(\begin{array}{cc}
		0 & -1\\
		1 & 0
	\end{array}\right)\mathscr{D}_{\mu}\Phi,\nonumber \\
	O_{\mu}^{-}\left(\Phi\right) & = & \left(O_{\mu}^{+}\left(\Phi\right)\right)^{\ast},\label{eq:thooft_operators}
\end{eqnarray}
where, for the sake of convenience, we have introduced the $SU\left(2\right)$
covariant derivative
\begin{eqnarray}
	\mathscr{D}_{\mu} & = & \partial_{\mu}-ig\frac{\tau^{a}}{2}W_{\mu}^{a}.
\end{eqnarray}
If we replace $\varphi$ by (\ref{eq:phi_rewitten}) in (\ref{eq:traditional_gauge_inv_op})
and (\ref{eq:thooft_operators}) and keep only the linear terms in
the field, we get
\begin{eqnarray}
	\left.O\left(\varphi\right)\right|_{\textrm{linear}} & = & \vartheta H,\nonumber \\
	\left.O_{\mu}\left(\varphi\right)\right|_{\textrm{linear}} & = & -\frac{1}{2}\vartheta\partial_{\mu}\rho^{3}+\frac{1}{4}g\vartheta^{2}W_{\mu}^{3}-\frac{1}{4}g'\vartheta^{2}X_{\mu},\label{eq:traditional_gauge_inv_op-1}
\end{eqnarray}
and
\begin{eqnarray}
	\left.O_{\mu}^{3}\left(\Phi\right)\right|_{\textrm{linear}} & = & -\frac{g'\vartheta^{2}}{4}\partial_{\mu}\frac{\partial_{\nu}X_{\nu}}{\partial^{2}}-\frac{\vartheta}{2}\partial_{\mu}\rho^{3}+\frac{g\vartheta^{2}}{4}W_{\mu}^{3},\nonumber \\
	\left.O_{\mu}^{\pm}\left(\Phi\right)\right|_{\textrm{linear}} & = & \mp\frac{ig\vartheta^{2}}{4}\left(W_{\mu}^{1}\pm iW_{\mu}^{2}\right)\pm\frac{i\vartheta}{2}\left(\partial_{\mu}\rho^{1}\pm i\partial_{\mu}\rho^{2}\right).\label{eq:thooft_operators-1}
\end{eqnarray}
This amazing property tells us that, at least at the tree level, these
operators describe the same particles of the elementary fields, but
with the additional advantage of being gauge-invariant.

\subsection{Localization of $\Phi$}

The non-locality of $\Phi\left(x\right)$ can be handle by introducing
an auxiliary Stueckelberg-type of field, $\xi\left(x\right)$.
By imposing the condition
\begin{eqnarray}
	\partial^{2}\xi & = & \partial_{\mu}X_{\mu},\label{eq:constraint_stueckelberg}
\end{eqnarray}
we obtain
\begin{eqnarray}
	\Phi\left(x\right) & = & e^{-\frac{ig'}{2}\xi\left(x\right)}\varphi\left(x\right).
\end{eqnarray}
Following the path integral quantization, the constraint (\ref{eq:constraint_stueckelberg}) can be implemented
at the quantum level by the general formula
\begin{eqnarray}
	F\left[\Phi_{non-local}\right] & = & \int D\xi\det\left(-\partial^{2}\right)\delta\left(\partial^{2}\xi-\partial_{\mu}X_{\mu}\right)F\left[\Phi_{local}\right]\label{eq:F_loval}
\end{eqnarray}
where $F$ is some functional of the fields. We easily see that (\ref{eq:F_loval})
holds due to the relation \begin{gather}
	\delta\left(\partial^{2}\xi-\partial_{\mu}X_{\mu}\right)=\frac{\delta\left(\xi-\frac{1}{\partial^{2}}\partial_{\mu}X_{\mu}\right)}{\det\left(-\partial^{2}\right)}.
\end{gather} 

The Dirac delta function and the determinant in (\ref{eq:F_loval})
can be exponentiated to obtain an effective local action. We can rewrite $\det\left(-\partial^{2}\right)$
as
\begin{eqnarray}
	\det\left(-\partial^{2}\right) & = & \int D\overline{\eta}D\eta\,e^{-\int d^{4}x\overline{\eta}\partial^{2}\eta},\label{eq:exponentiated_nonlocal}
\end{eqnarray}
where $\eta$ and $\overline{\eta}$ are ghost fields. The delta function
can be expressed as 
\begin{eqnarray}
	\delta\left(\partial^{2}\xi-\partial_{\mu}X_{\mu}\right) & = & \int D\theta\,e^{-\int d^{4}xi\theta\left(\partial^{2}\xi-\partial_{\mu}X_{\mu}\right)},\label{eq:exponentiated_nonlocal-1}
\end{eqnarray}
where $\theta$ is a Nakanishi-Lautrup-type of field. Thus, the effect
of localizing $\Phi$ is to introduce the following term in the starting
action:
\begin{eqnarray}
	S_{\xi} & = & \int d^{4}x\left[\overline{\eta}\partial^{2}\eta+i\theta\left(\partial^{2}\xi-\partial_{\mu}X_{\mu}\right)\right].
\end{eqnarray}

The local versions of the composite operators (\ref{eq:thooft_operators})
are
\begin{eqnarray}
	O_{\mu}^{3}\left(\Phi\right) & = & O_{\mu}^{3}\left(\varphi\right)-\frac{g'}{2}\left(O+\frac{v^{2}}{2}\right)\partial_{\mu}\xi,\nonumber \\
	O_{\mu}^{+}\left(\Phi\right) & = & e^{-ig'\xi}O_{\mu}^{+}\left(\varphi\right),\nonumber \\
	O_{\mu}^{-}\left(\Phi\right) & = & e^{ig'\xi}O_{\mu}^{-}\left(\varphi\right).\label{eq:thooft_operators_local}
\end{eqnarray}
Because of the residual Abelian symmetry, it also convenient to work
with the combination 
\begin{eqnarray*}
	O_{\mu}^{1} & = & \frac{O_{\mu}^{+}-O_{\mu}^{-}}{2i},\\
	O_{\mu}^{2} & = & \frac{O_{\mu}^{+}+O_{\mu}^{-}}{2},
\end{eqnarray*}
instead of $O_{\mu}^{+}$ and $O_{\mu}^{-}$. Thus, it follows that
\begin{eqnarray}
	O'{}_{\mu}^{\alpha} & = & R^{\alpha\beta}O_{\mu}^{\beta},\label{eq:rotarion_O_alpha}
\end{eqnarray}
with respect to the transformations (\ref{eq:SO1_tranf}).

The reader may wonder about the fundamental differences between $O_{\mu}$
and $O_{\mu}^{3}$, since they are quite similar. To clarify this
point, let us rewrite $O_{\mu}$ as
\begin{eqnarray}
	O_{\mu}\left(\varphi\right) & = & \frac{1}{2i}\left(\varphi^{\dagger}\left(\mathcal{D}_{\mu}-i\frac{g'}{2}X_{\mu}\right)\varphi-\left(\mathcal{D}_{\mu}\varphi-i\frac{g'}{2}X_{\mu}\varphi\right)^{\dagger}\varphi\right)\nonumber \\
	& = & \frac{1}{2i}\left(\varphi^{\dagger}\mathcal{D}_{\mu}\varphi-\left(\mathcal{D}_{\mu}\varphi\right)^{\dagger}\varphi\right)-\frac{g'}{2}\varphi^{\dagger}\varphi X_{\mu}\nonumber \\
	& = & O_{\mu}^{3}\left(\varphi\right)-\frac{g'}{2}\left(O+\frac{v^{2}}{2}\right)X_{\mu}.\label{eq:O_mu_exp}
\end{eqnarray}
Comparing (\ref{eq:O_mu_exp}) with (\ref{eq:thooft_operators_local}),
we see that $O_{\mu}$ has $X_{\mu}$ instead of $\partial_{\mu}\xi$.
On-shell, we have (\ref{eq:constraint_stueckelberg}), thus $\partial_{\mu}\xi=\frac{\partial_{\mu}\partial_{\nu}}{\partial^{2}}X_{\nu}$.
Therefore, $O_{\mu}^{3}$ is composed of the longitudinal part of
$X_{\mu}$ while $O_{\mu}$ has both components. This means that,
if we adopt the Landau gauge ($\partial_{\mu}X_{\mu}=0$), we expect
$\xi$ to decouple, and $O_{\mu}^{3}$ to be equivalent to $O_{\mu}^{3}\left(\varphi\right)=\frac{1}{2i}\left(\varphi^{\dagger}\mathcal{D}_{\mu}\varphi-\left(\mathcal{D}_{\mu}\varphi\right)^{\dagger}\varphi\right).$

\subsection{Additional gauge-invariant operators}

The need of renormalization forces us to consider all operators with
the same quantum numbers \cite{Collins:1984xc,Itzykson:1980rh}. Therefore, in addition to $O\left(\varphi\right)$,
we have to introduce 
\begin{gather}
	\vartheta^{2}.\label{eq:O_number}
\end{gather}
This is not a composite operator since $\vartheta$ is a parameter,
but it is also gauge-invariant and has the same dimension of $O$. 

Besides the operators $O_{\mu}\left(\varphi\right)$ and $O_{\mu}^{3}\left(\Phi\right)$,
we have the following singlet gauge-invariant vector operators
\begin{gather}
	\left(X_{\mu}-\partial_{\mu}\xi\right)O\left(\varphi\right),\,\left(X_{\mu}-\partial_{\mu}\xi\right)\vartheta^{2},\,\partial_{\mu}O\left(\varphi\right),\,\partial_{\nu}X_{\nu\mu}.\label{eq:O_mu_number}
\end{gather}
We called the operators singlets to differentiate them from $O_{\mu}^{\alpha}\left(\Phi\right)$,
which undergoes a special transformation with respect to the $SU\left(2\right)$
rotations, see Eq. (\ref{eq:rotarion_O_alpha}). It is worth noting
that $O_{\mu}^{3}\left(\Phi\right)$ can be written as a combination
of $O_{\mu}\left(\varphi\right)$, $\left(X_{\mu}-\partial_{\mu}\xi\right)O\left(\varphi\right)$
and $\left(X_{\mu}-\partial_{\mu}\xi\right)\vartheta^{2}$, see Eqs.
(\ref{eq:thooft_operators_local}) and (\ref{eq:O_mu_exp}). Thus,
we actually have only five independent operators. 

\subsection{BSRT structure of the localizing fields}

One remarkable aspect of this localization procedure is that there
is a BRST \cite{Becchi:1975nq,Tyutin:1975qk} structure behind it. We can introduce a BRST-type operator
$\mathscr{S}$, which generates the following transformations: 
\begin{gather}
	\mathscr{S}\overline{\eta}=i\theta,\quad\mathscr{S}\theta=0,\quad\mathscr{S}X_{\mu}=0,\quad\mathscr{S}\xi=-\eta,\quad\mathscr{S}\eta=0,\quad\mathscr{S}\left(\textrm{other fields}\right)=0.\label{eq:s_localizingg}
\end{gather}
Notice that $\mathscr{S}$ is nilpotent, \emph{i.e.}, 
\begin{eqnarray}
	\mathscr{S}^{2} & = & 0.
\end{eqnarray}

Now, the term that we introduced in the starting action to localize
$\Phi$ can be rewritten as a BRST-exact term, namely,
\begin{eqnarray}
	S_{\xi} & = & \mathscr{S}\int d^{4}x\left[\overline{\eta}\left(\partial^{2}\xi-\partial_{\mu}X_{\mu}\right)\right].
\end{eqnarray}
It easy to see that 
\begin{eqnarray}
	\mathscr{S}\left(S_{\textrm{Higgs}}+S_{\xi}\right) & = & 0.
\end{eqnarray}
From now on, we call this symmetry the $\mathscr{S}$-symmetry.
The action we will finish constructing has not this symmetry off-shell.
To calculate correlation functions of the composite operators
(\ref{eq:thooft_operators_local}), we have to introduce them in the
starting action coupled with external sources, and these operators
are not invariant under the transformations (\ref{eq:s_localizingg}).
We have that
\begin{gather}
	\mathscr{S}O_{\mu}^{3}\left(\Phi\right)=\frac{g'}{2}\left(O+\frac{v^{2}}{2}\right)\partial_{\mu}\eta,\quad\mathscr{S}O_{\mu}^{+}\left(\Phi\right)=ig'\eta e^{-ig'\xi}O_{\mu}^{+}\left(\varphi\right),\quad\mathscr{S}O_{\mu}^{-}\left(\Phi\right)=-ig'\eta e^{ig'\xi}O_{\mu}^{-}\left(\varphi\right).\label{eq:thooft_operators_local-1}
\end{gather}

\section{BRST symmetry and the gauge fixing\protect\label{sec:BRST-symmetry-and}}

\subsection{BRST operators}

To quantize this theory we have to fix the gauge. We can follow the
usual Faddeev-Popov method or, equivalently, the BRST quantization
procedure. We prefer the second way due to its simplicity and generality,
but first, in order to define the BRST transformation, let us identify
the infinitesimal gauge transformations of the fields we have so far.
From (\ref{eq:su2_gauge_tranf}), (\ref{eq:u1_gauge_transf}) and
(\ref{eq:constraint_stueckelberg}), we have for the $SU\left(2\right)$
symmetry
\begin{gather}
	\delta_{\omega}H=\frac{g}{2}\omega^{\alpha}\rho^{\alpha}+\frac{g}{2}\omega^{3}\rho^{3},\qquad\delta_{\omega}\rho^{a}=-\frac{g}{2}\omega^{a}\left(\vartheta+H\right)+\frac{g}{2}\varepsilon^{abc}\omega^{b}\rho^{c}\nonumber \\
	\delta_{\omega}W_{\mu}^{a}=-\partial_{\mu}\omega^{a}+g\varepsilon^{abc}\omega^{b}W_{\mu}^{c},\qquad\delta_{\omega}X_{\mu}=0,\nonumber \\
	\delta_{\omega}\xi=0,\qquad\delta_{\omega}\theta=0,\qquad\delta_{\omega}\eta=0,\qquad\delta_{\omega}\overline{\eta}=0,\label{eq:infinitesimal_gauge_transformations}
\end{gather}
and for the $U\left(1\right)$ symmetry
\begin{gather}
	\delta_{\omega}H=-\frac{g'}{2}\omega\rho^{3},\qquad\delta_{\omega}\rho^{3}=\frac{g'}{2}\omega\left(\vartheta+H\right),\qquad\delta_{\omega}\rho^{\alpha}=-\frac{g'}{2}\omega\varepsilon^{\alpha\beta}\rho^{\beta},\nonumber \\
	\delta_{\omega}W_{\mu}^{3}=0,\qquad\delta_{\omega}W_{\mu}^{\alpha}=0,\qquad\delta_{\omega}X_{\mu}=-\partial_{\mu}\omega,\qquad\delta_{\omega}\xi=-\omega,\nonumber \\
	\delta_{\omega}\theta=0,\qquad\delta_{\omega}\eta=0,\qquad\delta_{\omega}\overline{\eta}=0.\label{eq:infinitesimal_gauge_transformations-1}
\end{gather}
Now, let us introduce the Faddeev-Popov ghosts $\left\{ c^{\alpha},c^{3},c,\overline{c}^{\alpha},\overline{c}^{3},\overline{c}\right\} $
and the Nakanishi-Lautrup fields $\left\{ b^{\alpha},b^{3},b\right\} $.
Since there are two gauge symmetries, then it is possible to define
two sets of BRST transformations. For the $SU\left(2\right)$ group,
we have:
\begin{gather}
	s_{2}H=\frac{g}{2}c^{\alpha}\rho^{\alpha}+\frac{g}{2}c^{3}\rho^{3},\qquad s_{2}\rho^{a}=-\frac{g}{2}c^{a}\left(\vartheta+H\right)+\frac{g}{2}\varepsilon^{abc}c^{b}\rho^{c},\nonumber \\
	s_{2}W_{\mu}^{a}=-\partial_{\mu}c^{a}+g\varepsilon^{abc}c^{b}W_{\mu}^{c},\qquad s_{2}X_{\mu}=0,\nonumber \\
	s_{2}\xi=0,\qquad s_{2}\theta=0,\qquad s_{2}\eta=0,\qquad s_{2}\overline{\eta}=0,\nonumber \\
	s_{2}c^{a}=\frac{g}{2}\varepsilon^{abc}c^{b}c^{c},\qquad s_{2}\overline{c}^{a}=ib^{a},\qquad s_{2}b^{a}=0,\nonumber \\
	s_{2}c=0,\qquad s_{2}\overline{c}=0,\qquad s_{2}b=0,\label{eq:BRST_ransformation}
\end{gather}
where $s_{2}$ is the nilpotent $SU\left(2\right)$ BRST operator,
\emph{i.e.}
\begin{eqnarray}
	s_{2}^{2} & = & 0.\label{eq:BRTS_nilpotency}
\end{eqnarray}
Likewise, we have the following BRST transformations for the $U\left(1\right)$
group:
\begin{gather}
	s_{1}H=-\frac{g'}{2}c\rho^{3},\qquad s_{1}\rho^{3}=\frac{g'}{2}c\left(\vartheta+H\right),\qquad s_{1}\rho^{\alpha}=-\frac{g'}{2}\varepsilon^{\alpha\beta}c\rho^{\beta}\nonumber \\
	s_{1}W_{\mu}^{3}=0,\qquad s_{1}W_{\mu}^{\alpha}=0,\qquad s_{1}X_{\mu}=-\partial_{\mu}c,\nonumber \\
	s_{1}\xi=-c,\qquad s_{1}\theta=0,\qquad s_{1}\eta=0,\qquad s_{1}\overline{\eta}=0,\nonumber \\
	s_{1}c^{3}=0,\qquad s_{1}c^{\alpha}=0,\qquad s_{1}c=0,\nonumber \\
	s_{1}\overline{c}^{3}=0,\qquad s_{1}\overline{c}^{\alpha}=0,\qquad s_{1}\overline{c}=ib,\label{eq:BRST_ransformation-1}
\end{gather}
where $s_{1}$ is the nilpotent $U\left(1\right)$ BRST operator,
\emph{i.e.}
\begin{eqnarray}
	s_{1}^{2} & = & 0.\label{eq:BRTS_nilpotency-1}
\end{eqnarray}

Notice that the BRST transformations of the fields $\left\{ H,\rho^{3},\rho^{\alpha},W_{\mu}^{3},W_{\mu}^{\alpha},X_{\mu},\xi,\theta,\eta,\overline{\eta}\right\} $
are obtained by replacing $\omega^{\alpha}\rightarrow c^{\alpha}$,
$\omega^{3}\rightarrow c^{3}$ and $\omega\rightarrow c$ in (\ref{eq:infinitesimal_gauge_transformations})
and (\ref{eq:infinitesimal_gauge_transformations-1}). This observation
is very useful because it allows us to conclude that the gauge-invariant
operators presented in Section \ref{sec:Gauge-invariant-operators}
are also BRST invariant, as they do not contain Faddeev-Popov ghosts.
Also, we can conclude that the Higgs action (\ref{eq:higgs_action})
and the terms needed to localize $\Phi$ are BRST-invariant.

\subsection{Gauge fixing in the Landau gauge}

To fix the gauge, we employ here the Landau gauge, defined as
\begin{eqnarray}
	S_{gf} & = & s_{2}\int d^{4}x\left(\overline{c}^{a}\partial_{\mu}W_{\mu}^{a}\right)+s_{1}\int d^{4}x\left(\overline{c}\partial_{\mu}X_{\mu}\right)\nonumber \\
	& = & \int d^{4}x\left[ib^{a}\partial_{\mu}W_{\mu}^{a}-\overline{c}^{a}\partial_{\mu}\left(-\partial_{\mu}c^{a}+g\varepsilon^{abc}W_{\mu}^{c}c^{b}\right)+ib\partial_{\mu}X_{\mu}-\overline{c}\partial_{\mu}\left(-\partial_{\mu}c\right)\right].
\end{eqnarray}
Notice that the gauge fixing term is defined as a BRST-exact term,
\emph{i.e.} 
\begin{gather}
	S_{gf}=s_{1}\left(\int d^{4}x\ldots\right)+s_{2}\left(\int d^{4}x\ldots\right)
\end{gather}
Therefore, due to the nilpotency of $s_{i}$, it follows that
\begin{eqnarray}
	s_{i}S_{gf} & = & 0.
\end{eqnarray}
Thus, the BRST invariance of the theory is ensured. We could have
defined the gauge fixing in terms of $Z_{\mu}$ and $A_{\mu}$, but
this would amount to redefining the current Nakanishi-Lautrup fields
and ghosts. 

For theories with the Higgs mechanism, it is also convenient to use
the 't Hoof gauge \cite{tHooft:1971qjg}, which is renormalizable and free from
infrared divergencies. Although it introduces fictitious masses for
the Goldstones and the longitudinal components, the BRST quartet mechanism \cite{Kugo:1979gm}
guarantees that these unphysical modes are not present in the physical
space. Compared to the Landau gauge, the general 't Hoof gauge breaks
the global gauge symmetries. This is the main reason we decided to
use the Landau gauge. 

\section{Composite operators and the starting action\protect\label{sec:Composite-operators-and}}

\subsection{BRST-invariant operators\protect\label{subsec:BRST-invariant-operators}}

Our goal is to present a setup that allows us to compute finite correlation
functions of the composite operators $\left\{ O,\,O_{\mu},\,O_{\mu}^{3},\,O_{\mu}^{\alpha}\right\} $.
To obtain this, we introduce each one of these operators and the operators
of Eqs. (\ref{eq:O_number}), (\ref{eq:O_mu_number}) in the starting
action coupled with an external source \footnote{The reader may consult the textbooks \cite{Collins:1984xc} or \cite{Zinn-Justin:1989rgp} to understand why a composite operator like \(\varphi^2(x)\) cannot consistently be considered as \(\lim_{x \rightarrow y} \varphi(x) \varphi(y)\). Essentially, this occurs because new UV divergences are introduced in the theory. Hence, the only systematic way to define correlation functions of composite operators is by introducing them coupled with an external source.}
:
\begin{eqnarray}
	S_{O} & = & \int d^{4}x\left[JO\left(\varphi\right)+\Xi\vartheta^{2}+\Omega_{\mu}O_{\mu}\left(\varphi\right)+\Omega_{\mu}^{3}O_{\mu}^{3}\left(\Phi\right)\right.\nonumber \\
	&  & \left.+\Omega_{\mu}^{4}\partial_{\mu}O\left(\varphi\right)+\Omega_{\mu}^{5}\left(X_{\mu}-\partial_{\mu}\xi\right)\vartheta^{2}+\Omega_{\mu}^{6}\partial_{\nu}X_{\nu\mu}+\Omega_{\mu}^{\alpha}O_{\mu}^{\alpha}\left(\Phi\right)\right].
\end{eqnarray}
This implies that, for example, the generating functional of the connected
Green's functions
is also a functional of the external sources
$J_{O}=\left\{ J,\,\Xi,\,\Omega_{\mu},\,\Omega_{\mu}^{3},\,\Omega_{\mu}^{4},\,\Omega_{\mu}^{5},\,\Omega_{\mu}^{6},\,\Omega_{\mu}^{\alpha}\right\} $,  namely,
\begin{equation}
	W[J_{i},J_{O}]=- \ln \frac {\int D\phi \exp (-S_{\text{Higgs}}-S_{\xi}-S_{gf}-\int d^4 x \, J_i \phi_i -S_O  )}{\int D\phi \exp (-S_{\text{Higgs}}-S_{\xi}-S_{gf})},
\end{equation}
where $D\phi\equiv DW_{\mu}^aDX_{\mu}DHD\rho^a Dc^aD\bar{c}^a Dc D\bar{c}Db^aDb D\eta D\bar{\eta} D\xi D\theta$. One insertion of a certain composite operator is obtained by deriving
$W$ with respect to the corresponding source and taking all external
sources to zero in the end. For instance, one insertion of the composite
operator $O\left(\varphi\right)$ at the point $x$ is obtained by
\begin{eqnarray}
	\left\langle O\left(x\right)\ldots\right\rangle  & = & \left.\frac{\delta^{n}W}{\delta J\left(x\right)\ldots}\right|_{\textrm{all sources=0}}.
\end{eqnarray}

In the next section, we discuss the Ward identities of this model.
There, we present a very interesting non-integrated Ward identity
associated with the residual $U_{EM}\left(1\right)$ symmetry. Since
this is a global symmetry, we would only expect an integrated identity.
However, if we ignore the gauge fixing, the operator $O_{\mu}\left(\varphi\right)$
is precisely the current associated with this symmetry, \emph{i.e}.,
\begin{eqnarray}
	\varepsilon^{\alpha\beta}\left(\rho^{\beta}\frac{\delta S_{\textrm{Higgs}}}{\delta\rho^{\alpha}}+W_{\mu}^{\beta}\frac{\delta S_{\textrm{Higgs}}}{\delta W_{\mu}^{\alpha}}\right) & = & -\partial_{\mu}O_{\mu}+\frac{1}{g}\partial_{\rho}\frac{\delta S_{\textrm{Higgs}}}{\delta W_{\rho}^{3}}.
\end{eqnarray}
Therefore, by introducing $O_{\mu}$ coupled with a source in the
starting action, we may get a local symmetry and, consequently, a
non-integrated Ward identity. If we also take into account the gauge
fixing term, in addition to the gauge-invariant current $O_{\mu}$,
we obtain the BRST-exact current
\begin{gather}
	s_{2}\left(\varepsilon^{\alpha\beta}\overline{c}^{\alpha}W_{\mu}^{\beta}\right)=\varepsilon^{\alpha\beta}ib^{\alpha}W_{\mu}^{\beta}-\varepsilon^{\alpha\beta}\overline{c}^{\alpha}\left(-\partial_{\mu}c^{\beta}+g\varepsilon^{\beta\gamma}c^{\gamma}W_{\mu}^{3}-g\varepsilon^{\beta\gamma}c^{3}W_{\mu}^{\gamma}\right).
\end{gather}
This composite operator has the same quantum numbers of $O_{\mu}$,
so we could expect it to appear in the renormalization of this operator.
This does not happen because, differently than $s_{2}\left(\varepsilon^{\alpha\beta}\overline{c}^{\alpha}W_{\mu}^{\beta}\right)$,
$O_{\mu}$ is not a BRST exact term. Despite this, as will soon become
clear, we introduce $s_{2}\left(\varepsilon^{\alpha\beta}\overline{c}^{\alpha}W_{\mu}^{\beta}\right)$
in the starting action, also coupled with a source. Giving the BRST-exact
nature of this operator, it is convenient to take the BRST doublet
of sources:
\begin{gather}
	s_{2}\Upsilon_{\mu}^{a}=\zeta_{\mu}^{a},\qquad s_{2}\zeta^{a}=0,\qquad s_{1}\Upsilon_{\mu}^{a}=0,\qquad s_{1}\zeta_{\mu}^{a}=0,\label{eq:doublet_sources}
\end{gather}
and then introduce the composite operators
\begin{eqnarray}
	S_{\Upsilon} & = & s_{2}\int d^{4}x\left(\varepsilon^{abc}\Upsilon_{\mu}^{a}\overline{c}^{b}W_{\mu}^{c}\right)\nonumber \\
	& = & \int d^{4}x\left(\varepsilon^{abc}\zeta_{\mu}^{a}\overline{c}^{b}W_{\mu}^{c}+i\varepsilon^{abc}\Upsilon_{\mu}^{a}b^{b}W_{\mu}^{c}+\varepsilon^{abc}\Upsilon_{\mu}^{a}\overline{c}^{b}D_{\mu}^{cd}c^{d}\right).\label{eq:s_upsilon}
\end{eqnarray}
We choose this particular combination to keep the global rigid $SU\left(2\right)$
symmetry. 

\subsection{Non-linear BRST transformations}

We also want to extend the BRST symmetries to the level of Green's
functions. In other words, we wish to obtain Ward identities associated
with the BRST symmetries, which are the Slavnov-Taylor identities.
To do that, we have to account for the non-linear transformations
of the fields. The right procedure is also to treat these transformations
as composite operators and introduce each one of them coupled with
an external source \cite{Piguet:1995er}. Here, we define this term as
\begin{eqnarray}
	S_{s} & = & \int d^{4}x\left(K_{\mu}^{\alpha}s_{2}W^{\alpha}+K_{\mu}^{3}s_{2}W^{3}+Ys_{2}H+R^{\alpha}s_{2}\rho^{\alpha}+R^{3}s_{2}\rho^{3}+L^{\alpha}s_{2}c^{\alpha}+L^{3}s_{2}c^{3}\right)\nonumber \\
	&  & +\int d^{4}x\left(ys_{1}H+r^{\alpha}s_{1}\rho^{\alpha}+r^{3}s_{1}\rho^{3}\right),
\end{eqnarray}
where $\left\{ K_{\mu}^{\alpha},\,K_{\mu}^{3},\,Y,\,R^{\alpha},\,R^{3},\,L^{\alpha},\,L^{3},\,y,\,r^{\alpha},\,r^{3}\right\} $
are the external sources.

\subsection{The starting action $\Sigma$}

Therefore for now we will consider the classical starting action as
being
\begin{eqnarray}
	\Sigma & = & S_{\textrm{Higgs}}+S_{gf}+S_{s}+S_{\xi}+S_{O}+S_{\Upsilon}.\label{eq:starting_action-1}
\end{eqnarray}
This action is BRST-invariant if, in addition to the BRST transformations
(\ref{eq:BRST_ransformation}) and (\ref{eq:doublet_sources}), we
have
\begin{gather}
	s_{i}J=s_{i}\Xi=s_{i}\Omega_{\mu}=s_{i}\Omega_{\mu}^{3}=s_{i}\Omega_{\mu}^{4}=s_{i}\Omega_{\mu}^{5}=s_{i}\Omega_{\mu}^{6}=s_{i}\Omega_{\mu}^{\alpha}=0,\nonumber \\
	s_{i}K_{\mu}^{\alpha}=s_{i}K_{\mu}^{3}=s_{i}Y=s_{i}R^{\alpha}=s_{i}R^{3}=s_{i}L^{\alpha}=s_{i}L^{3}=s_{i}y=s_{i}r^{\alpha}=s_{i}r^{3}=0.
\end{gather}

\subsection{Partially restoring the local $SU\left(2\right)$ Symmetry\protect\label{subsec:Local--Symmetry}}

In Subsection \ref{subsec:BRST-invariant-operators}, we introduced
the term $S_{\Upsilon}$ without providing a compelling motivation.
We ruled out the renormalization of operator $O_{\mu}$, because it
has a different BRST nature. Therefore, here we aim to fill this gap.
What occurs is that, when combined with the gauge fixing term $S_{gf}$,
$S_{\Upsilon}$ yields additional, very interesting results, similar
to the Background Field Method (BFM) \cite{DeWitt:1964mxt,DeWitt:1980jv}. This is because the
local $SU\left(2\right)$ symmetry is partially restored when $S_{\Upsilon}$
is included. The source $\Upsilon_{\mu}^{a}$ shares similarities
with the background field, as both follow the same $SU\left(2\right)$
gauge transformations necessary to recover the $SU(2)$ gauge symmetry. 

Summing $S_{gf}$ and $S_{\Upsilon}$, we obtain
\begin{eqnarray}
	S_{gf}+S_{\Upsilon} & = & \int d^{4}x\left(ib^{a}\partial_{\mu}W_{\mu}^{a}-\left(\partial_{\mu}\overline{c}^{a}\right)D_{\mu}^{ab}c^{b}+ib\partial_{\mu}X_{\mu}+\overline{c}\partial^{2}c\right)\nonumber \\
	&  & +\int d^{4}x\left(\varepsilon^{abc}\zeta_{\mu}^{a}\overline{c}^{b}W_{\mu}^{c}+i\varepsilon^{abc}\Upsilon_{\mu}^{a}b^{b}W_{\mu}^{c}+\varepsilon^{abc}\Upsilon_{\mu}^{a}\overline{c}^{b}D_{\mu}^{cd}c^{d}\right)\nonumber \\
		& = & \int d^{4}x\left(ib^{a}\tilde{D}_{\mu}^{ab}W_{\mu}^{b}-\tilde{D}_{\mu}^{ac}\overline{c}^{c}D_{\mu}^{ab}c^{b}+ib\partial_{\mu}X_{\mu}+\overline{c}\partial^{2}c+\varepsilon^{abc}\zeta_{\mu}^{a}\overline{c}^{b}W_{\mu}^{c}\right).\label{eq:gf_upsilon}
\end{eqnarray}
To write (\ref{eq:gf_upsilon}), we introduced a new covariant derivative
with the help of the source $\Upsilon_{\mu}^{a}$, namely,
\begin{eqnarray}
	\tilde{D}_{\mu}^{ab} & = & \left(\partial_{\mu}\delta^{ab}+\varepsilon^{abc}\Upsilon_{\mu}^{c}\right).
\end{eqnarray}
Notice that the source $\Upsilon_{\mu}^{a}$ plays the role of a \emph{connection}
in $\tilde{D}_{\mu}^{ab}$. This way of expressing $S_{gf}+S_{\Upsilon}$
suggests that if we add to (\ref{eq:infinitesimal_gauge_transformations})
and (\ref{eq:infinitesimal_gauge_transformations-1}) the gauge transformation
\begin{eqnarray}
	\delta_{\omega}\Upsilon_{\mu}^{a} & = & g\left(\partial_{\mu}\omega^{a}+\varepsilon^{abc}\Upsilon_{\mu}^{c}\omega^{b}\right),
\end{eqnarray}
in addition to 
\begin{gather*}
	\delta_{\omega}c^{a}=g\varepsilon^{abc}\omega^{b}c^{c},\\
	\delta_{\omega}\overline{c}^{a}=g\varepsilon^{abc}\omega^{b}\overline{c}^{c},\\
	\delta_{\omega}b^{a}=g\varepsilon^{abc}\omega^{b}b^{c},\\
	\delta_{\omega}\zeta_{\mu}^{a}=g\varepsilon^{abc}\omega^{b}\zeta_{\mu}^{c},\\
	\delta_{\omega}\left(\textrm{other fields and sources}\right)=0,
\end{gather*}
the local $SU\left(2\right)$ gauge symmetry is \emph{almost }restored.
We have that
\begin{eqnarray}
	\delta_{\omega}\left(S_{\textrm{Higgs}}+S_{gf}+S_{\xi}+S_{O}+S_{\Upsilon}\right) & = & \int d^{4}x\,\omega^{a}\left[-i\partial^{2}b^{a}+i\varepsilon^{abc}\partial_{\mu}\left(\Upsilon_{\mu}^{b}b^{c}\right)+\varepsilon^{abc}\partial_{\mu}\left(\zeta_{\mu}^{b}\overline{c}^{b}\right)\right]\nonumber \\
	& = & \int d^{4}x\,\omega^{a}s\left[-\partial^{2}\overline{c}^{a}+\varepsilon^{abc}\partial_{\mu}\left(\Upsilon_{\mu}^{b}\overline{c}^{c}\right)\right],
\end{eqnarray}
which means that the symmetry is linearly broken. A linearly broken
symmetry at the quantum level can still be very relevant and useful
for establishing relationships between correlation functions \cite{Piguet:1995er}. We discuss
this symmetry further in Subsection \ref{subsec:Ghost-equations-and}. \par 
In a future article, we will present a slightly different approach to addressing symmetry breaking \emph{via} composite operators, which is equivalent to the BFM. In this approach, the local symmetry can be fully restored. 

\section{Symmetries and Ward identities\protect\label{sec:Symmetries-and-Ward}}

The next step towards renormalizing these BRST-invariant operators
is to determine the symmetry content of the starting action. Once
we have done this, we can employ the Algebraic Renormalization approach
\cite{Piguet:1995er} based on the Quantum Action principle and the general theory
of renormalization to establish the counterterms needed to renormalize
any correlation function. 

In the following subsections we present the Ward identities of the
theory defined by the action (\ref{eq:starting_action-1}).

\subsection{$SU\left(2\right)$ BRST symmetry and its Slavnov-Taylor identity}

The action $\Sigma$ is invariant with respect to the $SU\left(2\right)$
BRST symmetry transformations:
\begin{gather}
	s_{2}H=\frac{g}{2}c^{\alpha}\rho^{\alpha}+\frac{g}{2}c^{3}\rho^{3},\qquad s_{2}\rho^{a}=-\frac{g}{2}c^{a}\left(\vartheta+H\right)+\frac{g}{2}\varepsilon^{abc}c^{b}\rho^{c},\nonumber \\
	s_{2}W_{\mu}^{a}=-\partial_{\mu}c^{a}+g\varepsilon^{abc}c^{b}W_{\mu}^{c},\qquad s_{2}c^{a}=\frac{g}{2}\varepsilon^{abc}c^{b}c^{c},\nonumber \\
	s_{2}\overline{c}^{a}=ib^{a},\qquad s_{2}b^{a}=0,\nonumber \\
	s_{2}\Upsilon_{\mu}^{a}=\zeta_{\mu}^{a},\qquad s_{2}\zeta_{\mu}^{a}=0,\nonumber \\
	s_{2}\left(\textrm{other fields and sources}\right)=0.
\end{gather}
The $SU\left(2\right)$ BRST symmetry can be expressed by the functional
Slavnov-Taylor identity
\begin{eqnarray}
	\mathcal{S}_{2}\left(\Sigma\right) & = & 0,
\end{eqnarray}
where the $SU\left(2\right)$ Slavnov-Taylor operator is defined by
\begin{eqnarray}
	\mathcal{S}_{2}\left(\Sigma\right) & = & \int d^{4}x\left(\frac{\delta\Sigma}{\delta K_{\mu}^{\alpha}}\frac{\delta\Sigma}{\delta W_{\mu}^{\alpha}}+\frac{\delta\Sigma}{\delta K_{\mu}^{3}}\frac{\delta\Sigma}{\delta W_{\mu}^{3}}+\frac{\delta\Sigma}{\delta Y}\frac{\delta\Sigma}{\delta H}+\frac{\delta\Sigma}{\delta R^{\alpha}}\frac{\delta\Sigma}{\delta\rho^{\alpha}}+\frac{\delta\Sigma}{\delta R^{3}}\frac{\delta\Sigma}{\delta\rho^{3}}\right.\nonumber \\
	&  & \left.+\frac{\delta\Sigma}{\delta L^{\alpha}}\frac{\delta\Sigma}{\delta c^{\alpha}}+\frac{\delta\Sigma}{\delta L^{3}}\frac{\delta\Sigma}{\delta c^{3}}+ib^{\alpha}\frac{\delta\Sigma}{\delta\overline{c}^{\alpha}}+ib^{3}\frac{\delta\Sigma}{\delta\overline{c}^{3}}+\zeta_{\mu}^{\alpha}\frac{\delta\Sigma}{\delta\Upsilon_{\mu}^{\alpha}}+\zeta_{\mu}^{3}\frac{\delta\Sigma}{\delta\Upsilon_{\mu}^{3}}\right).\label{eq:slavnov_taylor}
\end{eqnarray}

\subsection{$U\left(1\right)$ BRST symmetry and its Slavnov-Taylor identity}

The action $\Sigma$ is invariant with respect to the $U\left(1\right)$
BRST symmetry transformations:
\begin{gather}
	s_{1}H=-\frac{g'}{2}c\rho^{3},\qquad s_{1}\rho^{3}=\frac{g'}{2}c\left(\vartheta+H\right),\qquad s_{1}\rho^{\alpha}=-\frac{g'}{2}\varepsilon^{\alpha\beta}c\rho^{\beta}\nonumber \\
	s_{1}X_{\mu}=-\partial_{\mu}c,\qquad s_{1}\xi=-c,\nonumber \\
	s_{1}\theta=0,\qquad s_{1}\eta=0,\nonumber \\
	s_{1}\overline{\eta}=0,\qquad s_{1}c=0,\nonumber \\
	s_{1}\overline{c}=ib,\qquad s_{1}b=0,\nonumber \\
	s_{1}\left(\textrm{other fields and sources}\right)=0.
\end{gather}
Likewise the $SU\left(2\right)$ BRST symmetry, the $U\left(1\right)$
BRST symmetry can be expressed through a functional Slavnov-Taylor
identity, namely,
\begin{eqnarray}
	\mathcal{S}_{1}\left(\Sigma\right) & = & 0,
\end{eqnarray}
where the $U\left(1\right)$ Slavnov-Taylor operator is defined by
\begin{eqnarray}
	\mathcal{S}_{1}\left(\Sigma\right) & = & \int d^{4}x\left(-\partial_{\mu}c\frac{\delta\Sigma}{\delta X_{\mu}}+\frac{\delta\Sigma}{\delta y}\frac{\delta\Sigma}{\delta H}+\frac{\delta\Sigma}{\delta r^{\alpha}}\frac{\delta\Sigma}{\delta\rho^{\alpha}}+\frac{\delta\Sigma}{\delta r^{3}}\frac{\delta\Sigma}{\delta\rho^{3}}+ib\frac{\delta\Sigma}{\delta\overline{c}}-c\frac{\delta\Sigma}{\delta\xi}\right).\label{eq:slavnov_taylor-1}
\end{eqnarray}

\subsection{Equation of motion of $X_{\mu}$}

As $\left[s_{i},\frac{\delta}{\delta X_{\mu}}\right]=0$ and $\Sigma$
is BRST-invariant, the equation of motion of the Abelian gauge field
should also be BRST-invariant, \emph{i.e.},
\begin{eqnarray}
	s_{i}\frac{\delta\Sigma}{\delta X_{\mu}} & = & 0.
\end{eqnarray}
Indeed, we do find it to be the case:
\begin{eqnarray}
	\frac{\delta\Sigma}{\delta X_{\mu}} & = & -\partial_{\nu}X_{\nu\mu}+i\partial_{\mu}\theta-g'O_{\mu}-i\partial_{\mu}b-\frac{g'}{2}\Omega_{\mu}O-\frac{g'\vartheta^{2}}{4}\Omega_{\mu}\nonumber \\
	&  & +\Omega_{\mu}^{5}\vartheta^{2}+\partial^{2}\Omega_{\mu}^{6}-\partial_{\mu}\partial_{\nu}\Omega_{\nu}^{6}.\label{eq:X_equation_motion}
\end{eqnarray}
Note on the righ hand side of (\ref{eq:X_equation_motion}), the presence
of the BRST-invariant operators $O_{\mu}$ and $O$. Therefore, we
can write the non-integrated Ward identity
\begin{eqnarray}
	\frac{\delta\Sigma}{\delta X_{\mu}}+g'\frac{\delta\Sigma}{\delta\Omega_{\mu}}+\frac{g'}{2}\Omega_{\mu}\frac{\delta\Sigma}{\delta J} & = & -\partial_{\nu}X_{\nu\mu}+i\partial_{\mu}\theta-i\partial_{\mu}b-\frac{g'\vartheta^{2}}{4}\Omega_{\mu}+\Omega_{\mu}^{5}\vartheta^{2}+\partial^{2}\Omega_{\mu}^{6}-\partial_{\mu}\partial_{\nu}\Omega_{\nu}^{6}.\label{eq:X_ward_identity}
\end{eqnarray}
Notice that, in the right hand side of (\ref{eq:X_ward_identity}),
we have only linear terms in the fields. This linear terms do not
receive quantum corrections, so this is a truly Ward identity \cite{Piguet:1995er}. 

\subsection{Equation of motion of the localizing fields and the on-shell $\mathscr{S}$-symmetry}
\begin{itemize}
	\item Equation of $\xi$
\end{itemize}
In the absence of the composite operators ($\Omega_{\mu}^{\alpha}=\Omega_{\mu}^{3}=0$),
we would have
\begin{eqnarray}
	\frac{\delta\Sigma}{\delta\xi} & = & i\partial^{2}\theta.
\end{eqnarray}
Such Ward identity says that $\xi$ is a non-interacting field. Due
to the particular form of the composite operators $O_{\mu}^{3}\left(\Phi\right)$
and $O_{\mu}^{\alpha}\left(\Phi\right)$, we can also write down a
Ward identity, namely,
\begin{eqnarray}
	\frac{\delta\Sigma}{\delta\xi}-\frac{g'}{2}\partial_{\mu}\left(\Omega_{\mu}^{3}\frac{\delta\Sigma}{\delta J}\right)+g'\varepsilon^{\alpha\beta}\Omega_{\mu}^{\alpha}\frac{\delta\Sigma}{\delta\Omega_{\mu}^{\beta}} & = & i\partial^{2}\theta+g'\frac{\vartheta^{2}}{4}\partial_{\mu}\Omega_{\mu}^{3}+\vartheta^{2}\partial_{\mu}\Omega_{\mu}^{5}.\label{eq:xi_equation}
\end{eqnarray}
This Ward identity imposes important limitations on the possible combinations
of $\xi$ we find in the counterterm, as we will show. 
\begin{itemize}
	\item Equation of motion of $\theta$
\end{itemize}
The equation of motion of $\theta$ leads to the constraint (\ref{eq:constraint_stueckelberg}),
\emph{i.e.,}
\begin{eqnarray}
	\frac{\delta\Sigma}{\delta\theta} & = & i\left(\partial^{2}\xi-\partial_{\mu}X_{\mu}\right).\label{eq:teta_equation}
\end{eqnarray}
so it is also a Ward identity with a linear breaking. It states that
$\theta$ is a non-interacting field.
\begin{itemize}
	\item Equations of motion of $\eta$ and $\overline{\eta}$
\end{itemize}
The ghost fields $\eta$ and $\overline{\eta}$ introduced to localize
$\det\left(-\partial^{2}\right)$ are non-interacting fields. Indeed,
their equations of motions have linear breaking terms, namely,
\begin{eqnarray}
	\frac{\delta\Sigma}{\delta\eta} & = & -\partial^{2}\overline{\eta},\nonumber \\
	\frac{\delta\Sigma}{\delta\overline{\eta}} & = & \partial^{2}\eta,\label{eq:eta_equations}
\end{eqnarray}
so they are Ward identities.
\begin{itemize}
	\item Ghost number 
\end{itemize}
The ghost fields $\eta$ and $\overline{\eta}$ also appear in pair,
thus it is possible to define a ghost number to them. The conservation
of this ghost number can be expressed by the functional equation
\begin{eqnarray}
	\int d^{4}x\left(\eta\frac{\delta\Sigma}{\delta\eta}-\overline{\eta}\frac{\delta\Sigma}{\delta\overline{\eta}}\right) & = & 0.
\end{eqnarray}

\begin{itemize}
	\item On-shell $\mathscr{S}$-symmetry
\end{itemize}
The $\mathscr{S}$ operator can be written in a functional way, namely,
\begin{eqnarray}
	\mathscr{S} & = & \int d^{4}x\left(i\theta\frac{\delta}{\delta\overline{\eta}}-\eta\frac{\delta}{\delta\xi}\right).
\end{eqnarray}
Thus, the on-shell $\mathscr{S}$-symmetry can be expressed as
\begin{eqnarray*}
	\mathscr{S}\left.\Sigma\right|_{\Omega_{\mu}^{3}=\Omega_{\mu}^{\alpha}=0} & = & 0.
\end{eqnarray*}

\subsection{Equations of motion of the Nakanishi-Lautrup fields }

As we are using the Landau gauge, which is a linear gauge, the equations
of motion of the Nakanishi-Lautrup fields are Ward identities:
\begin{eqnarray}
	\frac{\delta\Sigma}{\delta b^{\alpha}} & = & i\partial_{\mu}W_{\mu}^{\alpha}-i\varepsilon^{\alpha\beta}\Upsilon_{\mu}^{\beta}W_{\mu}^{3}+i\varepsilon^{\alpha\beta}\Upsilon_{\mu}^{3}W_{\mu}^{\beta},\nonumber \\
	\frac{\delta\Sigma}{\delta b^{3}} & = & i\partial_{\mu}W_{\mu}^{3}-i\varepsilon^{\alpha\beta}\Upsilon_{\mu}^{\alpha}W_{\mu}^{\beta},\nonumber \\
	\frac{\delta\Sigma}{\delta b} & = & i\partial_{\mu}X_{\mu}.\label{eq:b_equations}
\end{eqnarray}
They establish that $b^{\alpha}$, $b^{3}$ and $b$ are non-interacting
fields. 

\subsection{Antighost equations}

The antighosts fields $\left\{ \overline{c}^{\alpha},\,\overline{c}^{3},\,\overline{c}\right\} $
and the Nakanishi-Lautrup fields $\left\{ b^{\alpha},\,b^{3},\,b\right\} $
have a doublet structure, 
\begin{gather}
	s_{2}\overline{c}^{\alpha}=ib^{\alpha},\quad s_{2}b^{\alpha}=0,\nonumber \\
	s_{2}\overline{c}^{3}=ib^{3},\quad s_{2}b^{3}=0,\nonumber \\
	s_{1}\overline{c}=ib,\quad s_{1}b=0.
\end{gather}
It implies in the following relations with the Slavnov-Taylor operators:
\begin{eqnarray}
	\frac{\delta}{\delta b^{a}}\mathcal{S}_{2}\left(\Sigma\right) & = & \mathcal{S}_{2\,\Sigma}\left(\frac{\delta F}{\delta b^{a}}\right)+i\frac{\delta\Sigma}{\delta\overline{c}^{a}},\nonumber \\
	\frac{\delta}{\delta b}\mathcal{S}_{1}\left(\Sigma\right) & = & \mathcal{S}_{1\,\Sigma}\left(\frac{\delta F}{\delta b}\right)+i\frac{\delta\Sigma}{\delta\overline{c}},\label{eq:commutator_b_slavnov}
\end{eqnarray}
where
\begin{eqnarray}
	\mathcal{S}_{2\,\Sigma} & = & \int d^{4}x\left(\frac{\delta\Sigma}{\delta K_{\mu}^{\alpha}}\frac{\delta}{\delta W_{\mu}^{\alpha}}+\frac{\delta\Sigma}{\delta W_{\mu}^{\alpha}}\frac{\delta}{\delta K_{\mu}^{\alpha}}+\frac{\delta\Sigma}{\delta K_{\mu}^{3}}\frac{\delta}{\delta W_{\mu}^{3}}+\frac{\delta\Sigma}{\delta W_{\mu}^{3}}\frac{\delta}{\delta K_{\mu}^{3}}\right.\nonumber \\
	&  & +\frac{\delta\Sigma}{\delta Y}\frac{\delta}{\delta H}+\frac{\delta\Sigma}{\delta H}\frac{\delta}{\delta Y}+\frac{\delta\Sigma}{\delta R^{\alpha}}\frac{\delta}{\delta\rho^{\alpha}}+\frac{\delta\Sigma}{\delta\rho^{\alpha}}\frac{\delta}{\delta R^{\alpha}}+\frac{\delta\Sigma}{\delta R^{3}}\frac{\delta}{\delta\rho^{3}}+\frac{\delta\Sigma}{\delta\rho^{3}}\frac{\delta}{\delta R^{3}}\nonumber \\
	&  & +\frac{\delta\Sigma}{\delta L^{\alpha}}\frac{\delta}{\delta c^{\alpha}}+\frac{\delta\Sigma}{\delta c^{\alpha}}\frac{\delta}{\delta L^{\alpha}}+\frac{\delta\Sigma}{\delta L^{3}}\frac{\delta}{\delta c^{3}}+\frac{\delta\Sigma}{\delta c^{3}}\frac{\delta}{\delta L^{3}}\nonumber \\
	&  & \left.+ib^{\alpha}\frac{\delta}{\delta\overline{c}^{\alpha}}+ib^{3}\frac{\delta}{\delta\overline{c}^{3}}+\zeta_{\mu}^{\alpha}\frac{\delta}{\delta\Upsilon_{\mu}^{\alpha}}+\zeta_{\mu}^{3}\frac{\delta}{\delta\Upsilon_{\mu}^{3}}\right)\label{eq:linearized_slavnov_taylor}
\end{eqnarray}
and
\begin{eqnarray}
	\mathcal{S}_{1\,\Sigma} & = & \int d^{4}x\left(-\partial_{\mu}c\frac{\delta}{\delta X_{\mu}}+\frac{\delta\Sigma}{\delta y}\frac{\delta}{\delta H}+\frac{\delta\Sigma}{\delta H}\frac{\delta}{\delta y}+\frac{\delta\Sigma}{\delta r^{\alpha}}\frac{\delta}{\delta\rho^{\alpha}}\right.\nonumber \\
	&  & \left.+\frac{\delta\Sigma}{\delta\rho^{\alpha}}\frac{\delta}{\delta r^{\alpha}}+\frac{\delta\Sigma}{\delta r^{3}}\frac{\delta}{\delta\rho^{3}}+\frac{\delta\Sigma}{\delta\rho^{3}}\frac{\delta}{\delta r^{3}}-c\frac{\delta}{\delta\xi}+ib\frac{\delta}{\delta\overline{c}}\right)\label{eq:linearized_slavnov_taylor-1}
\end{eqnarray}
are the linearized Slavnov-Taylor operators. Therefore, from (\ref{eq:b_equations}),
(\ref{eq:commutator_b_slavnov}) and (\ref{eq:slavnov_taylor}), we
have the antighost Ward identities
\begin{eqnarray}
	\frac{\delta\Sigma}{\delta\overline{c}^{\alpha}}+\partial_{\mu}\frac{\delta\Sigma}{\delta K_{\mu}^{\alpha}}-\varepsilon^{\alpha\beta}\Upsilon_{\mu}^{\beta}\frac{\delta\Sigma}{\delta K_{\mu}^{3}}+\varepsilon^{\alpha\beta}\Upsilon_{\mu}^{3}\frac{\delta\Sigma}{\delta K_{\mu}^{\beta}} & = & \varepsilon^{\alpha\beta}\zeta_{\mu}^{\beta}W_{\mu}^{3}-\varepsilon^{\alpha\beta}\zeta_{\mu}^{3}W_{\mu}^{\beta},\nonumber \\
	\frac{\delta\Sigma}{\delta\overline{c}^{3}}+\partial_{\mu}\frac{\delta\Sigma}{\delta K_{\mu}^{3}}-\varepsilon^{\alpha\beta}\Upsilon_{\mu}^{\alpha}\frac{\delta\Sigma}{\delta K_{\mu}^{\beta}} & = & \varepsilon^{\alpha\beta}\zeta_{\mu}^{\alpha}W_{\mu}^{\beta},\nonumber \\
	\frac{\delta\Sigma}{\delta\overline{c}} & = & \partial^{2}c.\label{eq:antighost_equation}
\end{eqnarray}

\subsection{Ghost equations and Rigid symmetries\protect\label{subsec:Ghost-equations-and}}
\begin{itemize}
	\item Non-integrated equation of $c^{\alpha}$
\end{itemize}
One of the great features of the Landau gauge is the existence of
a Ward identity associated with the ghost equation of motion. As we
will show, this is not an accident. The Landau gauge preserves the
global gauge symmetry of the theory, which is also called rigid symmetry.
Since this is a global symmetry, we expect to have only an integrated
Ward identity. However, we introduced a set of local composite operators
in the starting action, and some of them are currents, so we may have
a local symmetry too.

If we set \emph{all }external sources to zero, it follows that
\begin{gather}
	\frac{\delta\Sigma}{\delta c^{\alpha}}-ig\varepsilon^{\alpha\beta}\overline{c}^{\beta}\frac{\delta\Sigma}{\delta b^{3}}+ig\varepsilon^{\alpha\beta}\overline{c}^{3}\frac{\delta\Sigma}{\delta b^{\beta}}=\partial_{\mu}\mathcal{\mathscr{J}}_{g}^{\alpha},
\end{gather}
where
\begin{eqnarray}
	\mathcal{\mathscr{J}}_{g}^{\alpha} & = & -g\varepsilon^{\alpha\beta}\overline{c}^{3}W_{\mu}^{\beta}+g\varepsilon^{\alpha\beta}\overline{c}^{\beta}W_{\mu}^{3}-\partial_{\mu}\overline{c}^{\alpha}.
\end{eqnarray}
Therefore, $\mathcal{\mathscr{J}}_{g}^{\alpha}$ is the current associated
with the global symmetry
\begin{eqnarray}
	\delta_{g}c^{\alpha} & = & -\omega^{\alpha},\nonumber \\
	\delta_{g}b^{3} & = & ig\varepsilon^{\alpha\beta}\overline{c}^{\beta}\omega^{\alpha},\nonumber \\
	\delta_{g}b^{\alpha} & = & ig\varepsilon^{\beta\alpha}\overline{c}^{3}\omega^{\beta},\nonumber \\
	\delta_{g}\left(\textrm{other fields}\right) & = & 0.
\end{eqnarray}
Analyzing $S_{\Upsilon}$ in Eq. (\ref{eq:s_upsilon}), we note that
$\mathcal{\mathscr{J}}_{g}^{\alpha}+\partial_{\mu}\overline{c}^{\alpha}$
was already introduced in the starting action coupled with the source
$\zeta_{\mu}^{\alpha}$. Thus, we can indeed write a non-integrated
Ward identity.

For convenience, we define the off-diagonal ghost operator
\begin{eqnarray}
	\mathcal{G}^{\alpha} & = & \frac{\delta}{\delta c^{\alpha}}-ig\varepsilon^{\alpha\beta}\overline{c}^{\beta}\frac{\delta}{\delta b^{3}}+ig\varepsilon^{\alpha\beta}\overline{c}^{3}\frac{\delta}{\delta b^{\beta}}+g\varepsilon^{\alpha\beta}\Upsilon_{\mu}^{\beta}\frac{\delta}{\delta\zeta_{\mu}^{3}}-g\varepsilon^{\alpha\beta}\Upsilon_{\mu}^{3}\frac{\delta}{\delta\zeta_{\mu}^{\beta}}-g\partial_{\mu}\frac{\delta}{\delta\zeta_{\mu}^{\alpha}}.\label{eq:ghost_op_off_diag}
\end{eqnarray}
Therefore, the non-integrated off-diagonal ghost equation can be written
as
\begin{gather}
	\mathcal{G}^{\alpha}\Sigma=\Delta^{\alpha},\label{eq:off_diagonal_ghost_equation}
\end{gather}
where $\Delta^{\alpha}$ is a linear breaking, namely,
\begin{eqnarray}
	\Delta^{\alpha} & = & -\partial^{2}\overline{c}^{\alpha}-\partial_{\mu}K_{\mu}^{\alpha}+g\varepsilon^{\alpha\beta}W_{\mu}^{3}K_{\mu}^{\beta}-g\varepsilon^{\alpha\beta}W_{\mu}^{\beta}K_{\mu}^{3}+\partial_{\mu}\left(\varepsilon^{\alpha\beta}\Upsilon_{\mu}^{\beta}\overline{c}^{3}-\varepsilon^{\alpha\beta}\Upsilon_{\mu}^{3}\overline{c}^{\beta}\right)\nonumber \\
	&  & -\frac{g}{2}Y\rho^{\alpha}+\frac{g}{2}R^{\alpha}\left(\vartheta+H\right)+\frac{g}{2}\varepsilon^{\alpha\beta}R^{\beta}\rho^{3}-\frac{g}{2}\varepsilon^{\alpha\beta}R^{3}\rho^{\beta}-g\varepsilon^{\alpha\beta}L^{\beta}c^{3}+g\varepsilon^{\alpha\beta}L^{3}c^{\beta}.
\end{eqnarray}

\begin{itemize}
	\item Rigid off-diagonal symmetry
\end{itemize}
We have that
\begin{eqnarray}
	\mathcal{G}^{\alpha}\mathcal{S}_{2}\left(\Sigma\right) & = & -\mathcal{S}_{2\,\Sigma}\left(\mathcal{G}^{\alpha}\Sigma\right)+g\varepsilon^{\alpha\beta}\overline{c}^{\beta}\frac{\delta\Sigma}{\delta\overline{c}^{3}}-g\varepsilon^{\alpha\beta}\overline{c}^{3}\frac{\delta\Sigma}{\delta\overline{c}^{\beta}}+g\varepsilon^{\alpha\beta}\Upsilon_{\mu}^{\beta}\frac{\delta\Sigma}{\delta\Upsilon_{\mu}^{3}}-g\varepsilon^{\alpha\beta}\Upsilon_{\mu}^{3}\frac{\delta\Sigma}{\delta\Upsilon_{\mu}^{\beta}}-g\partial_{\mu}\frac{\delta\Sigma}{\delta\Upsilon_{\mu}^{\alpha}}\nonumber \\
	& = & i\partial^{2}b^{\alpha}-\partial_{\mu}\left(\varepsilon^{\alpha\beta}\zeta_{\mu}^{\beta}\overline{c}^{3}+i\varepsilon^{\alpha\beta}\Upsilon_{\mu}^{\beta}b^{3}-\varepsilon^{\alpha\beta}\zeta_{\mu}^{3}\overline{c}^{\beta}-i\varepsilon^{\alpha\beta}\Upsilon_{\mu}^{3}b^{\beta}\right)\nonumber \\
	&  & +\partial_{\mu}\frac{\delta\Sigma}{\delta W_{\mu}^{\alpha}}+g\varepsilon^{\alpha\beta}K_{\mu}^{\beta}\frac{\delta\Sigma}{\delta K_{\mu}^{3}}-g\varepsilon^{\alpha\beta}W_{\mu}^{3}\frac{\delta\Sigma}{\delta W_{\mu}^{\beta}}-g\varepsilon^{\alpha\beta}K_{\mu}^{3}\frac{\delta\Sigma}{\delta K_{\mu}^{\beta}}+g\varepsilon^{\alpha\beta}W_{\mu}^{\beta}\frac{\delta\Sigma}{\delta W_{\mu}^{3}}\nonumber \\
	&  & +\frac{g}{2}\rho^{\alpha}\frac{\delta\Sigma}{\delta H}-\frac{g}{2}Y\frac{\delta\Sigma}{\delta R^{\alpha}}-\frac{g}{2}\left(\vartheta+H\right)\frac{\delta\Sigma}{\delta\rho^{\alpha}}+\frac{g}{2}R^{\alpha}\frac{\delta\Sigma}{\delta Y}\nonumber \\
	&  & +\frac{g}{2}\varepsilon^{\alpha\beta}R^{\beta}\frac{\delta\Sigma}{\delta R^{3}}-\frac{g}{2}\varepsilon^{\alpha\beta}\rho^{3}\frac{\delta\Sigma}{\delta\rho^{\beta}}-\frac{g}{2}\varepsilon^{\alpha\beta}R^{3}\frac{\delta\Sigma}{\delta R^{\beta}}+\frac{g}{2}\varepsilon^{\alpha\beta}\rho^{\beta}\frac{\delta\Sigma}{\delta\rho^{3}}\nonumber \\
	&  & -g\varepsilon^{\alpha\beta}c^{3}\frac{\delta\Sigma}{\delta c^{\beta}}+g\varepsilon^{\alpha\beta}L^{\beta}\frac{\delta\Sigma}{\delta L^{3}}+g\varepsilon^{\alpha\beta}c^{\beta}\frac{\delta\Sigma}{\delta c^{3}}-g\varepsilon^{\alpha\beta}L^{3}\frac{\delta\Sigma}{\delta L^{\beta}}\nonumber \\
	&  & +g\varepsilon^{\alpha\beta}\overline{c}^{\beta}\frac{\delta\Sigma}{\delta\overline{c}^{3}}-g\varepsilon^{\alpha\beta}\overline{c}^{3}\frac{\delta\Sigma}{\delta\overline{c}^{\beta}}+g\varepsilon^{\alpha\beta}\Upsilon_{\mu}^{\beta}\frac{\delta\Sigma}{\delta\Upsilon_{\mu}^{3}}-g\varepsilon^{\alpha\beta}\Upsilon_{\mu}^{3}\frac{\delta\Sigma}{\delta\Upsilon_{\mu}^{\beta}}-g\partial_{\mu}\frac{\delta\Sigma}{\delta\Upsilon_{\mu}^{\alpha}},
\end{eqnarray}
then
\begin{eqnarray}
	\mathcal{R}^{\alpha}\Sigma & = & -i\partial^{2}b^{\alpha}+\partial_{\mu}\left(\varepsilon^{\alpha\beta}\zeta_{\mu}^{\beta}\overline{c}^{3}+i\varepsilon^{\alpha\beta}\Upsilon_{\mu}^{\beta}b^{3}-\varepsilon^{\alpha\beta}\zeta_{\mu}^{3}\overline{c}^{\beta}-i\varepsilon^{\alpha\beta}\Upsilon_{\mu}^{3}b^{\beta}\right)
\end{eqnarray}
where the off-diagonal rigid operator $\mathcal{R}^{\alpha}$ is defined
as
\begin{eqnarray}
	\mathcal{R}^{\alpha} & = & \partial_{\mu}\frac{\delta}{\delta W_{\mu}^{\alpha}}+g\varepsilon^{\alpha\beta}K_{\mu}^{\beta}\frac{\delta}{\delta K_{\mu}^{3}}-g\varepsilon^{\alpha\beta}W_{\mu}^{3}\frac{\delta}{\delta W_{\mu}^{\beta}}-g\varepsilon^{\alpha\beta}K_{\mu}^{3}\frac{\delta}{\delta K_{\mu}^{\beta}}+g\varepsilon^{\alpha\beta}W_{\mu}^{\beta}\frac{\delta}{\delta W_{\mu}^{3}}\nonumber \\
	&  & +\frac{g}{2}\rho^{\alpha}\frac{\delta}{\delta H}-\frac{g}{2}Y\frac{\delta}{\delta R^{\alpha}}-\frac{g}{2}\left(\vartheta+H\right)\frac{\delta}{\delta\rho^{\alpha}}+\frac{g}{2}R^{\alpha}\frac{\delta}{\delta Y}\nonumber \\
	&  & +\frac{g}{2}\varepsilon^{\alpha\beta}R^{\beta}\frac{\delta}{\delta R^{3}}-\frac{g}{2}\varepsilon^{\alpha\beta}\rho^{3}\frac{\delta}{\delta\rho^{\beta}}-\frac{g}{2}\varepsilon^{\alpha\beta}R^{3}\frac{\delta}{\delta R^{\beta}}+\frac{g}{2}\varepsilon^{\alpha\beta}\rho^{\beta}\frac{\delta}{\delta\rho^{3}}\nonumber \\
	&  & -g\varepsilon^{\alpha\beta}c^{3}\frac{\delta}{\delta c^{\beta}}+g\varepsilon^{\alpha\beta}L^{\beta}\frac{\delta}{\delta L^{3}}+g\varepsilon^{\alpha\beta}c^{\beta}\frac{\delta}{\delta c^{3}}-g\varepsilon^{\alpha\beta}L^{3}\frac{\delta}{\delta L^{\beta}}\nonumber \\
	&  & +g\varepsilon^{\alpha\beta}\overline{c}^{\beta}\frac{\delta}{\delta\overline{c}^{3}}-g\varepsilon^{\alpha\beta}\overline{c}^{3}\frac{\delta}{\delta\overline{c}^{\beta}}+g\varepsilon^{\alpha\beta}\Upsilon_{\mu}^{\beta}\frac{\delta}{\delta\Upsilon_{\mu}^{3}}-g\varepsilon^{\alpha\beta}\Upsilon_{\mu}^{3}\frac{\delta}{\delta\Upsilon_{\mu}^{\beta}}-g\partial_{\mu}\frac{\delta}{\delta\Upsilon_{\mu}^{\alpha}}\label{eq:rigid_op_off_diag}
\end{eqnarray}

\begin{itemize}
	\item Non-integrated equation of $c^{3}$
\end{itemize}
For the diagonal component $c^{3}$, we have 

\begin{eqnarray}
	\mathcal{G}^{3}\Sigma & = & \Delta^{3},\label{eq:diagonal_ghost_equation}
\end{eqnarray}
where
\begin{eqnarray}
	\mathcal{G}^{3} & = & \frac{\delta}{\delta c^{3}}-ig\varepsilon^{\alpha\beta}\overline{c}^{\alpha}\frac{\delta}{\delta b^{\beta}}+g\varepsilon^{\alpha\beta}\Upsilon_{\mu}^{\alpha}\frac{\delta}{\delta\zeta_{\mu}^{\beta}}-g\partial_{\mu}\frac{\delta}{\delta\zeta_{\mu}^{3}}\label{eq:ghost_op_3}
\end{eqnarray}
and
\begin{eqnarray}
	\Delta^{3} & = & -\partial^{2}\overline{c}^{3}-\partial_{\mu}K_{\mu}^{3}+g\varepsilon^{\alpha\beta}K_{\mu}^{\alpha}W_{\mu}^{\beta}+\partial_{\mu}\left(\varepsilon^{\alpha\beta}\Upsilon_{\mu}^{\alpha}\overline{c}^{\beta}\right)-\frac{g}{2}Y\rho^{3}+\frac{g}{2}R^{3}\left(\vartheta+H\right)\nonumber \\
	&  & +\frac{g}{2}\varepsilon^{\alpha\beta}R^{\alpha}\rho^{\beta}-g\varepsilon^{\alpha\beta}L^{\alpha}c^{\beta}.
\end{eqnarray}

\begin{itemize}
	\item Diagonal rigid symmetry
\end{itemize}
Since
\begin{eqnarray}
	\mathcal{G}^{3}\mathcal{S}_{2}\left(\Sigma\right) & = & -\mathcal{S}_{2\,\Sigma}\left(\mathcal{G}^{3}\Sigma\right)+g\varepsilon^{\alpha\beta}\overline{c}^{\alpha}\frac{\delta\Sigma}{\delta\overline{c}^{\beta}}+g\varepsilon^{\alpha\beta}\Upsilon_{\mu}^{\alpha}\frac{\delta\Sigma}{\delta\Upsilon_{\mu}^{\beta}}-g\partial_{\mu}\frac{\delta\Sigma}{\delta\Upsilon_{\mu}^{3}}\nonumber \\
	& = & i\partial^{2}b^{3}-\partial_{\mu}\left(\varepsilon^{\alpha\beta}\zeta_{\mu}^{\alpha}\overline{c}^{\beta}+i\varepsilon^{\alpha\beta}\Upsilon_{\mu}^{\alpha}b^{\beta}\right)+\partial_{\mu}\frac{\delta\Sigma}{\delta W_{\mu}^{3}}+g\varepsilon^{\alpha\beta}K_{\mu}^{\alpha}\frac{\delta\Sigma}{\delta K_{\mu}^{\beta}}\nonumber \\
	&  & -g\varepsilon^{\alpha\beta}W_{\mu}^{\beta}\frac{\delta\Sigma}{\delta W_{\mu}^{\alpha}}-\frac{g}{2}Y\frac{\delta\Sigma}{\delta R^{3}}+\frac{g}{2}\rho^{3}\frac{\delta\Sigma}{\delta H}+\frac{g}{2}R^{3}\frac{\delta\Sigma}{\delta Y}-\frac{g}{2}\left(\vartheta+H\right)\frac{\delta\Sigma}{\delta\rho^{3}}\nonumber \\
	&  & -\frac{g}{2}\varepsilon^{\alpha\beta}\rho^{\beta}\frac{\delta\Sigma}{\delta\rho^{\alpha}}+\frac{g}{2}\varepsilon^{\alpha\beta}R^{\alpha}\frac{\delta\Sigma}{\delta R^{\beta}}+g\varepsilon^{\alpha\beta}c^{\alpha}\frac{\delta\Sigma}{\delta c^{\beta}}+g\varepsilon^{\alpha\beta}L^{\alpha}\frac{\delta\Sigma}{\delta L^{\beta}}\nonumber \\
	&  & +g\varepsilon^{\alpha\beta}\overline{c}^{\alpha}\frac{\delta\Sigma}{\delta\overline{c}^{\beta}}+g\varepsilon^{\alpha\beta}\Upsilon_{\mu}^{\alpha}\frac{\delta\Sigma}{\delta\Upsilon_{\mu}^{\beta}}-g\partial_{\mu}\frac{\delta\Sigma}{\delta\Upsilon_{\mu}^{3}},
\end{eqnarray}
then
\begin{eqnarray}
	\mathcal{R}^{3}\Sigma & = & -i\partial^{2}b^{3}+\partial_{\mu}\left(\varepsilon^{\alpha\beta}\zeta_{\mu}^{\alpha}\overline{c}^{\beta}+i\varepsilon^{\alpha\beta}\Upsilon_{\mu}^{\alpha}b^{\beta}\right),
\end{eqnarray}
where
\begin{eqnarray}
	\mathcal{R}^{3} & = & \partial_{\mu}\frac{\delta}{\delta W_{\mu}^{3}}+g\varepsilon^{\alpha\beta}K_{\mu}^{\alpha}\frac{\delta}{\delta K_{\mu}^{\beta}}-g\varepsilon^{\alpha\beta}W_{\mu}^{\beta}\frac{\delta}{\delta W_{\mu}^{\alpha}}\nonumber \\
	&  & -\frac{g}{2}Y\frac{\delta}{\delta R^{3}}+\frac{g}{2}\rho^{3}\frac{\delta}{\delta H}+\frac{g}{2}R^{3}\frac{\delta}{\delta Y}-\frac{g}{2}\left(\vartheta+H\right)\frac{\delta}{\delta\rho^{3}}\nonumber \\
	&  & -\frac{g}{2}\varepsilon^{\alpha\beta}\rho^{\beta}\frac{\delta}{\delta\rho^{\alpha}}+\frac{g}{2}\varepsilon^{\alpha\beta}R^{\alpha}\frac{\delta}{\delta R^{\beta}}+g\varepsilon^{\alpha\beta}c^{\alpha}\frac{\delta}{\delta c^{\beta}}+g\varepsilon^{\alpha\beta}L^{\alpha}\frac{\delta}{\delta L^{\beta}}\nonumber \\
	&  & +g\varepsilon^{\alpha\beta}\overline{c}^{\alpha}\frac{\delta}{\delta\overline{c}^{\beta}}+g\varepsilon^{\alpha\beta}\Upsilon_{\mu}^{\alpha}\frac{\delta}{\delta\Upsilon_{\mu}^{\beta}}-g\partial_{\mu}\frac{\delta}{\delta\Upsilon_{\mu}^{3}}.\label{eq:rigid_op_3}
\end{eqnarray}

\begin{itemize}
	\item Non-integrated equation for $c$
\end{itemize}
The Abelian ghost $c$ is a non-interacting field, then

\begin{eqnarray}
	\frac{\delta\Sigma}{\delta c} & = & -\partial^{2}\overline{c}+\frac{g'}{2}\varepsilon^{\alpha\beta}R^{\alpha}\rho^{\beta}-\frac{g'}{2}R^{3}\left(\vartheta+H\right)+\frac{g'}{2}Y\rho^{3}\label{eq:ghost_eq_abelian}
\end{eqnarray}
is a Ward identity.
\begin{itemize}
	\item Abelian rigid symmetry
\end{itemize}
Since
\begin{eqnarray}
	\frac{\delta}{\delta c}\mathcal{S}_{1}\left(\Sigma\right) & = & -\mathcal{S}_{1\,\Sigma}\left(\frac{\delta\Sigma}{\delta c}\right)-\frac{\delta\Sigma}{\delta\xi}+\partial_{\mu}\frac{\delta\Sigma}{\delta X_{\mu}}\nonumber \\
	& = & i\partial^{2}b-\frac{g'}{2}\varepsilon^{\alpha\beta}\frac{\delta\Sigma}{\delta\rho^{\alpha}}\rho^{\beta}+\frac{g'}{2}\varepsilon^{\alpha\beta}R^{\alpha}\frac{\delta\Sigma}{\delta R^{\beta}}\nonumber \\
	&  & +\frac{g'}{2}\left(\vartheta+H\right)\frac{\delta\Sigma}{\delta\rho^{3}}-\frac{g'}{2}R^{3}\frac{\delta\Sigma}{\delta Y}-\frac{g'}{2}\rho^{3}\frac{\delta\Sigma}{\delta H}+\frac{g'}{2}Y\frac{\delta\Sigma}{\delta R^{3}}\nonumber \\
	&  & -\frac{\delta\Sigma}{\delta\xi}+\partial_{\mu}\frac{\delta\Sigma}{\delta X_{\mu}},
\end{eqnarray}
then
\begin{eqnarray}
	\mathcal{R}\Sigma & = & -i\partial^{2}b,\label{eq:rigid_abelian}
\end{eqnarray}
where the Abelian rigid operator $\mathcal{R}$ is defined by
\begin{eqnarray}
	\mathcal{R} & = & -\frac{g'}{2}\varepsilon^{\alpha\beta}\rho^{\beta}\frac{\delta}{\delta\rho^{\alpha}}+\frac{g'}{2}\varepsilon^{\alpha\beta}R^{\alpha}\frac{\delta}{\delta R^{\beta}}\nonumber \\
	&  & +\frac{g'}{2}\left(\vartheta+H\right)\frac{\delta}{\delta\rho^{3}}-\frac{g'}{2}R^{3}\frac{\delta}{\delta Y}-\frac{g'}{2}\rho^{3}\frac{\delta}{\delta H}+\frac{g'}{2}Y\frac{\delta}{\delta R^{3}}\nonumber \\
	&  & -\frac{\delta}{\delta\xi}+\partial_{\mu}\frac{\delta}{\delta X_{\mu}}.
\end{eqnarray}

\subsection{$U_{EM}\left(1\right)$ symmetry}

The $U_{EM}\left(1\right)$ global symmetry is defined by the transformations
(\ref{eq:U(1)EM}), and
\begin{eqnarray}
	\delta b^{\alpha} & = & \omega\varepsilon^{\alpha\beta}b^{\beta},\nonumber \\
	\delta c^{\alpha} & = & \omega\varepsilon^{\alpha\beta}c^{\beta},\nonumber \\
	\delta\overline{c}^{\alpha} & = & \omega\varepsilon^{\alpha\beta}\overline{c}^{\beta},\nonumber \\
	\delta K_{\mu}^{\alpha} & = & \omega\varepsilon^{\alpha\beta}K_{\mu}^{\beta},\nonumber \\
	\delta R^{\alpha} & = & \omega\varepsilon^{\alpha\beta}R^{\beta},\nonumber \\
	\delta L^{\alpha} & = & \omega\varepsilon^{\alpha\beta}L^{\beta},\nonumber \\
	\delta\xi & = & -\omega,\label{eq:u1_eletro_2}
\end{eqnarray}
and
\begin{gather}
	\delta\Omega_{\mu}=\delta\Omega_{\mu}^{\alpha}=\delta\Omega_{\mu}^{3}=\delta\Omega_{\mu}^{4}=\delta\Omega_{\mu}^{5}=\delta\Omega_{\mu}^{6}=\delta K_{\mu}^{3}=\delta R^{3}=\delta L^{3}=\delta\theta=\delta\eta=\delta\overline{\eta}=\delta\Upsilon_{\mu}=\delta\zeta_{\mu}=0.\label{eq:u1_eletro_3}
\end{gather}
In general, we would expect only an integrated Ward identity associated
with this symmetry, as it is global. However, one of the operators
we introduced is, essentially, the current of this symmetry, namely,
$O_{\mu}$. This lead us to a non-integrated identity, as we will
try to explain.

The local gauge invariance of $S_{\textrm{Higgs}}$ with respect to
the $SU\left(2\right)\times U\left(1\right)$ transformations (\ref{eq:infinitesimal_gauge_transformations})
and (\ref{eq:infinitesimal_gauge_transformations-1}) can be expressed
as
\begin{eqnarray}
	\mathcal{W}S_{\textrm{Higgs}} & = & 0,\nonumber \\
	\mathcal{W}^{a}S_{\textrm{Higgs}} & = & 0,
\end{eqnarray}
where the operators $\mathcal{W}$ and $\mathcal{W}^{a}$ are defined
by
\begin{eqnarray}
	\mathcal{W} & = & \frac{1}{g'}\partial_{\mu}\frac{\delta}{\delta X_{\mu}}-\frac{i}{2}\frac{\delta}{\delta\varphi}\varphi+\frac{i}{2}\varphi^{\dagger}\frac{\delta}{\delta\varphi^{\dagger}},\nonumber \\
	\mathcal{W}^{a} & = & \frac{1}{g}\partial_{\mu}\frac{\delta}{\delta W_{\mu}^{a}}-\varepsilon^{abc}W_{\mu}^{c}\frac{\delta}{\delta W_{\mu}^{b}}-\frac{i}{2}\frac{\delta}{\delta\varphi}\tau^{a}\varphi+\frac{i}{2}\varphi^{\dagger}\tau^{a}\frac{\delta}{\delta\varphi^{\dagger}}.
\end{eqnarray}
If we combine $\mathcal{W}$ and $\mathcal{W}^{3}$, we obtain part
of the operator associated with the $U_{EM}\left(1\right)$ transformations,
Eqs. (\ref{eq:U(1)EM}), (\ref{eq:u1_eletro_2}), (\ref{eq:u1_eletro_3}),
namely,
\begin{eqnarray}
	-\mathcal{W}-\mathcal{W}^{3} & = & -\frac{1}{g'}\partial_{\mu}\frac{\delta}{\delta X_{\mu}}-\frac{1}{g}\partial_{\mu}\frac{\delta}{\delta W_{\mu}^{3}}+i\frac{\delta}{\delta\varphi}Q\varphi-i\varphi^{\dagger}Q\frac{\delta}{\delta\varphi^{\dagger}}+\varepsilon^{\alpha\beta}W_{\mu}^{\beta}\frac{\delta}{\delta W_{\mu}^{\alpha}}\nonumber \\
	& = & -\frac{1}{g'}\partial_{\mu}\frac{\delta}{\delta X_{\mu}}-\frac{1}{g}\partial_{\mu}\frac{\delta}{\delta W_{\mu}^{3}}+\varepsilon^{\alpha\beta}\left(\rho^{\beta}\frac{\delta}{\delta\rho^{\alpha}}+W_{\mu}^{\beta}\frac{\delta}{\delta W_{\mu}^{\alpha}}\right).
\end{eqnarray}
If we include the other transformations, we get the rigid operators
$\mathcal{R}$ and $\mathcal{R}^{3}$ instead of $\mathcal{W}$ and
$\mathcal{W}^{3}$, respectively. Thus, the operator associated with
the $U_{EM}\left(1\right)$ transformations is
\begin{eqnarray}
	\hat{\mathcal{W}} & = & -\frac{1}{g'}\mathcal{R}-\frac{1}{g}\mathcal{R}^{3}\nonumber \\
	& = & \varepsilon^{\alpha\beta}\left(\rho^{\beta}\frac{\delta}{\delta\rho^{\alpha}}+W_{\mu}^{\beta}\frac{\delta}{\delta W_{\mu}^{\alpha}}+R^{\beta}\frac{\delta}{\delta R^{\alpha}}+c^{\beta}\frac{\delta}{\delta c^{\alpha}}+L^{\beta}\frac{\delta}{\delta L^{\alpha}}+K_{\mu}^{\beta}\frac{\delta}{\delta K_{\mu}^{\alpha}}+\overline{c}^{\beta}\frac{\delta}{\delta\overline{c}^{\alpha}}+\Upsilon_{\mu}^{\beta}\frac{\delta}{\delta\Upsilon_{\mu}^{\alpha}}\right)\nonumber \\
	&  & +\frac{1}{g'}\frac{\delta}{\delta\xi}-\frac{1}{g'}\partial_{\mu}\frac{\delta}{\delta X_{\mu}}-\frac{1}{g}\partial_{\mu}\frac{\delta}{\delta W_{\mu}^{3}}+\partial_{\mu}\frac{\delta}{\delta\Upsilon_{\mu}^{3}}.\label{eq:U_(1)_em_op}
\end{eqnarray}
Now, the global $U_{EM}\left(1\right)$ symmetry can be expressed
as
\begin{eqnarray}
	\int d^{4}x\hat{\mathcal{W}}\Sigma & = & 0.
\end{eqnarray}
We can also use the identity (\ref{eq:xi_equation}) to interchange
the transformation of $\xi$ for $\Omega_{\mu}^{\alpha}$. So, there
are two equivalent ways of expressing the $U_{EM}\left(1\right)$
symmetry. One with
\begin{eqnarray}
	\delta_{\omega}\xi & = & -\omega,\nonumber \\
	\delta_{\omega}\Omega_{\mu}^{\alpha} & = & 0,\nonumber \\
	\delta_{\omega}J & = & 0,
\end{eqnarray}
and the other one with
\begin{eqnarray}
	\delta_{\omega}\xi & = & 0,\nonumber \\
	\delta_{\omega}\Omega_{\mu}^{\alpha} & = & \omega\varepsilon^{\alpha\beta}\Omega_{\mu}^{\beta},\nonumber \\
	\delta_{\omega}J & = & \frac{g}{2}\Omega_{\mu}^{3}\partial_{\mu}\omega.
\end{eqnarray}

As we discussed in Subsection \ref{subsec:Local--Symmetry}, the local
$SU\left(2\right)$ symmetry can be restored when we add $S_{\Upsilon}$.
This is achieved due to the connection role played by the source $\Upsilon_{\mu}^{a}$.
Therefore, we expect to write down a non-integrated Ward identity
associated with the $U_{EM}\left(1\right)$ symmetry. Indeed, we have
the following Ward identity:
\begin{eqnarray}
	\hat{\mathcal{W}}\Sigma & = & \frac{i}{g'}\partial^{2}b+\frac{i}{g}\partial^{2}b^{3}-\frac{1}{g}\partial_{\mu}\left(\varepsilon^{\alpha\beta}\zeta_{\mu}^{\alpha}\overline{c}^{\beta}+i\varepsilon^{\alpha\beta}\Upsilon_{\mu}^{\alpha}b^{\beta}\right).
\end{eqnarray}

\subsection{Integrated equation of $H$ \protect\label{subsec:Integrated-equation-of}}

The Higgs field and $\vartheta$ appear in the combination $\vartheta+H$
most of the time, mainly because the Higgs action can be written in
term of the scalar field $\varphi$ where this combination is found.
One exception is the Higgs potential, where $\vartheta$ appears in
different combinations or alone. Thus the symmetry under the transformations
$H\rightarrow H+\epsilon$ and $\vartheta\rightarrow\vartheta-\epsilon$
is broken, where $\epsilon$ is a constant. The special thing about
this symmetry is that the breaking is proportional to the scalar operator
$O$. Therefore, we can write the very important Ward identity 
\begin{eqnarray}
	\int d^{4}x\left(\frac{\delta\Sigma}{\delta H}-2\lambda\vartheta\frac{\delta\Sigma}{\delta J}\right)-\frac{\partial\Sigma}{\partial\vartheta} & = & \int d^{4}x\left(2\vartheta J-2\vartheta\Xi-2\vartheta\Omega_{\mu}^{5}\left(X_{\mu}-\partial_{\mu}\xi\right)\right).\label{eq:H_equation}
\end{eqnarray}

\subsection{Source equations}

Some of the operators we introduced in Subsection \ref{subsec:BRST-invariant-operators}
are linear in the fields. Therefore, the derivatives of their respective
sources are Ward identities. It follows that
\begin{eqnarray}
	\frac{\delta\Sigma}{\delta\Omega_{\mu}^{4}}-\partial_{\mu}\frac{\delta\Sigma}{\delta J} & = & 0,\nonumber \\
	\frac{\delta\Sigma}{\delta\Omega_{\mu}^{5}} & = & \vartheta^{2}\left(X_{\mu}-\partial_{\mu}\xi\right),\nonumber \\
	\frac{\delta\Sigma}{\delta\Omega_{\mu}^{6}} & = & \partial_{\nu}X_{\nu\mu},\nonumber \\
	\frac{\delta\Sigma}{\delta\Xi} & = & \vartheta^{2},\label{eq:sources_ward}
\end{eqnarray}
are also Ward identities. 

\subsection{Ghost numbers}

Ghost and antighost fields appear in pairs, so we can define a ghost
number for each field and each source, as indicated in Table 1 and
Table 2 of the Appendix B, respectively. We have one ghost charge for the $SU\left(2\right)$
group and another one for the $U\left(1\right)$ group. Therefore,
we have the following functional equations:
\begin{eqnarray}
	\mathcal{N}_{2\,g}\Sigma & = & 0,\\
	\mathcal{N}_{1\,g}\Sigma & = & 0,
\end{eqnarray}
where the ghost number operators are defined by
\begin{eqnarray}
	\mathcal{N}_{2\,g} & = & \int d^{4}x\left(c^{\alpha}\frac{\delta}{\delta c^{\alpha}}-\overline{c}^{\alpha}\frac{\delta}{\delta\overline{c}^{\alpha}}+c^{3}\frac{\delta}{\delta c^{3}}-\overline{c}^{3}\frac{\delta}{\delta\overline{c}^{3}}-2L^{\alpha}\frac{\delta}{\delta L^{\alpha}}-2L^{3}\frac{\delta}{\delta L^{3}}\right.\nonumber \\
	&  & \left.-K_{\mu}^{\alpha}\frac{\delta}{\delta K_{\mu}^{\alpha}}-K_{\mu}^{3}\frac{\delta}{\delta K_{\mu}^{3}}-Y\frac{\delta}{\delta Y}-R^{\alpha}\frac{\delta}{\delta R^{\alpha}}-R^{3}\frac{\delta}{\delta R^{3}}+\zeta_{\mu}^{\alpha}\frac{\delta}{\delta\zeta_{\mu}^{\alpha}}+\zeta_{\mu}^{3}\frac{\delta}{\delta\zeta_{\mu}^{3}}\right)
\end{eqnarray}
and
\begin{eqnarray}
	\mathcal{N}_{1\,g} & = & \int d^{4}x\left(c\frac{\delta}{\delta c}-\overline{c}\frac{\delta}{\delta\overline{c}}-y\frac{\delta}{\delta y}-r^{\alpha}\frac{\delta}{\delta r^{\alpha}}-r^{3}\frac{\delta}{\delta r^{3}}\right).
\end{eqnarray}

\section{All orders algebraic analysis of the renormalizability\protect\label{sec:All-orders-renormalization}}

\subsection{Brief review  of the renormalization theory}

By following the standard  renormalization theory \footnote{There are different, yet equivalent, ways of presenting this topic. Here, we adopt the so-called infinite counterterm method. This method consists of introducing, at each perturbative order, a local counterterm that depends on the regulator parameters and cancels the divergent part of the regularized perturbative corrections. By adding these counterterms to the tree-level action, we define the bare action. Consequently, if we attempt to take the limit corresponding to the removal of the UV regulator, the bare quantities- such as the bare action- become divergent. At this point, some may feel uncertain about the true meaning of the bare action. To clarify this, we suggest viewing the bare action as part of a mathematical procedure designed to obtain finite correlation functions. Some may argue that this is a more mathematical rather than an intuitive approach, but it is important to remember that renormalization theory exists to resolve the issue caused by the product of distributions at the same spacetime point. For instance, in the approach developed by Epstein-Glaser-Scharf \cite{Epstein:1973gw,Scharf:1996zi}, where all mathematical operations are well-defined from the outset, UV divergences do not appear at any stage of the calculation. The crucial point is that, regardless of the approach we choose- including the latter- amplitudes or correlation functions remain unique up to an arbitrary polynomial of the external momenta and masses. These must be fixed by renormalization conditions at some arbitrary scale, say $\mu$. Since a physical observable, say $\mathcal{O}$, should not depend on the arbitrary scale $\mu$, we have $\mu\frac{d \mathcal{O}}{d\mu}=0$, which is the fundamental equation of the Renormalization Group.   }, we know that at
any given order, it is necessary to introduce counterterms to cancel
the divergences in loop diagrams. If the divergences of the previous
orders are correctly eliminated, then the most general counterterm
is a local polynomial of dimension four built with the fields, derivatives,
sources and masses present in the theory. Furthermore, this counterterm,
which we call $\Sigma_{ct}$, has the same form at any given order
from one-loop onward. Here, we want to find the \emph{bare action
}$\Sigma_{\textrm{bare}}$ that allow us to compute finite correlation
functions, and such that
\begin{eqnarray}
	\Sigma_{\textrm{bare}} & = & \Sigma+\epsilon\Sigma_{ct},\label{eq:bare_sigma_plus_counterterm}
\end{eqnarray}
where $\epsilon$ is the perturbative parameter.  

The $SU\left(2\right)\times U\left(1\right)$ Higgs model in the Landau
gauge is renormalizable \cite{tHooft:1972tcz,Becchi:1975nq,Kraus:1997bi}. The only thing that might prevent
us from saying the same about the model we constructed is the existence
of the auxiliary Stueckelberg field, $\xi$, which is a dimensionless
field. Thus, in principle, we would have an infinite number of counterterms
with $\xi$. However, if we set the external sources $\Omega_{\mu}^{\alpha}$,
$\Omega_{\mu}^{3}$ and $\Omega_{\mu}^{5}$ to zero, the integration
over $\xi$, $\theta$, $\eta$ and $\overline{\eta}$ gives the identity.
Then we recover the $SU\left(2\right)\times U\left(1\right)$ Higgs
model plus the composite operators $O$ and $O_{\mu}$, which certainly
is renormalizable. This observation leads us to the conclusion that
all counterterms with $\xi$, if they exist, come together with the
external sources $\Omega_{\mu}^{\alpha}$ and $\Omega_{\mu}^{3}$.
The propagators (see Eq. (\ref{eq:propagators_tree})) also indicate
that we should not worry about $\xi$. The only non-vanishing propagators
involving $\xi$ are 
\begin{gather}
	\left\langle \xi\left(p\right)b\left(-p\right)\right\rangle _{0}=\left\langle \xi\left(p\right)\theta\left(-p\right)\right\rangle _{0}=\frac{i}{p^{2}}.
\end{gather}
Given that $b$ and $\theta$ are non-interacting fields, then we
can conclude that $\xi$ completely decouples. This is a special property of the Landau gauge.  \par 

A similar situation arises in the context of the Refined Gribov-Zwanziger theory (RGZ), where a Stueckelberg-type field is also introduced to establish a BRST-invariant framework, see \cite{Capri:2016ovw,Capri:2017npq}. In this case, the Stueckelberg-type field is used to localize the gauge-invariant field $A_{\mu}^{h}(x)$, which is necessary to construct the BRST-invariant condensate $\langle (A^h_{\mu})^2 \rangle $. Using similar arguments, \cite{Capri:2016ovw,Capri:2017npq} demonstrated that the local BRST-invariant RGZ theory is renormalizable to all order in a broad class of gauges continuously connected to the Landau gauge.

\subsection{Applying the algebraic method to determine the invariant counterterm}

The Algebraic Renormalization method \cite{Piguet:1995er} consists of using the set of symmetries of the classical action $\Sigma$- expressed in terms of Ward identities- and renormalization theory to determine the most general counterterm required to make all correlation functions finite. Furthermore, the Algebraic Renormalization method helps us investigate whether a Ward identity is anomalous. Therefore, let us briefly outline the algebraic method. 
	 
The algebraic method is based on the set of classical Ward identities that can be extended at the quantum level. If the tree-level action $\Sigma$ satisfies the Ward identity
\begin{gather}
	\mathcal{W}\Sigma=0,
\end{gather}
where $\mathcal{W}$ is a local linear functional operator, we aim to find, at all perturbative orders, an effective quantum action $\Gamma$, such that 
\begin{gather}
		\mathcal{W}\Gamma=0.
\end{gather} 
However, the general mathematical solution of the renormalization problem leads us to a slightly different result. If the \emph{n}th-order  effective action $\Gamma^{\left(n\right)}$ satisfies this Ward identity, then for the next leading order
\begin{gather}
	\mathcal{W}\Gamma^{(n+1)}=\hbar^{n+1} \Delta,
\end{gather} 
where $\Delta$ is a local polynomial in fields, sources and masses, consistent with the dimension of the left-hand side. The identity is non-anomalous when it is possible to write $\Delta=\mathcal{W} \hat{\Delta}$. In this case, we can obtain an effective action that satisfies the Ward identity through the redefinition $\Gamma^{\left(n+1\right)}\rightarrow \Gamma^{\left(n+1\right)}-\hbar^{n+1}\hat{\Delta}$, where $\hbar^{n+1}\hat{\Delta}$ is called a non-invariant counterterm \cite{Piguet:1995er}. Consequently, we can also obtain a bare action $\Sigma_{\text{bare}}$ that obeys the same Ward identity, and this is the main result we will use in the following analysis. It is important to mention that these results also hold for non-linear symmetries, such as the BRST symmetry, as well as linearly broken symmetries, see \cite{Piguet:1995er}.

The Ward identities we presented in Section \ref{sec:Symmetries-and-Ward} are non-anomalous since there are no chiral fermions, and we are considering the fundamental representation of the $SU(2)\times U(1)$ group. Therefore, they can
be extended to $\Sigma_{\textrm{bare}}$, \emph{i.e.}, we can replace
$\Sigma\rightarrow\Sigma_{\textrm{bare}}$ in all the Ward identities
of Section \ref{sec:Symmetries-and-Ward}. Hence, we have that
the bare action also satisfies the Slavnov-Taylor identities,
\begin{eqnarray}
	\mathcal{S}_{1}\left(\Sigma_{\textrm{bare}}\right) & = & 0,
\end{eqnarray}
and
\begin{eqnarray}
	\mathcal{S}_{2}\left(\Sigma_{\textrm{bare}}\right) & = & 0.\label{eq:su_2_slavnov_bare}
\end{eqnarray}
Expanding the bare action as ($\ref{eq:bare_sigma_plus_counterterm}$),
it follows that
\begin{eqnarray}
	\mathcal{S}_{i}\left(\Sigma+\epsilon\Sigma_{ct}\right) & = & 0\nonumber \\
	\mathcal{S}_{i}\left(\Sigma\right)+\epsilon\mathcal{S}_{i\,\Sigma}\Sigma_{ct}+\mathcal{O}\left(\epsilon^{2}\right) & = & 0,
\end{eqnarray}
then
\begin{eqnarray}
	\mathcal{S}_{i\,\Sigma}\Sigma_{ct} & = & 0,\label{eq:linearized_st_counterterm}
\end{eqnarray}
where the linearized Slavnov-Taylor operator $\mathcal{S}_{i\,\Sigma}$
is defined by Eq. (\ref{eq:linearized_slavnov_taylor}) and (\ref{eq:linearized_slavnov_taylor-1}).
The linearized Slavnov-Taylor is nilpotent, \emph{i.e.}, 
\begin{eqnarray}
	\mathcal{S}_{i\,F}^{2} & = & 0,
\end{eqnarray}
for any functional $F$ such that $\mathcal{S}_{i}\left(F\right)=0$,
which is the case for $F=\Sigma$. This property implies that the
counterterm consistent with (\ref{eq:linearized_st_counterterm})
can be separated in two parts, namely,
\begin{eqnarray}
	\Sigma_{ct} & = & \Delta_{1}+\mathcal{S}_{1\,\Sigma}\Delta_{1}^{\left(-1\right)}
\end{eqnarray}
or
\begin{eqnarray}
	\Sigma_{ct} & = & \Delta_{2}+\mathcal{S}_{2\,\Sigma}\Delta_{2}^{\left(-1\right)}.\label{eq:cohom_s2}
\end{eqnarray}
$\Delta_{i}$ belongs to the non-trivial cohomology of $\mathcal{S}_{i\,\Sigma}$,
\emph{i.e.}, 
\begin{eqnarray}
	\mathcal{S}_{i\,\Sigma}\Delta_{i} & = & 0,\nonumber \\
	\Delta_{i} & \neq & \mathcal{S}_{i\,\Sigma}\left(\ldots\right),\label{eq:non_trivial_cohomology}
\end{eqnarray}
whereas $\mathcal{S}_{i\,\Sigma}\Delta_{i}^{\left(-1\right)}$ belongs
to the trivial cohomology of $\mathcal{S}_{i\,\Sigma}$. As $\mathcal{S}_{i\,\Sigma}$
inherits the ghost number of $s_{i}$ and
\begin{gather}
	\mathcal{N}_{g\,i}\Sigma_{\textrm{bare}}=0\Rightarrow\mathcal{N}_{g\,i}\Sigma_{ct}=0,\label{eq:ghost_number_couterterm}
\end{gather}
then $\Delta_{i}^{\left(-1\right)}$ is a polynomial with ghost number
$-1$. 

In addition to the Slavnov-Taylor identities and the ghost numbers,
we have the equations of motion (\ref{eq:teta_equation}), (\ref{eq:eta_equations}),
(\ref{eq:b_equations}), (\ref{eq:antighost_equation}) and (\ref{eq:ghost_eq_abelian}).
They lead to the following identities:
\begin{gather}
	\frac{\delta\Sigma_{\textrm{bare}}}{\delta\theta}=i\left(\partial^{2}\xi-\partial_{\mu}X_{\mu}\right), \quad \frac{\delta\Sigma_{\textrm{bare}}}{\delta\eta}=-\partial^{2}\overline{\eta},\quad \frac{\delta\Sigma_{\textrm{bare}}}{\delta\overline{\eta}}=\partial^{2}\eta,\nonumber \\
	\frac{\delta\Sigma_{\textrm{bare}}}{\delta b^{\alpha}}=i\partial_{\mu}W_{\mu}^{\alpha}+i\varepsilon^{\alpha\beta}\Upsilon_{\mu}W_{\mu}^{\beta}, \quad \frac{\delta\Sigma_{\textrm{bare}}}{\delta b^{3}}=i\partial_{\mu}W_{\mu}^{3}, \quad \frac{\delta\Sigma_{\textrm{bare}}}{\delta b}=i\partial_{\mu}X_{\mu}, \nonumber \\
	\frac{\delta\Sigma_{\textrm{bare}}}{\delta\overline{c}}=\partial^{2}c,, \quad\frac{\delta\Sigma_{\textrm{bare}}}{\delta c}=-\partial^{2}\overline{c}+\frac{g'}{2}\varepsilon^{\alpha\beta}R^{\alpha}\rho^{\beta}-\frac{g'}{2}R^{3}\left(\vartheta+H\right)+\frac{g'}{2}Y\rho^{3}.
\end{gather}
Therefore, it follows that
\begin{gather}
	\frac{\delta\Sigma_{ct}}{\delta\theta}=\frac{\delta\Sigma_{ct}}{\delta\eta}=\frac{\delta\Sigma_{ct}}{\delta\overline{\eta}}=\frac{\delta\Sigma_{ct}}{\delta b^{\alpha}}=\frac{\delta\Sigma_{ct}}{\delta b^{3}}=\frac{\delta\Sigma_{ct}}{\delta b}=\frac{\delta\Sigma_{ct}}{\delta\overline{c}}=\frac{\delta \Sigma_{ct}}{\delta c}=0.\label{eq:zero_motion}
\end{gather}
These equations tell us that $\Sigma_{ct}$ is independent of $\theta$,
$\eta$, $\overline{\eta}$, $b^{\alpha}$, $b^{3}$, $b$, $c$ and
$\overline{c}$. 

As a consequence of Eq. (\ref{eq:zero_motion}) , $\Sigma_{ct}$ is
also independent of $y$, $r^{\alpha}$ and $r^{3}$, because there
is no other field or source with Abelian ghost number equals to 1.
Thus, all counterterms belong to the non-trivial cohomology of $\mathcal{S}_{1\,\Sigma}$,
\emph{i.e.},
\begin{eqnarray}
	\Sigma_{ct} & = & \Delta_{1}\left(W_{\mu}^{a},X_{\mu},H,\rho^{a},\vartheta,K_{\mu}^{a},Y,R^{a},L^{a},c^{a},\overline{c}^{a},\left\{ \Omega\right\} ,J,\Xi,\Upsilon_{\mu}^{a},\zeta_{\mu}^{a}\right).\label{eq:ct_u1}
\end{eqnarray}
Then, $\mathcal{S}_{1\,\Sigma}$ is equivalent to $s_{1}$ when acting
on $\Sigma_{ct}$, \emph{i.e.},
\begin{gather}
	\mathcal{S}_{1\,\Sigma}\Sigma_{ct}=s_{1}\Sigma_{ct}=0.\label{eq:s_1_counterterm}
\end{gather}

The result we just derived for the $U\left(1\right)$ Slavnov-Taylor
identity simplifies our problem of finding the most general counterterm
that satisfies the $SU\left(2\right)$ Slavnov-Taylor identity (\ref{eq:su_2_slavnov_bare})
or, equivalently, 
\begin{gather*}
	\mathcal{S}_{2\,\Sigma}\Sigma_{ct}=0.
\end{gather*}
From (\ref{eq:cohom_s2}), (\ref{eq:ct_u1}) and the \emph{doublet theorem}\footnote{The doublet theorem establishes that all terms with doublet fields belong to the trivial cohomology.}
\cite{Piguet:1995er}, we have that
\begin{eqnarray}
	\Sigma_{ct} & = & \Delta_{2}\left(W_{\mu}^{a},X_{\mu},H,\rho^{a},\vartheta,K_{\mu}^{a},Y,R^{a},L^{a},c^{a},\left\{ \Omega\right\} ,J,\Xi\right)\nonumber \\
	&  & +\mathcal{S}_{2\,\Sigma}\Delta_{2}^{\left(-1\right)}\left(W_{\mu}^{a},X_{\mu},H,\rho^{a},\vartheta,K_{\mu}^{a},Y,R^{a},L^{a},c^{a},\overline{c}^{a},\left\{ \Omega\right\} ,J,\Xi,\Upsilon_{\mu}^{a},\zeta_{\mu}^{a}\right),
\end{eqnarray}
such that Eq. (\ref{eq:s_1_counterterm}) holds.

We also have the external sources equations (\ref{eq:sources_ward}).
They imply that
\begin{gather}
	\frac{\delta\Sigma_{\textrm{bare}}}{\delta\Omega_{\mu}^{4}}-\partial_{\mu}\frac{\delta\Sigma_{\textrm{bare}}}{\delta J} = 0, \quad
	\frac{\delta\Sigma_{\textrm{bare}}}{\delta\Omega_{\mu}^{5}} = \vartheta^{2}\left(X_{\mu}-\partial_{\mu}\xi\right), \quad
	\frac{\delta\Sigma_{\textrm{bare}}}{\delta\Omega_{\mu}^{6}} = \partial_{\nu}X_{\nu\mu},\quad
	\frac{\delta\Sigma_{\textrm{bare}}}{\delta\Xi} = \vartheta^{2}.\label{eq:sources_ward-1}
\end{gather}
Therefore, we obtain the following constraints for the counterterms:
\begin{gather}
	\frac{\delta\Sigma_{ct}}{\delta\Omega_{\mu}^{4}}-\partial_{\mu}\frac{\delta\Sigma_{ct}}{\delta J} = 0, \quad 
	\frac{\delta\Sigma_{ct}}{\delta\Omega_{\mu}^{5}} = 0, \quad
	\frac{\delta\Sigma_{ct}}{\delta\Omega_{\mu}^{6}} = 0, \quad 
	\frac{\delta\Sigma_{ct}}{\delta\Xi} = 0,\label{eq:sources_ward-1-1}
\end{gather}
These equations mean that $\Sigma_{ct}$ is independent of $\Omega_{\mu}^{5}$,
$\Omega_{\mu}^{6}$ and $\Xi$. Furthermore, the first one establishes
that $J$ and $\Omega_{\mu}^{4}$ appear always in the combination
\begin{eqnarray}
	\hat{J} & = & J+\partial_{\mu}\Omega_{\mu}^{4}.
\end{eqnarray}

By imposing the remaining Ward identities, we have the following constraints
for the counterterms:
\begin{itemize}
	\item $U_{EM}\left(1\right)$ symmetry
\end{itemize}
\begin{eqnarray}
	\hat{\mathcal{W}}\Sigma_{ct} & = & 0,
\end{eqnarray}
where the operator $\hat{\mathcal{W}}$ is defined by Eq. (\ref{eq:U_(1)_em_op}).
\begin{itemize}
	\item Equation of $\xi$
\end{itemize}
\begin{eqnarray}
	\frac{\delta\Sigma_{ct}}{\delta\xi}-\frac{g'}{2}\partial_{\mu}\left(\Omega_{\mu}^{3}\frac{\delta\Sigma_{ct}}{\delta J}\right)+g'\varepsilon^{\alpha\beta}\Omega_{\mu}^{\alpha}\frac{\delta\Sigma_{ct}}{\delta\Omega_{\mu}^{\beta}} & = & 0.\label{eq:xi_equation-1}
\end{eqnarray}

\begin{itemize}
	\item Equation of $X_{\mu}$
\end{itemize}
\begin{eqnarray}
	\frac{\delta\Sigma_{ct}}{\delta X_{\mu}}+g'\frac{\delta\Sigma_{ct}}{\delta\Omega_{\mu}}+\frac{g'}{2}\Omega_{\mu}\frac{\delta\Sigma_{ct}}{\delta J} & =0 & .
\end{eqnarray}

\begin{itemize}
	\item Antighost equations
\end{itemize}
\begin{eqnarray}
	\frac{\delta\Sigma_{ct}}{\delta\overline{c}^{\alpha}}+\partial_{\mu}\frac{\delta\Sigma_{ct}}{\delta K_{\mu}^{\alpha}}+\varepsilon^{\alpha\beta}\Upsilon_{\mu}^{3}\frac{\delta\Sigma_{ct}}{\delta K_{\mu}^{\beta}} & = & 0,\nonumber \\
	\frac{\delta\Sigma_{ct}}{\delta\overline{c}^{3}}+\partial_{\mu}\frac{\delta\Sigma_{ct}}{\delta K_{\mu}^{3}} & = & 0.\label{eq:antighost_equation-1}
\end{eqnarray}

\begin{itemize}
	\item Ghost equations 
\end{itemize}
\begin{eqnarray}
	\mathcal{G}^{\alpha}\Sigma_{ct} & = & 0, \nonumber\\
	\mathcal{G}^{3}\Sigma_{ct} & = & 0,
\end{eqnarray}
where $\mathcal{G}^{\alpha}$ and $\mathcal{G}^{3}$ are defined by
Eqs. (\ref{eq:ghost_op_off_diag}) and (\ref{eq:ghost_op_3}), respectively. 
\begin{itemize}
	\item Rigid symmetries
\end{itemize}
\begin{eqnarray}
	\mathcal{R}^{\alpha}\Sigma_{ct} & = & 0, \nonumber \\
	\mathcal{R}^{3}\Sigma_{ct} & = & 0,
\end{eqnarray}
where $\mathcal{R}^{\alpha}$ and $\mathcal{R}^{3}$ are defined by
Eqs. (\ref{eq:rigid_op_off_diag}) and (\ref{eq:rigid_op_3}), respectively.
\begin{itemize}
	\item Integrated equation of $H$
\end{itemize}
\begin{eqnarray}
	\int d^{4}x\left(\frac{\delta\Sigma_{ct}}{\delta H}-2\lambda\vartheta\frac{\delta\Sigma_{ct}}{\delta J}\right)-\frac{\partial\Sigma_{ct}}{\partial\vartheta} & = & 0.
\end{eqnarray}

After imposing all the Ward identities, we obtain the following invariant
counterterms for the non-trivial cohomology $\Delta$:
\begin{eqnarray}
	\Delta & = & \sum_{i=0}^{28}a_{i}\Delta_{i}\label{eq:non_trivial_cohomology_final}
\end{eqnarray}
where $a_{1},\ldots,a_{28}$ are arbitrary constants that need renormalization
conditions to be specified. Due to their sizes, we prefer to present the list of counterterms $\left\{\Delta_i\right\}$ in Appendix B.

The integrated polynomial with ghost number -1 consistent with the
Ward identities is 
\begin{eqnarray}
	\Delta^{\left(-1\right)} & = & \int d^{4}x\left\{ -b_{1}\left(K_{\mu}^{a}+\partial_{\mu}\overline{c}^{a}-\varepsilon^{abc}\Upsilon_{\mu}^{b}\overline{c}^{c}\right)\left(W_{\mu}^{a}+\frac{1}{g}\Upsilon_{\mu}^{a}\right)\right.\nonumber \\
	&  & \left.+b_{2}\left[Y\left(\vartheta+H\right)+R^{a}\rho^{a}\right]\right\}.
\end{eqnarray}
Therefore, the trivial cohomology of $\mathcal{S}_{\Sigma}$ is
\begin{eqnarray}
	\mathcal{S}_{2\,\Sigma}\Delta^{\left(-1\right)} & = & \int d^{4}x\left\{ -b_{1}\left(\frac{\delta\Sigma}{\delta W_{\mu}^{a}}+i\partial_{\mu}b^{a}-\varepsilon^{abc}\zeta_{\mu}^{b}\overline{c}^{c}-i\varepsilon^{abc}\Upsilon_{\mu}^{b}b^{c}\right)\left(W_{\mu}^{a}+\frac{1}{g}\Upsilon_{\mu}^{a}\right)\right.\nonumber \\
	&  & +b_{1}\left(K_{\mu}^{a}+\partial_{\mu}\overline{c}^{a}-\varepsilon^{abc}\Upsilon_{\mu}^{b}\overline{c}^{c}\right)\left(\frac{\delta\Sigma}{\delta K_{\mu}^{a}}+\frac{1}{g}\zeta_{\mu}^{a}\right)\nonumber \\
	&  & \left.+b_{2}\left[\frac{\delta\Sigma}{\delta H}\left(\vartheta+H\right)+Y\frac{\delta\Sigma}{\delta Y}+\rho^{a}\frac{\delta\Sigma}{\delta\rho^{a}}+R^{a}\frac{\delta\Sigma}{\delta R^{a}}\right]\right\} ,\label{eq:trivial_cohomology}
\end{eqnarray}
where $b_{1}$ and $b_{2}$ are arbitrary constants to be fixed by
renormalization conditions.

\subsection{Bare action and the renormalization factors}

We would like to find a stable bare action, such that Eq. (\ref{eq:bare_sigma_plus_counterterm})
holds through a renormalization of the fields, sources and parameter
of the starting action. In cases where we have spontaneous symmetry
breaking, this becomes a bit subtle. When the scalar field $\varphi$
is expanded around one of the minimums of the classical potential
$V\left(\varphi\right)$, the linear terms in the fields are eliminated
but, in the other hand, cubic interactions are generated. These cubic
interactions produce the famous tadpoles that contribute to the one
point function $\left\langle H\right\rangle $. To have $\left\langle H\right\rangle =0$,
we must to add a linear counterterm ($\sim\int d^{4}x\vartheta^{3}H$)
to cancel the tadpoles contribution. If we analyze the invariant counterterm
$\Delta_{2}$ , see Eq. ($\ref{eq:delta_2}$), we notice the presence
of the term
\begin{eqnarray*}
	\int d^{4}x\vartheta^{2}O\left(\varphi\right) & = & \int d^{4}x\frac{1}{2}\vartheta^{2}\left(H^{2}+2\vartheta H+\rho^{a}\rho^{a}\right),
\end{eqnarray*}
which contains the exact term we mentioned. Obviously, due to the
BRST symmetry, we get it in a BRST-invariant combination. Such BRST
invariant combination is not present in the tree level action, so
it is impossible to absorb this counterterm through a multiplicative
renormalization. We need an additive renormalization to account it.
Something similar happens with other terms which are non-linear in
the sources. They are generated from one loop onward. 

\subsubsection*{Bare action}

Before we present the bare action, let us establish the notation.
The subscript ``$0$'' in the fields, sources and parameters means
that they are bare quantities. Now, to reabsorb all the counterterms,
we propose the bare action
\begin{eqnarray}
	\Sigma_{\textrm{bare}} & = & \left(S_{\textrm{Higgs}}\right)_{\textrm{bare}}+\left(S_{gf}\right)_{\textrm{bare}}+\left(S_{s}\right)_{\textrm{bare}}+\left(S_{\xi}\right)_{\textrm{bare}}+\left(S_{O}\right)_{\textrm{bare}}+\left(S_{\Upsilon}\right)_{\textrm{bare}}\nonumber \\
	&  & +\sum_{i=1}^{27}\hat{a}_{i,0}\left(\hat{\Delta}_{i}\right)_{0},\label{eq:bare_pop}
\end{eqnarray}
where $\left\{ (\hat{\Delta}_{i})_{0}\right\}$ are listed in Appendix C and
\begin{eqnarray}
	\left(S_{\textrm{Higgs}}\right)_{\textrm{bare}} & = & \int d^{4}x\left\{ \frac{1}{4}W_{0\mu\nu}^{\alpha}W_{0\mu\nu}^{\alpha}+\frac{1}{4}W_{0\mu\nu}^{3}W_{0\mu\nu}^{3}+\frac{1}{4}X_{0\mu\nu}X_{0\mu\nu}\right.\nonumber \\
	&  & +\frac{1}{2}\left(\partial_{\mu}H_{0}\right)^{2}+\frac{1}{2}\left(\partial_{\mu}\rho_{0}^{\alpha}\right)^{2}+\frac{1}{2}\left(\partial_{\mu}\rho_{0}^{3}\right)^{2}\nonumber \\
	&  & +\frac{1}{2}g_{0}W_{0\mu}^{\alpha}\left[\left(\partial_{\mu}H_{0}\right)\rho_{0}^{\alpha}-\left(\partial_{\mu}\rho_{0}^{\alpha}\right)\left(\vartheta_{0}+H_{0}\right)-\varepsilon^{\alpha\beta}\left(\partial_{\mu}\rho_{0}^{\beta}\right)\rho_{0}^{3}+\varepsilon^{\alpha\beta}\left(\partial_{\mu}\rho_{0}^{3}\right)\rho_{0}^{\beta}\right]\nonumber \\
	&  & +\frac{1}{2}g_{0}W_{0\mu}^{3}\left[\left(\partial_{\mu}H_{0}\right)\rho_{0}^{3}-\left(\partial_{\mu}\rho_{0}^{3}\right)\left(\vartheta_{0}+H_{0}\right)-\varepsilon^{\alpha\beta}\left(\partial_{\mu}\rho_{0}^{\alpha}\right)\rho_{0}^{\beta}\right]\nonumber \\
	&  & +\frac{1}{8}g_{0}^{2}W_{0\mu}^{\alpha}W_{0\mu}^{\alpha}\left[\left(\vartheta_{0}+H_{0}\right)^{2}+\rho_{0}^{\beta}\rho_{0}^{\beta}+\rho_{0}^{3}\rho_{0}^{3}\right]\nonumber \\
	&  & +\frac{1}{8}g_{0}^{2}W_{0\mu}^{3}W_{0\mu}^{3}\left[\left(\vartheta_{0}+H_{0}\right)^{2}+\rho_{0}^{\alpha}\rho_{0}^{\alpha}+\rho_{0}^{3}\rho_{0}^{3}\right]\nonumber \\
	&  & +\frac{1}{4}g_{0}g'_{0}W_{0\mu}^{3}X_{0\mu}\left(\vartheta_{0}+H_{0}\right)^{2}-\frac{1}{2}g_{0}g'_{0}\left(\vartheta_{0}+H_{0}\right)\varepsilon^{\alpha\beta}W_{0\mu}^{\alpha}\rho_{0}^{\beta}X_{0\mu}\nonumber \\
	&  & -\frac{1}{4}g_{0}g'_{0}W_{0\mu}^{3}X_{0\mu}\rho_{0}^{\alpha}\rho_{0}^{\alpha}+\frac{1}{4}g_{0}g'_{0}W_{0\mu}^{3}X_{0\mu}\rho_{0}^{3}\rho_{0}^{3}+\frac{1}{2}g_{0}g'_{0}W_{0\mu}^{\alpha}X_{0\mu}\rho_{0}^{3}\rho_{0}^{\alpha}\nonumber \\
	&  & +\frac{1}{2}g'_{0}X_{0\mu}\left[-\left(\partial_{\mu}H_{0}\right)\rho_{0}^{3}-\left(\partial_{\mu}\rho_{0}^{3}\right)\left(\vartheta_{0}+H_{0}\right)+\varepsilon^{\alpha\beta}\left(\partial_{\mu}\rho_{0}^{\alpha}\right)\rho_{0}^{\beta}\right]\nonumber \\
	&  & +\frac{1}{8}g'{}_{0}^{2}X_{0\mu}X_{0\mu}\left[\left(\vartheta_{0}+H_{0}\right)^{2}+\rho_{0}^{\alpha}\rho_{0}^{\alpha}+\rho_{0}^{3}\rho_{0}^{3}\right]\nonumber \\
	&  & \left.+\frac{\lambda_{0}}{4}\left[\left(\vartheta_{0}+H_{0}\right)^{2}+\rho_{0}^{\alpha}\rho_{0}^{\alpha}+\rho_{0}^{3}\rho_{0}^{3}\right]^{2}\right\} ,
\end{eqnarray}
\begin{eqnarray}
	\left(S_{gf}\right)_{\textrm{bare}} & = & \int d^{4}x\left[ib_{0}^{\alpha}\partial_{\mu}W_{0\mu}^{\alpha}-\overline{c}_{0}^{\alpha}\partial_{\mu}\left(-\partial_{\mu}c_{0}^{\alpha}+g_{0}\varepsilon^{\alpha\beta}c_{0}^{\beta}W_{0\mu}^{3}-g_{0}\varepsilon^{\alpha\beta}c_{0}^{3}W_{0\mu}^{\beta}\right)\right.\nonumber \\
	&  & +ib_{0}^{3}\partial_{\mu}W_{0\mu}^{3}-\overline{c}_{0}^{3}\partial_{\mu}\left(-\partial_{\mu}c_{0}^{3}+g_{0}\varepsilon^{\alpha\beta}c_{0}^{\alpha}W_{0\mu}^{\beta}\right)\nonumber \\
	&  & \left.+ib_{0}\partial_{\mu}X_{0\mu}-\overline{c}_{0}\partial_{\mu}\left(-\partial_{\mu}c_{0}\right)\right],
\end{eqnarray}
\begin{eqnarray}
	\left(S_{O}\right)_{\textrm{bare}} & = & \int d^{4}x\left[J_{0}O_{0}\left(\varphi\right)+\Xi_{0}\vartheta_{0}^{2}+\Omega_{0\mu}O_{0\mu}\left(\varphi\right)+\Omega_{0\mu}^{3}O_{0\mu}^{3}\left(\Phi\right)\right.\nonumber \\
	&  & \left.+\Omega_{0\mu}^{4}\partial_{\mu}O_{0}\left(\varphi\right)+\Omega_{0\mu}^{5}\left(X_{0\mu}-\partial_{\mu}\xi_{0}\right)\vartheta_{0}^{2}+\Omega_{0\mu}^{6}\partial_{\nu}X_{0\nu\mu}+\Omega_{0\mu}^{\alpha}O_{0\mu}^{\alpha}\left(\Phi\right)\right],
\end{eqnarray}
\begin{gather}
	\left(S_{\Upsilon}\right)_{\textrm{bare}}=\int d^{4}x\left[-\varepsilon^{\alpha\beta}\zeta_{0\mu}^{\alpha}\overline{c}_{0}^{3}W_{0\mu}^{\beta}-i\varepsilon^{\alpha\beta}\Upsilon_{0\mu}^{\alpha}b_{0}^{3}W_{0\mu}^{\beta}-\varepsilon^{\alpha\beta}\Upsilon_{0\mu}^{\alpha}\overline{c}_{0}^{3}\left(\partial_{\mu}c_{0}^{\beta}-g_{0}\varepsilon^{\beta\gamma}W_{0\mu}^{3}c_{0}^{\gamma}+g_{0}\varepsilon^{\beta\gamma}W_{0\mu}^{\gamma}c_{0}^{3}\right)\right.\nonumber \\
	+\varepsilon^{\alpha\beta}\zeta_{0\mu}^{\alpha}\overline{c}_{0}^{\beta}W_{0\mu}^{3}+i\varepsilon^{\alpha\beta}\Upsilon_{0\mu}^{\alpha}b_{0}^{\beta}W_{0\mu}^{3}+\varepsilon^{\alpha\beta}\Upsilon_{0\mu}^{\alpha}\overline{c}_{0}^{\beta}\left(\partial_{\mu}c_{0}^{3}-g_{0}\varepsilon^{\gamma\delta}W_{0\mu}^{\delta}c_{0}^{\gamma}\right)\nonumber \\
	\left.+\varepsilon^{\alpha\beta}\zeta_{0\mu}^{3}\overline{c}_{0}^{\alpha}W_{0\mu}^{\beta}+i\varepsilon^{\alpha\beta}\Upsilon_{0\mu}^{3}b_{0}^{\alpha}W_{0\mu}^{\beta}+\varepsilon^{\alpha\beta}\Upsilon_{0\mu}^{3}\overline{c}_{0}^{\alpha}\left(g_{0}\varepsilon^{\beta\gamma}W_{0\mu}^{\gamma}c_{0}^{3}+\partial_{\mu}c_{0}^{\beta}-g_{0}\varepsilon^{\beta\gamma}W_{0\mu}^{3}c_{0}^{\gamma}\right)\right],\label{eq:s_upsilon-1}
\end{gather}
\begin{eqnarray}
	\left(S_{\xi}\right)_{\textrm{bare}} & = & \int d^{4}x\left[\overline{\eta}_{0}\partial^{2}\eta_{0}+i\theta_{0}\left(\partial^{2}\xi_{0}-\partial_{\mu}X_{0\mu}\right)\right],
\end{eqnarray}
\begin{eqnarray}
	\left(S_{s}\right)_{\textrm{bare}} & = & \int d^{4}x\left[K_{0\mu}^{\alpha}\left(s_{2}W^{\alpha}\right)_{0}+K_{0\mu}^{3}\left(s_{2}W^{3}\right)_{0}+Y_{0}\left(s_{2}H\right)_{0}+R_{0}^{\alpha}\left(s_{2}\rho^{\alpha}\right)_{0}+R_{0}^{3}\left(s_{2}\rho^{3}\right)_{0}\right.\nonumber \\
	&  & \left.+L_{0}^{\alpha}\left(s_{2}c^{\alpha}\right)_{0}+L_{0}^{3}\left(s_{2}c^{3}\right)_{0}+y_{0}\left(s_{1}H\right)_{0}+r_{0}^{\alpha}\left(s_{1}\rho^{\alpha}\right)_{0}+r_{0}^{3}\left(s_{1}\rho^{3}\right)_{0}\right],
\end{eqnarray}
For convenience, we introduced the short notation
\begin{eqnarray}
	W_{0\mu\nu}^{\alpha} & = & \partial_{\mu}W_{0\nu}^{a}-\partial_{\nu}W_{0\mu}^{a}+g_{0}\varepsilon^{abc}W_{0\mu}^{b}W_{0\nu}^{c},\nonumber \\
	X_{0\mu\nu} & = & \partial_{\mu}X_{0\nu}-\partial_{\nu}X_{0\mu},\nonumber \\
	O_{0} & = & \frac{1}{2}\left(H_{0}^{2}+2\vartheta_{0}H_{0}+\rho_{0}^{a}\rho_{0}^{a}\right),\nonumber \\
	O_{0\mu} & = & O_{\mu}^{3}\left(\varphi_{0}\right)-\frac{g'_{0}}{2}\left(O_{0}+\frac{\vartheta_{0}^{2}}{2}\right)X_{0\mu},\nonumber \\
	O_{0\mu}^{3} & = & O_{\mu}^{3}\left(\varphi_{0}\right)-\frac{g'_{0}}{2}\left(O_{0}+\frac{\vartheta_{0}^{2}}{2}\right)\partial_{\mu}\xi_{0},\nonumber \\
	\hat{J}_{0} & = & J_{0}+\partial_{\mu}\Omega_{0\mu}^{4}.
\end{eqnarray}
$\left(s\phi\right)_{0}$ stands for the BRST transformation of the
field $\phi$ , where all renormalized quantities are replaced by
their bare counterparts. The next step is to propose a renormalization
for the bare quantities. 

\subsubsection*{Renormalization factors}

For the fields we consider the following multiplicative renormalizations:
\begin{gather}
	W_{0\mu}^{a}=Z_{WW}^{\frac{1}{2}}W_{\mu}^{a}+Z_{W\Upsilon}^{\frac{1}{2}}\Upsilon_{\mu}^{a},\qquad X_{0\mu}=Z_{X}^{\frac{1}{2}}X_{\mu},\qquad H_{0}=Z_{H}^{\frac{1}{2}}H,\qquad\rho_{0}^{a}=Z_{\rho}^{\frac{1}{2}}\rho^{a},\nonumber \\
	c_{0}^{a}=Z_{C}^{\frac{1}{2}}c^{a},\qquad\overline{c}_{0}^{a}=Z_{C}^{\frac{1}{2}}\overline{c}^{a},\qquad b_{0}^{a}=Z_{B}^{\frac{1}{2}}b^{a},\qquad c_{0}=Z_{c}^{\frac{1}{2}}c,\qquad\overline{c}_{0}=Z_{c}^{\frac{1}{2}}\overline{c},\nonumber \\
	b_{0}=Z_{b}^{\frac{1}{2}}b,\qquad\xi_{0}=Z_{\xi}^{\frac{1}{2}}\xi,\qquad\eta_{0}=Z_{\eta}^{\frac{1}{2}}\eta,\qquad\overline{\eta}_{0}=Z_{\eta}^{\frac{1}{2}}\overline{\eta},\qquad\theta_{0}=Z_{\theta}^{\frac{1}{2}}\theta.
\end{gather}
The renormalization of $W_{\mu}^{a}$ might called the reader's attention,
since it involves also the source $\Upsilon_{\mu}^{a}$. This is what
the trivial cohomology of $\mathcal{S}_{2\,\Sigma}$ suggests, see
Eq. (\ref{eq:trivial_cohomology}). In fact, $W_{\mu}^{a}$ and $\Upsilon_{\mu}^{a}$
have the same dimension and the same $SU\left(2\right)$ transformation.
The renormalization of a field can involve a source, but the reverse
is not true, \emph{i.e.}, a source cannot acquire the nature of a
field through renormalization. 

We also propose a multiplicative renormalization for the parameter
$\vartheta$, $g$, $g'$ and $\lambda$, namely,
\begin{gather}
	\vartheta_{0}=Z_{\vartheta}^{\frac{1}{2}}\vartheta,\qquad g_{0}=Z_{g}g,\qquad g'_{0}=Z_{g'}g',\qquad\lambda_{0}=Z_{\lambda}\lambda.
\end{gather}
The other parameters, $\hat{a}_{1,0},\ldots,\hat{a}_{26,0}$ need
additive renormalizations that start from one loop onward. Therefore,
they can be written as
\begin{eqnarray}
	\hat{a}_{i,0} & = & \epsilon\hat{a}_{i}.
\end{eqnarray}

The sources introduced to compute correlation functions of the gauge
invariant operators require a \emph{matricial renormalization}, as
we have different operators with the same quantum numbers. For the
sources $J$ and $\Xi$, we define a 2x2 matrix, namely,
\begin{eqnarray}
	\left[\begin{array}{c}
		J_{0}\\
		\Xi_{0}
	\end{array}\right] & = & \left[\begin{array}{cc}
		Z_{JJ} & Z_{J\Xi}\\
		Z_{\Xi J} & Z_{\Xi\Xi}
	\end{array}\right]\left[\begin{array}{c}
		J\\
		\Xi
	\end{array}\right].
\end{eqnarray}
For $\left\{ \Omega_{\mu},\,\Omega_{\mu}^{3},\,\Omega_{\mu}^{4},\,\Omega_{\mu}^{5},\,\Omega_{\mu}^{6}\right\} $,
we have a 5x5 matrix, namely,
\begin{eqnarray}
	\left[\begin{array}{c}
		\Omega_{0\mu}\\
		\Omega_{0\mu}^{3}\\
		\Omega_{0\mu}^{4}\\
		\Omega_{0\mu}^{5}\\
		\Omega_{0\mu}^{6}
	\end{array}\right] & = & \left[\begin{array}{ccccc}
		Z_{00} & Z_{03} & Z_{04} & Z_{05} & Z_{06}\\
		Z_{30} & Z_{33} & Z_{34} & Z_{35} & Z_{36}\\
		Z_{40} & Z_{43} & Z_{44} & Z_{45} & Z_{46}\\
		Z_{50} & Z_{53} & Z_{54} & Z_{55} & Z_{56}\\
		Z_{60} & Z_{63} & Z_{64} & Z_{65} & Z_{66}
	\end{array}\right]\left[\begin{array}{c}
		\Omega_{\mu}\\
		\Omega_{\mu}^{3}\\
		\Omega_{\mu}^{4}\\
		\Omega_{\mu}^{5}\\
		\Omega_{\mu}^{6}
	\end{array}\right].
\end{eqnarray}
The other sources, such as those coupled with the BRST transformations
of the fields, do not require a special renormalization. A simple
multiplicative renormalization is sufficient, \emph{i.e.},
\begin{gather}
	K_{0\mu}^{a}=Z_{K}K_{\mu}^{a},\qquad L_{0}^{a}=Z_{L}L^{a},\qquad Y_{0}=Z_{Y}Y,\qquad R_{0}^{a}=Z_{R}R^{a},\nonumber \\
	\Upsilon_{0\mu}^{a}=Z_{\Upsilon}\Upsilon_{\mu}^{a},\qquad\zeta_{0\mu}^{a}=Z_{\zeta}\zeta_{\mu}^{a},\qquad y_{0}=Z_{y}y,\qquad r_{0}^{a}=Z_{r}r^{a}.
\end{gather}

Equating (\ref{eq:bare_pop}) with $\Sigma+\epsilon\left(\Delta_{2}+\mathcal{S}_{2\,\Sigma}\Delta^{\left(-1\right)}\right)$,
where $\Delta_{2}$ and $\mathcal{S}_{2\,\Sigma}\Delta^{\left(-1\right)}$
are given by Eqs. (\ref{eq:non_trivial_cohomology_final}) and (\ref{eq:trivial_cohomology}),
and comparing both sides of the equation, we conclude that
\begin{gather}
	Z_{WW}^{\frac{1}{2}}=Z_{B}^{-\frac{1}{2}}=1+\epsilon\frac{1}{2}\left(a_{0}-2b_{1}\right),\nonumber \\
	Z_{W\Upsilon}^{\frac{1}{2}}=-\epsilon\frac{b_{1}}{g},\nonumber \\
	Z_{g}=1-\epsilon\frac{a_{0}}{2},\nonumber \\
	Z_{H}^{\frac{1}{2}}=Z_{\rho}^{\frac{1}{2}}=Z_{\vartheta}^{\frac{1}{2}}=1+\epsilon b_{2},\nonumber \\
	Z_{X}^{\frac{1}{2}}=Z_{b}^{-\frac{1}{2}}=Z_{\theta}^{-\frac{1}{2}}=Z_{\xi}^{-\frac{1}{2}}=Z_{g'}^{-1}=1+\epsilon\frac{a_{1}}{2},\nonumber \\
	Z_{C}^{\frac{1}{2}}=Z_{\overline{C}}^{\frac{1}{2}}=1+\epsilon\frac{b_{1}}{2}=Z_{g}^{-\frac{1}{2}}Z_{WW}^{-\frac{1}{4}},\nonumber \\
	Z_{c}^{\frac{1}{2}}=Z_{\overline{c}}^{\frac{1}{2}}=1,\nonumber \\
	Z_{\eta}^{\frac{1}{2}}=Z_{\overline{\eta}}^{\frac{1}{2}}=1,\nonumber \\
	Z_{\lambda}=1+\epsilon\frac{a_{3}}{\lambda},\label{eq:field_factors}
\end{gather}
are the renormalization factors for the fields and parameters. For
the sources that introduce the BRST transformations, we get
\begin{gather}
	Z_{K}=Z_{g}^{-\frac{1}{2}}Z_{WW}^{-\frac{1}{4}},\nonumber \\
	Z_{L}=Z_{g}^{-1}Z_{C}^{-1}=Z_{WW}^{\frac{1}{2}},\nonumber \\
	Z_{Y}=Z_{R}=Z_{g}^{-\frac{1}{2}}Z_{WW}^{\frac{1}{4}}Z_{H}^{-\frac{1}{2}},\nonumber \\
	Z_{y}=Z_{r}=Z_{g'}^{-1}Z_{c}^{-\frac{1}{2}}Z_{\rho}^{-\frac{1}{2}}=1+\epsilon\left(\frac{a_{0}}{2}-b_{2}\right).
\end{gather}
The renormalization factor of the sources that introduce the BRST
invariant fields, we obtain
\begin{gather}
	Z_{\zeta}=Z_{\Upsilon}=Z_{C}^{-\frac{1}{2}}Z_{WW}^{-\frac{1}{2}},
\end{gather}
\begin{gather}
	Z_{JJ}=1+\epsilon\left(-\frac{a_{2}}{\lambda}+\frac{a_{3}}{\lambda}\right),\nonumber \\
	Z_{J\Xi}=Z_{\Xi\Xi}=0,\nonumber \\
	Z_{\Xi J}=1+\epsilon\left(\frac{a_{2}}{\lambda}-2\frac{a_{4}}{\lambda}\right),
\end{gather}
and
\begin{gather}
	Z_{00}=1,\,Z_{03}=\epsilon\left(\frac{a_{2}}{\lambda}-\frac{a_{3}}{\lambda}+a_{5}\right),\,Z_{04}=Z_{05}=Z_{06}=0,\nonumber \\
	Z_{30}=0,\,Z_{33}=1+\epsilon\left(\frac{a_{2}}{\lambda}-\frac{a_{3}}{\lambda}\right),\,Z_{34}=Z_{35}=Z_{36}=0,\nonumber \\
	Z_{40}=0,\,Z_{43}=\epsilon a_{7},\,Z_{44}=1+\epsilon\left(\frac{a_{2}}{\lambda}-\frac{a_{3}}{\lambda}\right),\,Z_{45}=Z_{46}=0,\nonumber \\
	Z_{50}=0,\,Z_{53}=\epsilon\frac{g'}{\lambda}\left(\frac{a_{2}}{2}-\frac{a_{3}}{4}-a_{4}\right),\,Z_{54}=Z_{55}=Z_{56}=0,\nonumber \\
	Z_{60}=\epsilon\frac{a_{1}}{g'},\,Z_{63}=-\epsilon g'a_{10},\,Z_{64}=Z_{65}=Z_{66}=0.
\end{gather}

The additive renormalization of the parameters $\hat{a}_{i,0}$ can
be expressed as
\begin{eqnarray}
	\hat{a}_{i,0} & = & \begin{cases}
		\epsilon a_{i}, & i=1,\ldots,5\\
		\epsilon a_{i+2}, & i=6,\ldots,26\\
		\epsilon\frac{b_{1}}{g}, & i=27.
	\end{cases}
\end{eqnarray}
Therefore, we obtain the following relation for the first five coefficients
\begin{eqnarray}
	\hat{a}_{1,0} & = & Z_{X}-1,\nonumber \\
	\hat{a}_{2,0} & = & \lambda\left(2-Z_{\lambda}-Z_{JJ}\right),\nonumber \\
	\hat{a}_{3,0} & = & \lambda\left(1-Z_{\lambda}\right),\nonumber \\
	\hat{a}_{4,0} & = & \frac{\lambda}{2}\left(3-Z_{\lambda}-Z_{JJ}-Z_{\Xi J}\right),\nonumber \\
	\hat{a}_{5,0} & = & -3+Z_{03}+2Z_{\lambda}+Z_{JJ}.
\end{eqnarray}
We also have a relation for $\hat{a}_{8,0}$ and $Z_{63}$, namely,
\begin{eqnarray*}
	Z_{63} & = & -g'_{0}a_{8,0}.
\end{eqnarray*}

\section{Comments on the Ward identities implications\protect\label{sec:Exploring-some-of}}

In Section \ref{sec:Symmetries-and-Ward}, we presented the set of
symmetries and Ward identities of the model defined by the action
$\Sigma$, Eq. (\ref{eq:starting_action-1}), but we did not discuss
their consequences beyond the implications for renormalization. In
Appendix A, a few results for the two-point functions are derived
in more detail with the help of the BRST symmetries and the equations
of motion of the Nakanishi-Lautrup fields. In this section, we underline
what we think are the most important results. 

\subsection{Non-renormalization Theorem and the transversality of the gauge propagators}

Certainly, the Landau gauge is fundamental for deriving some of the
important results we just found in the previous section. We would
like to point out the \emph{Non-renormalization Theorem
	\begin{gather}
		Z_{c}Z_{WW}^{\frac{1}{2}}Z_{g}=1,\label{eq:non_renorm_theorem}
	\end{gather}
}which the reader can explicitly verify by using the renormalization
factors from Eq. (\ref{eq:field_factors}). Only in the Landau gauge
do we have the ghost equations (\ref{eq:off_diagonal_ghost_equation})
and (\ref{eq:diagonal_ghost_equation}), whose main direct consequence
is the result (\ref{eq:non_renorm_theorem}). This is well-known feature
of the Landau gauge and the ghost equation, see \cite{Piguet:1995er}. Through
(\ref{eq:non_renorm_theorem}), it is possible to determine $Z_{c}^{\frac{1}{2}}$
without even computing a correlation function with a ghost field,
which is certainly remarkable convenient. 

Another important aspect of the Landau gauge is the transversality
of the gauge fields propagators. At the quantum level, the gauge condition
is implemented by the $b$-field equations of motion, see Eq. (\ref{eq:b_equations}).
By using this equations of motion, it is rather straightforward to
demonstrate (see Appendix A) that 
\begin{eqnarray}
	0  &=&  \partial_{\mu}^{x}\left\langle W_{\mu}^{a}\left(x\right)W_{\nu}^{a}\left(y\right)\right\rangle ,\nonumber \\
	0 &=&  \partial_{\mu}^{x}\left\langle X_{\mu}\left(x\right)X_{\nu}\left(y\right)\right\rangle ,\nonumber \\
	0  &=&  \partial_{\mu}^{x}\left\langle W_{\mu}^{3}\left(x\right)X_{\nu}\left(y\right)\right\rangle ,
\end{eqnarray}
meaning that all these propagators are transverse.

The last direct consequence of the Landau gauge we would like to point
out is the relation
\begin{gather}
	Z_{H}^{\frac{1}{2}}=Z_{\rho}^{\frac{1}{2}}=Z_{\vartheta}^{\frac{1}{2}}.\label{eq:Z_H_Z_rho}
\end{gather}
The Landau gauge preserves the global gauge symmetry, which plays
a key role in deriving Eq. (\ref{eq:Z_H_Z_rho}). It follows from
(\ref{eq:Z_H_Z_rho}) that
\begin{gather}
	\varphi_{0}=Z_{H}^{\frac{1}{2}}\varphi,
\end{gather}
a very important result for constructing a consistent bare action
without requiring significant changes to the original ideas. 

\subsection{Integrated $H$-equation, tadpoles and effective potential}

In general, when we casually add some composite operator to a certain
theory, we do not expect to obtain a new Ward identity. However, our
experience with the $U\left(1\right)$ and $SU\left(2\right)$ Higgs
models has shown that this is possible \cite{Capri:2020ppe,Dudal:2021pvw}. In both cases, we
found an integrated Ward identity involving the equation of motion
of the Higgs field $H\left(x\right)$. Here, in the $SU\left(2\right)\times U\left(1\right)$
Higgs model, we have a similar Ward identity, namely, the Eq. (\ref{eq:H_equation}).
As we discussed in Subsection \ref{subsec:Integrated-equation-of},
the symmetry breaking of the global transformations $H\rightarrow H+\epsilon$
and $\vartheta\rightarrow\vartheta-\epsilon$ is proportional to the
$O\left(\varphi\right)$ operator. That is why introducing this operator
into the starting action produces a new Ward identity. We do not expect
this identity to affect any aspect of the original theory, such as
the renormalization factors of the fundamental fields, for instance.
However, we do anticipate an effect on the correlation functions of
the gauge-invariant composite operators. Indeed, this was already
observed it the previous section, when we determine the invariant
counterterms. 

The Ward identity (\ref{eq:H_equation}) is non-anomalous, thus we
can find an effective action $\Gamma$ (the generating functional
of the 1PI functions) such that
\begin{eqnarray}
	\int d^{4}x\left(\frac{\delta\Gamma}{\delta H}-2\lambda\vartheta\frac{\delta\Gamma}{\delta J}\right)-\frac{\partial\Gamma}{\partial\vartheta} & = & \int d^{4}x\left(2\vartheta J-2\vartheta\Xi-2\vartheta\Omega_{\mu}^{5}\left(X_{\mu}-\partial_{\mu}\xi\right)\right).
\end{eqnarray}
Therefore, we have the following relationship for the 1PI functions:
\begin{eqnarray}
	\left\langle H\right\rangle _{1PI}-2\lambda\vartheta\left\langle O\right\rangle _{1PI}-\frac{\partial E_{vacuum}}{\partial\vartheta} & = & 0, \label{eq:H_O_vacuum}
\end{eqnarray}
where $E_{vacuum}$ is the vacuum energy, and the 1PI functions are
defined by
\begin{eqnarray}
	\left\langle H\right\rangle _{1PI} & = & \left.\frac{\delta\Gamma}{\delta H}\right|_{\textrm{fields=sources=0}},\nonumber \\
	\left\langle O\right\rangle _{1PI} & = & \left.\frac{\delta\Gamma}{\delta J}\right|_{\textrm{fields=sources=0}}.
\end{eqnarray}
To compute $E_{vacuum}$, we first determine the effective potential
$V\left(\phi_{o}\right)$, which is defined as
\begin{eqnarray}
	\Gamma\left[\phi_{o}\right] & = & V\left(\phi_{o}\right)\left(\int d^{4}x\right),
\end{eqnarray}
where $\phi_{o}$ is a position independent field. Then, the vacuum
energy is simply the minimum of the effective potential, \emph{i.e.},
$E_{vacuum}=V\left(\phi_{o}\right)_{min}$.

If we cancel the tadpoles and maintain $\vartheta$ always as the
minimum of the effective potential- \emph{i.e.}, by setting $\left\langle H\right\rangle _{1PI}=0$
and $\frac{\partial E_{vacuum}}{\partial\vartheta}=0$-then the condensate
of $O$ vanishes, 
\begin{eqnarray}
	\left\langle O\right\rangle _{1PI} & = & 0.\label{eq:O_condensate}
\end{eqnarray}
Actually, we hope not to find a condensate other than $\left\langle \varphi^{\dagger}\varphi\right\rangle \sim\vartheta^{2}$,
so it is very important to have a Ward identity that controls this and, under consistent
renormalization conditions, ensures it along with (\ref{eq:O_condensate}).

\subsection{Non-integrated $X$-equation, anomalous dimension of $O_{\mu}$ and
	the transversality of $\left\langle O_{\mu}\left(x\right)O_{\nu}\left(y\right)\right\rangle $\protect\label{subsec:Non-integrated--equation-and}}

Throughout the article we have repeatedly emphasized that the Landau
gauge preserves the global gauge symmetries, also known as rigid symmetries.
In particular, the $U\left(1\right)$ local symmetry is always linearly
broken, so we have a non-integrated Ward identity associated with
it, see Eq. (\ref{eq:rigid_abelian}). For non-abelian symmetries
this is not possible because the ghost fields are not free anymore.
As we saw in Subsection \ref{subsec:Local--Symmetry}, to write a
non-integrated Ward identity for the $SU\left(2\right)$ symmetry,
we had to introduce the source $\Upsilon_{\mu}^{a}$ and the term
$S_{\Upsilon}$.

The $U\left(1\right)$ global gauge symmetry of $\Sigma$ can be expressed
as
\begin{gather}
	-\frac{g'}{2}\varepsilon^{\alpha\beta}\rho^{\beta}\frac{\delta\Sigma}{\delta\rho^{\alpha}}+\frac{g'}{2}\varepsilon^{\alpha\beta}R^{\alpha}\frac{\delta\Sigma}{\delta R^{\beta}}+\frac{g'}{2}\left(\vartheta+H\right)\frac{\delta\Sigma}{\delta\rho^{3}}-\frac{g'}{2}R^{3}\frac{\delta\Sigma}{\delta Y}-\frac{g'}{2}\rho^{3}\frac{\delta\Sigma}{\delta H}+\frac{g'}{2}Y\frac{\delta\Sigma}{\delta R^{3}}-\frac{\delta\Sigma}{\delta\xi}\nonumber \\
	=\partial_{\mu}\mathscr{J}_{\mu},
\end{gather}
where $\mathscr{J}_{\mu}$ is the current associated with it. By doing
an explicit computation, we find that
\begin{eqnarray}
	\mathscr{J}_{\mu} & = & -i\partial_{\mu}b-\frac{\delta\Sigma}{\delta X_{\mu}}.\label{eq:J_U1_global}
\end{eqnarray}
The way we can express $\mathscr{J}_{\mu}$, through a total derivative
of $X_{\mu}$, is the main reason why it is possible to write a non-integrated
Ward identity. So, the non-trivial part of the current should be contained
in $\frac{\delta\Sigma}{\delta X_{\mu}}$. Indeed, we do have that
\begin{eqnarray}
	\frac{\delta\Sigma}{\delta X_{\mu}} & = & -g'O_{\mu}-\frac{g'}{2}\Omega_{\mu}O-\partial_{\nu}X_{\nu\mu}+i\partial_{\mu}\theta\nonumber \\
	&  & -i\partial_{\mu}b-\frac{g'\vartheta^{2}}{4}\Omega_{\mu}+\Omega_{\mu}^{5}\vartheta^{2}+\partial^{2}\Omega_{\mu}^{6}-\partial_{\mu}\partial_{\nu}\Omega_{\nu}^{6}.
\end{eqnarray}
As the $U\left(1\right)$ current $O_{\mu}$ and the composite operator
$O$ were introduced in the starting action, we were able to derive
the Ward identity (\ref{eq:X_ward_identity}). This result is familiar,
as we also obtained an analogous identity in the $U\left(1\right)$
Higgs model \cite{Capri:2020ppe}. This appears to be a general feature of the
Abelian symmetries. 

Similar to the $H$-equation, the $X$-equation establishes relationships
between correlation functions of composite operators and elementary
fields. Consequently, it also provides relations to the renormalization
factors, such as
\begin{eqnarray}
	\hat{a}_{1,0} & = & Z_{X}-1.
\end{eqnarray}
Notably,
\begin{eqnarray}
	Z_{00} & = & 1,
\end{eqnarray}
which implies that the anomalous dimension
\begin{eqnarray}
	\gamma_{O_{\mu}O_{\mu}} & = & \mu\frac{d\log Z_{00}}{d\mu}, \label{eq:anomalous_dimension}
\end{eqnarray}
is zero. This result is expected for a conserved current \cite{Itzykson:1980rh}.
Therefore, confirming it \emph{via }Ward identity is a significant
achievement. Since the renormalization of $O_\mu$ is matricial, we actually have an anomalous dimension matrix, and the other elements may not be null. In fact, this is expected according to  \cite{Collins:2005nj}.

In the Appendix E, we explicitly show that the two-point function of $O_{\mu}$ is related to the two-point function of $X_{\mu}$ and $\langle O\left(x\right) \rangle$, namely,
\begin{eqnarray}
	g'{}^{2}\left\langle O_{\mu}\left(x\right)O_{\nu}\left(y\right)\right\rangle  & = & \left(\partial^{4}\right)^{x}\left\langle X_{\mu}\left(x\right)X_{\nu}\left(y\right)\right\rangle +\left(\partial^{2}\delta_{\mu\nu}-\partial_{\mu}\partial_{\nu}\right)^{x}\delta^{4}\left(x-y\right)\nonumber \\
	&  & +\frac{g'{}^{2}}{2}\delta_{\mu\nu}\delta^{4}\left(x-y\right)\left(\left\langle O\left(x\right)\right\rangle +\frac{\vartheta^{2}}{2}\right). \label{eq:OOvector1}
\end{eqnarray}
This result also follows from the equation of motion of $X$. Such a result is a special feature of Abelian symmetries, since in the $U\left(1\right)$ Higgs model, we have found a similar result \cite{Dudal:2021pvw}.  We particularly would like to emphasize the fact that the transverse part of the two-point function of
$O_{\mu}$ is determined by $\left\langle X_{\mu}\left(x\right)X_{\nu}\left(y\right)\right\rangle $.
Furthermore, notice that the longitudinal part is given by
\begin{gather}
	\frac{\partial_\mu \partial_\nu}{\partial^2}\left\langle O_{\mu}\left(x\right)O_{\nu}\left(y\right)\right\rangle =\frac{1}{2}\left(\left\langle O\left(x\right)\right\rangle +\frac{\vartheta^{2}}{2}\right). \label{eq:OO_transv1}
\end{gather}
Independently of the renormalization conditions we decide to adopt,
this is a constant term, thus, no propagating mode can be associated
to it.

\section{Conclusions\label{sec:Conclusions}}

For any gauge field theory in continuous space-time, we must choose a gauge condition and implement it at the quantum level. Besides the difficulties involved in this implementation, we must also ensure that physical observables remain independent of the gauge condition. These are truly non-trivial aspects to consider, especially given the existence of Gribov-Singer copies and the fact that we typically employ gauge-dependent objects to calculate physical quantities, such as masses, cross-sections, etc. Although we do not yet have a definitive solution to the problem of copies, composite gauge-invariant operators provide a potentially explicit and elegant gauge-independent formalism. Previously, we constructed this formalism for the fundamental $U(1)$ and $SU(2)$ Higgs models, and more recently, in \cite{Dudal:2023jsu}, we extended it to the fundamental $SU(2) \times U(1)$ case. In this article, we establish the renormalization of the composite operators proposed in the latter case, completing an important technical step necessary for obtaining consistent perturbative results. The next step will be to explicitly verify some of these results at one-loop order. This is one of the projects we intend to undertake in the near future, particularly to confirm that the  K\"{a}ll\'{e}n-Lehmann representations of the two-point functions of the operators $\left\{ O,\,O_{\mu},\,O_{\mu}^3,\,O_{\mu}^{\alpha}\right\}$ are indeed positive. \par

Due to the special features of the Higgs model and the gauge-invariant operators, we were able to derive important Ward identities that establish relationships between the correlation functions of elementary fields and those of composite operators. For instance, we obtained exact results for $\langle O_{\mu}(x) O_{\nu}(y) \rangle$ and $\langle O(x) \rangle$, as shown in Eqs. (\ref{eq:OOvector}), (\ref{eq:OO_transv}), and (\ref{eq:H_O_vacuum}). Other significant results derivable from the Ward identities include the non-renormalization theorem (\ref{eq:non_renorm_theorem}) and the anomalous dimension of $O_{\mu}$. These results are expected to hold beyond the perturbative level, making them valuable constraints for any non-perturbative methods that might eventually be employed. Unfortunately, we cannot derive analogous Ward identities for the other gauge-invariant operators, which prevents us from making exact statements about, for example, the pole structure of their correlation functions. A diagrammatic analysis, as performed in \cite{Maas:2020kda}, may provide useful insights into these questions.  \par

As we discussed, a gauge-invariant framework has conceptual advantages over the usual gauge-dependent approach, but we can also emphasize its technical and practical benefits. By employing gauge-invariant operators, in principle, it would be possible to navigate through different gauges without changing the physical content of the underlying theory. This feature would allow us to compare distinct methods and approaches for computing correlation functions. In particular, we can mention the lattice simulations, which have been an important source of non-perturbative insights into gauge field theories.. As lattice gauge theory is naturally formulated in terms of gauge-invariant objects, we have a compelling motivation to keep studying gauge-invariant operators in the continuous space-time. \par

Now that we have developed a gauge-invariant setup for the fundamental $SU(2)\times U(1)$ Higgs model, the natural next step would be to include fermions of the Standard Model. This is something we are looking forward to. Certainly, a gauge-invariant framework for the Standard Model is important for gaining a clear understanding of the fundamental structure it describes, which will eventually allow us to incorporate some of the missing elements of the Standard Model, such as the dark matter, neutrino's masses, etc, in a consistent manner.

\section*{Acknowledgments}

The author is grateful to Antonio Duarte Pereira Junior for helpful discussions. The author would like to thank the Brazilian agency FAPERJ for financial support. G.~Peruzzo is a FAPERJ postdoctoral fellow in the P{\'O}S-DOUTORADO NOTA 10 program under the contracts E-26/205.924/2022 and E-26/205.925/2022.

\section*{Appendix A: Propagators}

\subsection*{Tree-level propagators}

The propagators of the theory are determined by the quadratic part
of $\Sigma$, namely 
\begin{eqnarray}
	\Sigma_{quad} & = & \int d^{4}x\left[\frac{1}{4}\left(\partial_{\mu}W_{\nu}^{\alpha}-\partial_{\nu}W_{\mu}^{\alpha}\right)^{2}+\frac{1}{4}\left(\partial_{\mu}W_{\nu}-\partial_{\nu}W_{\mu}\right)^{2}+\frac{1}{4}\left(\partial_{\mu}X_{\nu}-\partial_{\nu}X_{\mu}\right)^{2}\right.\nonumber \\
	&  & +\frac{1}{2}\left(\partial_{\mu}H\right)^{2}+\lambda\vartheta^{2}H^{2}+\frac{1}{2}\left(\partial_{\mu}\rho^{\alpha}\right)^{2}+\frac{1}{2}\left(\partial_{\mu}\rho^{3}\right)^{2}\nonumber \\
	&  & -\frac{1}{2}\vartheta gW_{\mu}^{\alpha}\left(\partial_{\mu}\rho^{\alpha}\right)-\frac{1}{2}\vartheta gW_{\mu}^{3}\left(\partial_{\mu}\rho^{3}\right)-\frac{1}{2}\vartheta g'X_{\mu}\left(\partial_{\mu}\rho^{3}\right)\nonumber \\
	&  & +\frac{1}{8}\vartheta^{2}g^{2}W_{\mu}^{\alpha}W_{\mu}^{\alpha}+\frac{1}{8}\vartheta^{2}g^{2}W_{\mu}^{3}W_{\mu}^{3}+\frac{1}{4}\vartheta^{2}gg'W_{\mu}^{3}X_{\mu}+\frac{1}{8}\vartheta^{2}g'^{2}X_{\mu}X_{\mu}\nonumber \\
	&  & +ib^{\alpha}\partial_{\mu}W_{\mu}^{\alpha}+ib^{3}\partial_{\mu}W_{\mu}^{3}+ib\partial_{\mu}X_{\mu}+\overline{c}^{\alpha}\partial^{2}c^{\alpha}+\overline{c}^{3}\partial^{2}c^{3}+\overline{c}\partial^{2}c\nonumber \\
	&  & \left.+\overline{\eta}\partial^{2}\eta+i\theta\left(\partial^{2}\xi-\partial_{\mu}X_{\mu}\right)\right].
\end{eqnarray}
We can rewrite it as
\begin{eqnarray}
	\Sigma_{quad} & = & \int d^{4}xd^{4}y\left[\frac{1}{2}\phi_{i}\left(x\right)\Delta_{ij}\left(x-y\right)\phi_{j}\left(y\right)+\frac{1}{2}\Phi_{I}\left(x\right)D_{IJ}\left(x-y\right)\Phi_{J}\left(y\right)\right]\nonumber \\
	&  & +\int d^{4}x\left[\frac{1}{2}\left(\partial_{\mu}H\right)^{2}+\lambda\vartheta^{2}H^{2}\right.\nonumber \\
	&  & \left.+\overline{c}^{\alpha}\partial^{2}c^{\alpha}+\overline{c}^{3}\partial^{2}c^{3}+\overline{c}\partial^{2}c+\overline{\eta}\partial^{2}\eta\right],
\end{eqnarray}
where the fields $\phi$ and $\Phi$ are defined as
\begin{eqnarray}
	\phi & = & \left(\begin{array}{c}
		X_{\mu}\\
		W_{\mu}^{3}\\
		b\\
		b^{3}\\
		\rho^{3}\\
		\theta\\
		\xi
	\end{array}\right),\qquad\Phi=\left(\begin{array}{c}
		W_{\mu}^{\alpha}\\
		b^{\alpha}\\
		\rho^{\alpha}
	\end{array}\right),
\end{eqnarray}
and the mixing matrices, $\Delta$ and $D$, are defined by
\begin{eqnarray}
	\Delta\left(x-y\right) & = & \left(\begin{array}{ccccccc}
		-\delta_{\mu\nu}\partial^{2}+\partial_{\mu}\partial_{\nu}+\frac{g'^{2}\vartheta^{2}}{4}\delta_{\mu\nu} & \frac{\vartheta^{2}gg'}{4}\delta_{\mu\nu} & -i\partial_{\mu} & 0 & -\frac{\vartheta g'}{2}\partial_{\mu} & i\partial_{\mu} & 0\\
		\frac{\vartheta^{2}gg'}{4}\delta_{\mu\nu} & -\delta_{\mu\nu}\partial^{2}+\partial_{\mu}\partial_{\nu}+\frac{g^{2}\vartheta^{2}}{4}\delta_{\mu\nu} & 0 & -i\partial_{\mu} & -\frac{\vartheta g}{2}\partial_{\mu} & 0 & 0\\
		i\partial_{\nu} & 0 & 0 & 0 & 0 & 0 & 0\\
		0 & i\partial_{\nu} & 0 & 0 & 0 & 0 & 0\\
		\frac{\vartheta g'}{2}\partial_{\nu} & \frac{\vartheta g}{2}\partial_{\nu} & 0 & 0 & -\partial^{2} & 0 & 0\\
		-i\partial_{\nu} & 0 & 0 & 0 & 0 & 0 & i\partial^{2}\\
		0 & 0 & 0 & 0 & 0 & i\partial^{2} & 0
	\end{array}\right)\delta\left(x-y\right)\nonumber \\
\end{eqnarray}
and
\begin{eqnarray}
	D\left(x-y\right) & = & \delta^{\alpha\beta}\left(\begin{array}{ccc}
		-\delta_{\mu\nu}\partial^{2}+\partial_{\mu}\partial_{\nu}+\frac{g^{2}\vartheta^{2}}{4}\delta_{\mu\nu} & -i\partial_{\mu} & -\frac{\vartheta g}{2}\partial_{\mu}\\
		i\partial_{\nu} & 0 & 0\\
		\frac{\vartheta g}{2}\partial_{\nu} & 0 & -\partial^{2}
	\end{array}\right)\delta\left(x-y\right).
\end{eqnarray}

Immediately, we can say that the propagators of $H$ and the ghost
fields are:
\begin{eqnarray}
	\left\langle H\left(p\right)H\left(-p\right)\right\rangle _{0} & = & \frac{1}{p^{2}+2\lambda\vartheta^{2}},\nonumber \\
	\left\langle \overline{c}^{\alpha}\left(p\right)c^{\beta}\left(-p\right)\right\rangle _{0} & = & \frac{1}{p^{2}},\nonumber \\
	\left\langle \overline{c}^{3}\left(p\right)c^{3}\left(-p\right)\right\rangle _{0} & = & \frac{1}{p^{2}},\nonumber \\
	\left\langle \overline{c}\left(p\right)c\left(-p\right)\right\rangle _{0} & = & \frac{1}{p^{2}},\nonumber \\
	\left\langle \overline{\eta}\left(p\right)\eta\left(-p\right)\right\rangle _{0} & = & \frac{1}{p^{2}}.
\end{eqnarray}
To find the other propagators we need to find $\Delta^{-1}$ and $D^{-1}$.
To invert $\Delta$ and $D$, it is convenient to work in the momentum
space. We define the Fourier transform of $\Delta$ and $D$ as
\begin{eqnarray}
	\Delta\left(x-y\right) & = & \int\frac{d^{4}p}{\left(2\pi\right)^{4}}e^{-ip\cdot\left(x-y\right)}\tilde{\Delta}\left(p\right),\nonumber \\
	D\left(x-y\right) & = & \int\frac{d^{4}p}{\left(2\pi\right)^{4}}e^{-ip\cdot\left(x-y\right)}\tilde{D}\left(p\right),
\end{eqnarray}
respectively. Thus, we have that

\begin{eqnarray}
	\tilde{\Delta}\left(p\right) & = & \left(\begin{array}{ccccccc}
		\delta_{\mu\nu}p^{2}-p_{\mu}p_{\nu}+\frac{g'^{2}\vartheta^{2}}{4}\delta_{\mu\nu} & \frac{\vartheta^{2}gg'}{4}\delta_{\mu\nu} & -p_{\mu} & 0 & \frac{\vartheta g'i}{2}p_{\mu} & p_{\mu} & 0\\
		\frac{\vartheta^{2}gg'}{4}\delta_{\mu\nu} & \delta_{\mu\nu}p^{2}-p_{\mu}p_{\nu}+\frac{g^{2}\vartheta^{2}}{4}\delta_{\mu\nu} & 0 & -p_{\mu} & \frac{\vartheta gi}{2}p_{\mu} & 0 & 0\\
		p_{\nu} & 0 & 0 & 0 & 0 & 0 & 0\\
		0 & p_{\nu} & 0 & 0 & 0 & 0 & 0\\
		-\frac{\vartheta g'i}{2}p_{\nu} & -\frac{\vartheta gi}{2}p_{\nu} & 0 & 0 & p^{2} & 0 & 0\\
		-p_{\nu} & 0 & 0 & 0 & 0 & 0 & -ip^{2}\\
		0 & 0 & 0 & 0 & 0 & -ip^{2} & 0
	\end{array}\right)\nonumber \\
\end{eqnarray}
and

\begin{eqnarray}
	\tilde{D}\left(p\right) & = & \delta^{\alpha\beta}\left(\begin{array}{ccc}
		\delta_{\mu\nu}p^{2}-p_{\mu}p_{\nu}+\frac{g^{2}\vartheta^{2}}{4}\delta_{\mu\nu} & -p_{\mu} & \frac{\vartheta gi}{2}p_{\mu}\\
		p_{\nu} & 0 & 0\\
		-\frac{\vartheta gi}{2}p_{\nu} & 0 & p^{2}
	\end{array}\right).
\end{eqnarray}
After some algebraic work, we obtain that

\begin{eqnarray}
	\tilde{\Delta}^{-1}\left(p\right) & = & \left(\begin{array}{ccccccc}
		\frac{p^{2}+\frac{g^{2}\vartheta^{2}}{4}}{p^{2}\left(p^{2}+\frac{\vartheta^{2}\left(g^{2}+g'^{2}\right)}{4}\right)}P_{\mu\nu} & -\frac{\vartheta^{2}gg'}{4}\frac{1}{p^{2}\left(p^{2}+\frac{\vartheta^{2}\left(g^{2}+g'^{2}\right)}{4}\right)}P_{\mu\nu} & \frac{1}{p^{2}}p_{\mu} & 0 & 0 & 0 & 0\\
		-\frac{\vartheta^{2}gg'}{4}\frac{1}{p^{2}\left(p^{2}+\frac{\vartheta^{2}\left(g^{2}+g'^{2}\right)}{4}\right)}P_{\mu\nu} & \frac{p^{2}+\frac{g'^{2}\vartheta^{2}}{4}}{p^{2}\left(p^{2}+\frac{\vartheta^{2}\left(g^{2}+g'^{2}\right)}{4}\right)}P_{\mu\nu} & 0 & \frac{1}{p^{2}}p_{\mu} & 0 & 0 & 0\\
		-\frac{1}{p^{2}}p_{\nu} & 0 & 0 & 0 & \frac{\vartheta g'i}{2}\frac{1}{p^{2}} & 0 & \frac{i}{p^{2}}\\
		0 & -\frac{1}{p^{2}}p_{\nu} & 0 & 0 & \frac{\vartheta gi}{2}\frac{1}{p^{2}} & 0 & 0\\
		0 & 0 & \frac{\vartheta g'i}{2}\frac{1}{p^{2}} & \frac{\vartheta gi}{2}\frac{1}{p^{2}} & \frac{1}{p^{2}} & 0 & 0\\
		0 & 0 & 0 & 0 & 0 & 0 & \frac{i}{p^{2}}\\
		0 & 0 & \frac{i}{p^{2}} & 0 & 0 & \frac{i}{p^{2}} & 0
	\end{array}\right)\nonumber \\
\end{eqnarray}
and
\begin{eqnarray}
	\tilde{D}^{-1}\left(p\right) & = & \delta^{\alpha\beta}\left(\begin{array}{ccc}
		\frac{1}{p^{2}+\frac{\vartheta^{2}g^{2}}{4}}P_{\mu\nu} & \frac{1}{p^{2}}p_{\mu} & \frac{\vartheta gi}{2}p_{\mu}\\
		-\frac{1}{p^{2}}p_{\nu} & 0 & \frac{\vartheta gi}{2}\frac{1}{p^{2}}\\
		0 & \frac{\vartheta gi}{2}\frac{1}{p^{2}} & \frac{1}{p^{2}}
	\end{array}\right).
\end{eqnarray}
Therefore, we have the following non-vanishing propagators
\begin{eqnarray}
	\left\langle X_{\mu}\left(p\right)X_{\nu}\left(-p\right)\right\rangle _{0} & = & \frac{p^{2}+\frac{g^{2}\vartheta^{2}}{4}}{p^{2}\left(p^{2}+\frac{\vartheta^{2}\left(g^{2}+g'^{2}\right)}{4}\right)}P_{\mu\nu},\nonumber \\
	\left\langle X_{\mu}\left(p\right)W_{\nu}^{3}\left(-p\right)\right\rangle _{0} & = & -\frac{\vartheta^{2}gg'}{4}\frac{1}{p^{2}\left(p^{2}+\frac{\vartheta^{2}\left(g^{2}+g'^{2}\right)}{4}\right)}P_{\mu\nu},\nonumber \\
	\left\langle X_{\mu}\left(p\right)b\left(-p\right)\right\rangle _{0} & = & \frac{1}{p^{2}}p_{\mu},\nonumber \\
	\left\langle W_{\mu}^{3}\left(p\right)W_{\nu}^{3}\left(-p\right)\right\rangle _{0} & = & \frac{p^{2}+\frac{g'^{2}\vartheta^{2}}{4}}{p^{2}\left(p^{2}+\frac{\vartheta^{2}\left(g^{2}+g'^{2}\right)}{4}\right)}P_{\mu\nu}\nonumber \\
	\left\langle W_{\mu}^{3}\left(p\right)b\left(-p\right)\right\rangle _{0} & = & \frac{1}{p^{2}}p_{\mu},\nonumber \\
	\left\langle b\left(p\right)\rho^{3}\left(-p\right)\right\rangle _{0} & = & \frac{\vartheta g'i}{2}\frac{1}{p^{2}},\nonumber \\
	\left\langle b\left(p\right)\xi\left(-p\right)\right\rangle _{0} & = & \frac{i}{p^{2}},\nonumber \\
	\left\langle b^{3}\left(p\right)\rho^{3}\left(-p\right)\right\rangle _{0} & = & \frac{\vartheta gi}{2}\frac{1}{p^{2}},\nonumber \\
	\left\langle \theta\left(p\right)\xi\left(-p\right)\right\rangle _{0} & = & \frac{i}{p^{2}},\nonumber \\
	\left\langle W_{\mu}^{\alpha}\left(p\right)W_{\nu}^{\beta}\left(-p\right)\right\rangle _{0} & = & \frac{\delta^{\alpha\beta}}{p^{2}+\frac{\vartheta^{2}g^{2}}{4}}P_{\mu\nu},\nonumber \\
	\left\langle W_{\mu}^{\alpha}\left(p\right)b^{\beta}\left(-p\right)\right\rangle _{0} & = & \frac{\delta^{\alpha\beta}}{p^{2}}p_{\mu},\nonumber \\
	\left\langle W_{\mu}^{\alpha}\left(p\right)\rho^{\beta}\left(-p\right)\right\rangle _{0} & = & \frac{\vartheta gi}{2}\frac{\delta^{\alpha\beta}}{p^{2}}\nonumber \\
	\left\langle \rho^{\alpha}\left(p\right)\rho^{\beta}\left(-p\right)\right\rangle _{0} & = & \frac{\delta^{\alpha\beta}}{p^{2}}.\label{eq:propagators_tree}
\end{eqnarray}

We introduced the fields $Z_{\mu}$ and $A_{\mu}$ in Section \ref{sec:Review-of-the}
in order to diagonalize the mass matrix. However, we ignored the mixing
terms. Therefore, it is also important to write their propagators.
From (\ref{eq:propagators_tree}), we have that

\begin{eqnarray}
	\left\langle Z_{\mu}\left(p\right)Z_{\nu}\left(-p\right)\right\rangle _{0} & = & \cos^{2}\theta_{W}\left\langle W_{\mu}^{3}\left(p\right)W_{\nu}^{3}\left(-p\right)\right\rangle _{0}+2\cos\theta_{W}\sin\theta_{W}\left\langle W_{\mu}^{3}\left(p\right)X_{\nu}\left(-p\right)\right\rangle _{0}\nonumber \\
	&  & +\sin^{2}\theta_{W}\left\langle X_{\mu}\left(p\right)X_{\nu}\left(-p\right)\right\rangle _{0}\nonumber \\
	& = & \frac{1}{p^{2}+\frac{\vartheta^{2}\left(g^{2}+g'^{2}\right)}{4}}P_{\mu\nu}
\end{eqnarray}
and

\begin{eqnarray}
	\left\langle A_{\mu}\left(p\right)A_{\nu}\left(-p\right)\right\rangle _{0} & = & \sin^{2}\theta_{W}\left\langle W_{\mu}^{3}\left(p\right)W_{\nu}^{3}\left(-p\right)\right\rangle _{0}-2\cos\theta_{W}\sin\theta_{W}\left\langle W_{\mu}^{3}\left(p\right)X_{\nu}\left(-p\right)\right\rangle _{0}\nonumber \\
	&  & +\cos^{2}\theta_{W}\left\langle X_{\mu}\left(p\right)X_{\nu}\left(-p\right)\right\rangle _{0}\nonumber \\
	& = & \frac{1}{p^{2}}P_{\mu\nu}.
\end{eqnarray}
Notice that, in fact, $A_{\mu}$ is a massless field, whereas $Z_{\mu}$
has a mass, namely, $m_{Z}=\frac{\vartheta\sqrt{g^{2}+g'^{2}}}{2}=\frac{\vartheta g}{2\cos\theta_{W}}$.

\subsection*{Exact results}

The explicit calculation showed that
\begin{gather}
	\left\langle b\left(p\right)b\left(-p\right)\right\rangle _{0}=\left\langle b\left(p\right)b^{3}\left(-p\right)\right\rangle _{0}=\left\langle b^{3}\left(p\right)b^{3}\left(-p\right)\right\rangle _{0}\nonumber \\
	=\left\langle b\left(p\right)\theta\left(-p\right)\right\rangle _{0}=\left\langle b^{3}\left(p\right)\theta\left(-p\right)\right\rangle _{0}=\left\langle b^{\alpha}\left(p\right)b^{\beta}\left(-p\right)\right\rangle _{0}=0.
\end{gather}
This is not a simple coincidence. These results hold at all perturbative
orders. They follow from the BRST symmetry and the fact that these
are correlation functions of BRST-exact terms, \emph{i.e.},
\begin{gather}
	\left\langle b\left(x\right)b\left(y\right)\right\rangle =\left\langle s\left(-i\overline{c}\left(x\right)b\left(y\right)\right)\right\rangle =0,\nonumber \\
	\left\langle b\left(x\right)b^{3}\left(y\right)\right\rangle =\left\langle s\left(-i\overline{c}\left(x\right)b^{3}\left(y\right)\right)\right\rangle =0,\nonumber \\
	\left\langle b^{3}\left(x\right)b^{3}\left(y\right)\right\rangle =\left\langle s\left(-i\overline{c}^{3}\left(x\right)b^{3}\left(y\right)\right)\right\rangle =0,\nonumber \\
	\left\langle b\left(x\right)\theta\left(y\right)\right\rangle =\left\langle s\left(-i\overline{c}\left(x\right)\theta\left(y\right)\right)\right\rangle =0,\nonumber \\
	\left\langle b^{3}\left(x\right)\theta\left(y\right)\right\rangle =\left\langle s\left(-i\overline{c}^{3}\left(x\right)\theta\left(y\right)\right)\right\rangle =0,\nonumber \\
	\left\langle b^{\alpha}\left(x\right)b^{\beta}\left(y\right)\right\rangle =\left\langle s\left(-i\overline{c}^{\alpha}\left(x\right)b^{\beta}\left(y\right)\right)\right\rangle =0.
\end{gather}

We can derive exact results from the Ward identities (\ref{eq:b_equations}).
These are non-anomalous identities, thus we can find an effective
action, $\Gamma$, such that 
\begin{eqnarray}
	\frac{\delta\Gamma}{\delta b^{\alpha}} & = & i\partial_{\mu}W_{\mu}^{\alpha}+i\varepsilon^{\alpha\beta}\Upsilon_{\mu}W_{\mu}^{\beta},\nonumber \\
	\frac{\delta\Gamma}{\delta b^{3}} & = & i\partial_{\mu}W_{\mu}^{3},\nonumber \\
	\frac{\delta\Gamma}{\delta b} & = & i\partial_{\mu}X_{\mu}.\label{eq:b_equations-1}
\end{eqnarray}
In terms of the generator function of connected function,
\begin{eqnarray*}
	W\left[J\right] & = & \Gamma\left[\phi\right]+\int d^{4}xJ_{i}\phi_{i},
\end{eqnarray*}
Eq. (\ref{eq:b_equations-1}) can be rewritten as
\begin{eqnarray}
	-J_{b}^{\alpha} & = & i\partial_{\mu}\frac{\delta W}{\delta J_{\mu}^{\alpha}}+i\varepsilon^{\alpha\beta}\Upsilon_{\mu}\frac{\delta W}{\delta J_{\mu}^{\beta}},\nonumber \\
	-J_{b}^{3} & = & i\partial_{\mu}\frac{\delta W}{\delta J_{\mu}^{3}},\nonumber \\
	-J_{b} & = & i\partial_{\mu}\frac{\delta W}{\delta j_{\mu}},\label{eq:b_equations-1-1}
\end{eqnarray}
where $J_{\mu}^{\alpha}$, $J_{\mu}^{3}$ and $j_{\mu}$ are the sources
of $W_{\mu}^{\alpha}$, $W_{\mu}^{3}$ and $X_{\mu}$, respectively. 

The propagator of $W_{\mu}^{\alpha}$ is given by
\begin{eqnarray}
	\left\langle W_{\mu}^{\alpha}\left(x\right)W_{\nu}^{\beta}\left(y\right)\right\rangle  & = & -\left.\frac{\delta^{2}W}{\delta J_{\nu}^{\beta}\left(y\right)\delta J_{\mu}^{\alpha}\left(x\right)}\right|_{\textrm{all sources}=0}.
\end{eqnarray}
Then, from the first line of (\ref{eq:b_equations-1-1}), we have
that
\begin{eqnarray}
	0 & = & i\partial_{\mu}^{x}\left\langle W_{\mu}^{\alpha}\left(x\right)W_{\nu}^{\beta}\left(y\right)\right\rangle ,
\end{eqnarray}
which means that $\left\langle W_{\mu}^{\alpha}\left(x\right)W_{\nu}^{\beta}\left(y\right)\right\rangle $
is transverse. This result agrees with the tree-level propagator,
see Eq. (\ref{eq:propagators_tree}). Now, differentiating the first
line of (\ref{eq:b_equations-1-1}) with respect to $J_{b}^{\beta}\left(y\right)$,
we get 
\begin{eqnarray}
	\partial_{\mu}^{x}\left\langle W_{\mu}^{\alpha}\left(x\right)b^{\beta}\left(y\right)\right\rangle  & = & -i\delta\left(x-y\right),
\end{eqnarray}
where
\begin{eqnarray}
	\left\langle W_{\mu}^{\alpha}\left(x\right)b^{\beta}\left(y\right)\right\rangle  & = & -\left.\frac{\delta^{2}W}{\delta J_{b}^{\beta}\left(y\right)\delta J_{\mu}^{\alpha}\left(x\right)}\right|_{\textrm{all sources}=0}.
\end{eqnarray}
Therefore, we have the exact result
\begin{eqnarray}
	\left\langle W_{\mu}^{\alpha}\left(x\right)b^{\beta}\left(y\right)\right\rangle  & = & \int\frac{d^{4}p}{\left(2\pi\right)^{4}}e^{-ip\cdot\left(x-y\right)}\frac{p_{\mu}}{p^{2}}.\label{eq:exact_W_b}
\end{eqnarray}
Notice that the tree-level result (\ref{eq:propagators_tree}) agrees
with (\ref{eq:exact_W_b}), as it should. In the same way, we can
derive similar results for the diagonal component $W_{\mu}^{3}$ and
the Abelian component $X_{\mu}$.

The correlation function $\left\langle b\left(x\right)\xi\left(y\right)\right\rangle $
can be related to the Abelian ghost propagator $\left\langle \overline{c}\left(x\right)c\left(y\right)\right\rangle $
as follows: 

\begin{eqnarray}
	\left\langle b\left(x\right)\xi\left(y\right)\right\rangle  & = & \left\langle s\left(-i\overline{c}\left(x\right)\xi\left(y\right)\right)\right\rangle +i\left\langle \overline{c}\left(x\right)s\xi\left(y\right)\right\rangle \nonumber \\
	& = & -i\left\langle \overline{c}\left(x\right)c\left(y\right)\right\rangle .
\end{eqnarray}
From the ghost equation (\ref{eq:ghost_eq_abelian}), we have that
\begin{eqnarray}
	\frac{\delta\Gamma}{\delta c} & = & -\partial^{2}\overline{c}+\frac{g'}{2}\varepsilon^{\alpha\beta}R^{\alpha}\rho^{\beta}-\frac{g'}{2}R^{3}\left(\vartheta+H\right)+\frac{g'}{2}Y\rho^{3},
\end{eqnarray}
which is equivalent to

\begin{eqnarray}
	J_{c} & = & \partial^{2}\frac{\delta W}{\delta J_{\overline{c}}}+\frac{g'}{2}\varepsilon^{\alpha\beta}R^{\alpha}\frac{\delta W}{\delta J_{\rho}^{\beta}}-\frac{g'}{2}R^{3}\left(\vartheta+\frac{\delta W}{\delta J_{H}}\right)+\frac{g'}{2}Y\frac{\delta W}{\delta J_{\rho}^{3}},
\end{eqnarray}
where $J_{c}$, $J_{\overline{c}}$, $J_{\rho}^{\beta}$, $J_{H}$
and $J_{\rho}^{3}$ are the sources of $c$, $\overline{c}$, $\rho^{\beta}$,
$H$ and $\rho^{3}$, respectively. This last equation tells us that
the exact ghost propagator does not receive perturbative corrections,
\emph{i.e.}, 
\begin{eqnarray}
	\left\langle \overline{c}\left(x\right)c\left(y\right)\right\rangle  & = & \left.\frac{\delta^{2}W}{\delta J_{\overline{c}}\left(x\right)\delta J_{c}\left(y\right)}\right|_{\textrm{all sources}=0}\nonumber \\
	& = & \int\frac{d^{4}p}{\left(2\pi\right)^{4}}e^{-ip\cdot\left(x-y\right)}\frac{1}{p^{2}}.
\end{eqnarray}
Consequently, it follows that
\begin{eqnarray}
	\left\langle b\left(x\right)\xi\left(y\right)\right\rangle  & = & -\int\frac{d^{4}p}{\left(2\pi\right)^{4}}e^{-ip\cdot\left(x-y\right)}\frac{i}{p^{2}},
\end{eqnarray}
which agrees with the tree-level result (\ref{eq:propagators_tree}).

To calculate correlation functions we set the external sources $\Omega_{\mu}^{3}$
and $\Omega_{\mu}^{\alpha}$ to zero, then the $\mathscr{S}$-symmetry
is restored on-shell. This means that correlation functions of $\mathscr{S}$-exact
terms are zero, \emph{i.e.}, $\left\langle \mathscr{S}\left(\ldots\right)\right\rangle =0$.
Therefore, it follows that
\begin{gather}
	\left\langle \theta\left(x\right)X_{\mu}\left(y\right)\right\rangle =\left\langle \mathscr{S}\left(-i\overline{\eta}\left(x\right)X_{\mu}\left(y\right)\right)\right\rangle =0,\nonumber \\
	\left\langle \theta\left(x\right)W_{\mu}^{3}\left(y\right)\right\rangle =\left\langle \mathscr{S}\left(-i\overline{\eta}\left(x\right)W_{\mu}^{3}\left(y\right)\right)\right\rangle =0,\nonumber \\
	\left\langle \theta\left(x\right)\theta\left(y\right)\right\rangle =\left\langle \mathscr{S}\left(-i\overline{\eta}\left(x\right)\theta\left(y\right)\right)\right\rangle =0,\nonumber \\
	\left\langle \theta\left(x\right)\rho^{3}\left(y\right)\right\rangle =\left\langle \mathscr{S}\left(-i\overline{\eta}\left(x\right)\rho^{3}\left(y\right)\right)\right\rangle =0.\label{eq:teta_corr}
\end{gather}
We can also use the on-shell $\mathscr{S}$-symmetry to relate correlation
functions, such as
\begin{eqnarray}
	\left\langle \theta\left(x\right)\xi\left(y\right)\right\rangle  & = & \left\langle \mathscr{S}\left(-i\overline{\eta}\left(x\right)\xi\left(y\right)\right)\right\rangle -i\left\langle \overline{\eta}\left(x\right)\mathscr{S}\xi\left(y\right)\right\rangle \nonumber \\
	& = & i\left\langle \overline{\eta}\left(x\right)\eta\left(y\right)\right\rangle .\label{eq:teta_xi}
\end{eqnarray}
We could also used the Ward identity that follows from Eq. (\ref{eq:xi_equation})
to derivate (\ref{eq:teta_corr}) and (\ref{eq:teta_xi}), namely,
\begin{eqnarray}
	\frac{\delta\Gamma}{\delta\xi} & = & i\partial^{2}\theta+ig'\partial_{\mu}\left(\Omega_{\mu}^{3}\frac{\delta\Gamma}{\delta J}\right)+ig'\frac{\vartheta^{2}}{2}\partial_{\mu}\Omega_{\mu}^{3}-2g'\varepsilon^{\alpha\beta}\Omega_{\mu}^{\alpha}\frac{\delta\Gamma}{\delta\Omega_{\mu}^{\beta}}.
\end{eqnarray}
From the ghost equation (\ref{eq:eta_equations}), we have that
\begin{eqnarray}
	\left\langle \overline{\eta}\left(x\right)\eta\left(y\right)\right\rangle  & = & \int\frac{d^{4}p}{\left(2\pi\right)^{4}}e^{-ip\cdot\left(x-y\right)}\frac{1}{p^{2}}.\label{eq:ghost_eta_prop}
\end{eqnarray}
Thus, it follows that
\begin{eqnarray}
	\left\langle \theta\left(x\right)\xi\left(y\right)\right\rangle  & = & \int\frac{d^{4}p}{\left(2\pi\right)^{4}}e^{-ip\cdot\left(x-y\right)}\frac{i}{p^{2}}.\label{eq:teta_xi_exact}
\end{eqnarray}
Notice that the tree-level results obtained in the previous subsection
agree with (\ref{eq:teta_corr}), (\ref{eq:ghost_eta_prop}) and (\ref{eq:teta_xi_exact}).

\newpage

\newpage

\section*{Appendix B: Dimension and ghost number of the fields and sources}

The dimension and ghost number of fields and sources are shown in Table 1 and Table 2.

\begin{table} [H]
	\begin{centering}
		\label{table_fields}%
		\begin{tabular}{|c|c|c|c|c|c|c|c|c|c|c|c|c|c|c|c|c|}
			\hline 
			field & $W_{\mu}^{\alpha}$ & $W_{\mu}^{3}$ & $X_{\mu}$ & $\rho^{\alpha}$ & $\rho^{3}$ & $H$ & $\xi$ & $\theta$ & $\overline{\eta}$ & $\eta$ & $c^{\alpha}$ & $\overline{c}^{\alpha}$ & $c^{3}$ & $\overline{c}^{3}$ & $c$ & $\overline{c}$\tabularnewline
			\hline 
			\hline 
			$SU\left(2\right)$ ghost number & 0 & 0 & 0 & 0 & 0 & 0 & 0 & 0 & 0 & 0 & 1 & -1 & 1 & -1 & 0 & 0\tabularnewline
			\hline 
			$U\left(1\right)$ ghost number & 0 & 0 & 0 & 0 & 0 & 0 & 0 & 0 & 0 & 0 & 0 & 0 & 0 & 0 & 1 & -1\tabularnewline
			\hline 
			dimension & 1 & 1 & 1 & 1 & 1 & 1 & 0 & 2 & 2 & 0 & 0 & 2 & 0 & 2 & 0 & 2\tabularnewline
			\hline 
		\end{tabular}
		\par\end{centering}
	\caption{Ghost number and dimension of the fields.}
\end{table}

\begin{table} [H]
	\begin{centering}
		\label{table_sources}%
		\begin{tabular}{|c|c|c|c|c|c|c|c|c|c|c|c|c|c|c|c|c|}
			\hline 
			source & $\Omega_{\mu}^{\alpha}$ & $\Omega_{\mu}^{3}$ & $\Omega_{\mu}$ & $K_{\mu}^{\alpha}$ & $K_{\mu}^{3}$ & $Y$ & $R^{\alpha}$ & $R^{3}$ & $L^{\alpha}$ & $L^{3}$ & $y$ & $r^{\alpha}$ & $r^{3}$ & $J$ & $\Upsilon_{\mu}^{\alpha}$ & $\zeta_{\mu}^{\alpha}$\tabularnewline
			\hline 
			\hline 
			$SU\left(2\right)$ ghost number & 0 & 0 & 0 & -1 & -1 & -1 & -1 & -1 & -2 & -2 & 0 & 0 & 0 & 0 & 0 & 1\tabularnewline
			\hline 
			$U\left(1\right)$ ghost number & 0 & 0 & 0 & 0 & 0 & 0 & 0 & 0 & 0 & 0 & -1 & -1 & -1 & 0 & 0 & 0\tabularnewline
			\hline 
			dimension & 1 & 1 & 1 & 3 & 3 & 3 & 3 & 3 & 4 & 4 & 3 & 3 & 3 & 2 & 1 & 1\tabularnewline
			\hline 
		\end{tabular}
		\par\end{centering}
	\caption{Ghost number and dimension of the sources.}
\end{table}

	\section*{Appendix C: List of counterterms $\Delta_{i}$}

In this appendix, we present the set of non-trivial BRST-invariant counterterms $\left\{\Delta_{i}\right\}$consistent with the Ward identities from Section \ref{sec:Symmetries-and-Ward}. We have the following list:
\begin{gather}
	\Delta_{0}=\int d^{4}x\frac{1}{4}W_{\mu\nu}^{a}W_{\mu\nu}^{a},
\end{gather}
\begin{gather}
	\Delta_{1}=\int d^{4}x\left[\frac{1}{4}X_{\nu\mu}X_{\nu\mu}-\frac{1}{2g'{}^{2}}\Omega_{\mu}\partial^{2}\Omega_{\mu}-\frac{1}{2g'{}^{2}}\left(\partial_{\mu}\Omega_{\mu}\right)\left(\partial_{\nu}\Omega_{\nu}\right)+\frac{1}{g'}\left(\partial_{\mu}X_{\mu\nu}\right)\Omega_{\nu}\right],
\end{gather}
\begin{gather}
	\Delta_{2}=\int d^{4}x\left[\vartheta^{2}O-\frac{1}{\lambda}\hat{J}O+\frac{1}{2\lambda}\hat{J}\vartheta^{2}-\frac{1}{2\lambda^{2}}\hat{J}^{2}-\frac{g'{}^{2}}{8\lambda^{2}}\left(X_{\mu}-\partial_{\mu}\xi\right)\left(X_{\nu}-\partial_{\nu}\xi\right)\Omega_{\mu}^{3}\Omega_{\nu}^{3}\right.\nonumber \\
	-\frac{g'}{2\lambda}\Omega_{\mu}^{3}\left(X_{\mu}-\partial_{\mu}\xi\right)\left(O+\frac{1}{2\lambda}\hat{J}\right)+\frac{g'}{4\lambda}\Omega_{\mu}^{3}\left(X_{\mu}-\partial_{\mu}\xi\right)\left(\vartheta^{2}-\frac{1}{\lambda}\hat{J}\right)\nonumber \\
	+\frac{1}{4\lambda}\left(\Omega_{\mu}\Omega_{\mu}\left(O+\frac{1}{2\lambda}\hat{J}\right)+\frac{g'}{4\lambda}\Omega_{\mu}^{3}\left(X_{\mu}-\partial_{\mu}\xi\right)\Omega_{\nu}\Omega_{\nu}\right)\nonumber \\
	+\frac{1}{2\lambda}\left(\Omega_{\mu}^{3}\Omega_{\mu}\left(O+\frac{1}{2\lambda}\hat{J}\right)+\frac{g'}{4\lambda}\Omega_{\mu}^{3}\left(X_{\mu}-\partial_{\mu}\xi\right)\Omega_{\nu}\Omega_{\nu}^{3}\right)\nonumber \\
	-\frac{1}{4\lambda}\left(\Omega_{\mu}^{3}\Omega_{\mu}\left(\vartheta^{2}-\frac{1}{\lambda}\hat{J}\right)-\frac{g'}{2\lambda}\Omega_{\mu}^{3}\left(X_{\mu}-\partial_{\mu}\xi\right)\Omega_{\nu}\Omega_{\nu}^{3}\right)\nonumber \\
	-\frac{1}{8\lambda}\left(\Omega_{\mu}\Omega_{\mu}\left(\vartheta^{2}-\frac{1}{\lambda}\hat{J}\right)-\frac{g'}{2\lambda}\Omega_{\mu}^{3}\left(X_{\mu}-\partial_{\mu}\xi\right)\Omega_{\nu}\Omega_{\nu}\right)\nonumber \\
	\left.-\frac{1}{32\lambda^{2}}\left(\Omega_{\mu}\Omega_{\mu}\Omega_{\nu}\Omega_{\nu}+4\Omega_{\mu}\Omega_{\mu}\Omega_{\nu}\Omega_{\nu}^{3}+4\Omega_{\mu}\Omega_{\nu}\Omega_{\nu}^{3}\Omega_{\mu}^{3}\right)\right],\label{eq:delta_2}
\end{gather}
\begin{gather}
	\Delta_{3}=\int d^{4}x\left[O^{2}+\frac{1}{\lambda}\hat{J}O+\frac{1}{4\lambda^{2}}\hat{J}^{2}+\frac{g'{}^{2}}{16\lambda^{2}}\left(X_{\mu}-\partial_{\mu}\xi\right)\left(X_{\nu}-\partial_{\nu}\xi\right)\Omega_{\mu}^{3}\Omega_{\nu}^{3}\right.\nonumber \\
	+\frac{g'}{2\lambda}\Omega_{\mu}^{3}\left(X_{\mu}-\partial_{\mu}\xi\right)\left(O+\frac{1}{2\lambda}\hat{J}\right)-\frac{1}{2\lambda}\left(\Omega_{\mu}^{3}\Omega_{\mu}\left(O+\frac{1}{2\lambda}\hat{J}\right)+\frac{g'}{4\lambda}\Omega_{\mu}^{3}\left(X_{\mu}-\partial_{\mu}\xi\right)\Omega_{\nu}\Omega_{\nu}^{3}\right)\nonumber \\
	-\frac{1}{4\lambda}\left(\Omega_{\mu}\Omega_{\mu}\left(O+\frac{1}{2\lambda}\hat{J}\right)+\frac{g'}{4\lambda}\Omega_{\mu}^{3}\left(X_{\mu}-\partial_{\mu}\xi\right)\Omega_{\nu}\Omega_{\nu}\right)\nonumber \\
	\left.+\frac{1}{64\lambda^{2}}\left(\Omega_{\mu}\Omega_{\mu}\Omega_{\nu}\Omega_{\nu}+4\Omega_{\mu}\Omega_{\mu}\Omega_{\nu}\Omega_{\nu}^{3}+4\Omega_{\mu}\Omega_{\nu}\Omega_{\nu}^{3}\Omega_{\mu}^{3}\right)\right],
\end{gather}
\begin{gather}
	\Delta_{4}=\int d^{4}x\left[\vartheta^{4}-2\frac{1}{\lambda}\hat{J}\vartheta^{2}+\frac{1}{\lambda^{2}}\hat{J}^{2}+\frac{g'{}^{2}}{4\lambda^{2}}\left(X_{\mu}-\partial_{\mu}\xi\right)\left(X_{\nu}-\partial_{\nu}\xi\right)\Omega_{\mu}^{3}\Omega_{\nu}^{3}\right.\nonumber \\
	-\frac{g'}{\lambda}\Omega_{\mu}^{3}\left(X_{\mu}-\partial_{\mu}\xi\right)\left(\vartheta^{2}-\frac{1}{\lambda}\hat{J}\right)+\frac{1}{\lambda}\left(\Omega_{\mu}^{3}\Omega_{\mu}\left(\vartheta^{2}-\frac{1}{\lambda}\hat{J}\right)-\frac{g'}{2\lambda}\Omega_{\mu}^{3}\left(X_{\mu}-\partial_{\mu}\xi\right)\Omega_{\nu}\Omega_{\nu}^{3}\right)\nonumber \\
	+\frac{1}{2\lambda}\left(\Omega_{\mu}\Omega_{\mu}\left(\vartheta^{2}-\frac{1}{\lambda}\hat{J}\right)-\frac{g'}{2\lambda}\Omega_{\mu}^{3}\left(X_{\mu}-\partial_{\mu}\xi\right)\Omega_{\nu}\Omega_{\nu}\right)\nonumber \\
	\left.+\frac{1}{16\lambda^{2}}\left(\Omega_{\mu}\Omega_{\mu}\Omega_{\nu}\Omega_{\nu}+4\Omega_{\mu}\Omega_{\mu}\Omega_{\nu}\Omega_{\nu}^{3}+4\Omega_{\mu}\Omega_{\nu}\Omega_{\nu}^{3}\Omega_{\mu}^{3}\right)\right],
\end{gather}
\begin{gather}
	\Delta_{5}=\int d^{4}x\left[\Omega_{\mu}^{3}O_{\mu}+\frac{1}{2}\left(\Omega_{\mu}^{3}\Omega_{\mu}\left(O+\frac{1}{2\lambda}\hat{J}\right)+\frac{g'}{4\lambda}\Omega_{\mu}^{3}\left(X_{\mu}-\partial_{\mu}\xi\right)\Omega_{\nu}\Omega_{\nu}^{3}\right)+\frac{1}{4}\left(\Omega_{\mu}^{3}\Omega_{\mu}\left(\vartheta^{2}-\frac{1}{\lambda}\hat{J}\right)\right)\right.\nonumber \\
	\left.-\frac{g'}{2\lambda}\Omega_{\mu}^{3}\left(X_{\mu}-\partial_{\mu}\xi\right)\Omega_{\nu}\Omega_{\nu}^{3}\right],
\end{gather}
\begin{gather}
	\Delta_{6}=\int d^{4}x\Omega_{\mu}^{\alpha}O_{\mu}^{\alpha},
\end{gather}
\begin{gather}
	\Delta_{7}=\int d^{4}x\Omega_{\mu}^{3}\partial_{\mu}O,
\end{gather}
\begin{gather}
	\Delta_{8}=\int d^{4}x\Omega_{\mu}^{3}\partial^{2}\Omega_{\mu}^{3},
\end{gather}
\begin{gather}
	\Delta_{9}=\int d^{4}x\left(\partial_{\mu}\Omega_{\mu}^{3}\right)\left(\partial_{\nu}\Omega_{\nu}^{3}\right),
\end{gather}
\begin{gather}
	\Delta_{10}=\int d^{4}x\left[\Omega_{\mu}^{3}\partial^{2}\Omega_{\mu}+\left(\partial_{\mu}\Omega_{\mu}^{3}\right)\left(\partial_{\nu}\Omega_{\nu}\right)-g'\left(\partial_{\mu}X_{\mu\nu}\right)\Omega_{\nu}^{3}\right],
\end{gather}
\begin{gather}
	\Delta_{11}=\int d^{4}x\left[\Omega_{\mu}^{\alpha}\Omega_{\mu}^{\alpha}\left(O+\frac{1}{2\lambda}\hat{J}\right)+\frac{g'}{4\lambda}\Omega_{\mu}^{3}\left(X_{\mu}-\partial_{\mu}\xi\right)\Omega_{\nu}^{\alpha}\Omega_{\nu}^{\alpha}-\frac{1}{4\lambda}\left(\Omega_{\mu}^{\alpha}\Omega_{\mu}^{\alpha}\Omega_{\nu}^{3}\Omega_{\nu}+\frac{1}{2}\Omega_{\mu}^{\alpha}\Omega_{\mu}^{\alpha}\Omega_{\nu}\Omega_{\nu}\right)\right],
\end{gather}
\begin{gather}
	\Delta_{12}=\int d^{4}x\left[\Omega_{\mu}^{3}\Omega_{\mu}^{3}\left(O+\frac{1}{2\lambda}\hat{J}\right)+\frac{g'}{4\lambda}\Omega_{\mu}^{3}\left(X_{\mu}-\partial_{\mu}\xi\right)\Omega_{\nu}^{3}\Omega_{\nu}^{3}-\frac{1}{8\lambda}\left(\Omega_{\mu}\Omega_{\mu}\Omega_{\nu}^{3}\Omega_{\nu}^{3}+2\Omega_{\mu}\Omega_{\mu}^{3}\Omega_{\nu}^{3}\Omega_{\nu}^{3}\right)\right],
\end{gather}
\begin{gather}
	\Delta_{13}=\int d^{4}x\left[\Omega_{\mu}^{\alpha}\Omega_{\mu}^{\alpha}\left(\vartheta^{2}-\frac{1}{\lambda}\hat{J}\right)-\frac{g'}{2\lambda}\Omega_{\mu}^{3}\left(X_{\mu}-\partial_{\mu}\xi\right)\Omega_{\nu}^{\alpha}\Omega_{\nu}^{\alpha}+\frac{1}{2\lambda}\left(\Omega_{\mu}^{\alpha}\Omega_{\mu}^{\alpha}\Omega_{\nu}^{3}\Omega_{\nu}+\frac{1}{2}\Omega_{\mu}^{\alpha}\Omega_{\mu}^{\alpha}\Omega_{\nu}\Omega_{\nu}\right)\right],
\end{gather}
\begin{gather}
	\Delta_{14}=\int d^{4}x\left[\Omega_{\mu}^{3}\Omega_{\mu}^{3}\left(\vartheta^{2}-\frac{1}{\lambda}\hat{J}\right)-\frac{g'}{2\lambda}\Omega_{\mu}^{3}\left(X_{\mu}-\partial_{\mu}\xi\right)\Omega_{\nu}^{3}\Omega_{\nu}^{3}+\frac{1}{4\lambda}\left(\Omega_{\mu}\Omega_{\mu}\Omega_{\nu}^{3}\Omega_{\nu}^{3}+2\Omega_{\mu}\Omega_{\mu}^{3}\Omega_{\nu}^{3}\Omega_{\nu}^{3}\right)\right],
\end{gather}
\begin{gather}
	\Delta_{15}=\int d^{4}x\left[\left(\partial_{\mu}\Omega_{\mu}^{3}\right)\hat{J}+\frac{g'}{2}\left(X_{\mu}-\partial_{\mu}\xi\right)\Omega_{\mu}^{3}\partial_{\nu}\Omega_{\nu}^{3}+\frac{1}{2}\Omega_{\mu}^{3}\Omega_{\nu}^{3}\left(\partial_{\mu}\Omega_{\nu}\right)+\frac{1}{2}\Omega_{\mu}^{3}\left(\partial_{\mu}\Omega_{\nu}^{3}\right)\Omega_{\nu}-\frac{1}{4}\Omega_{\mu}\Omega_{\mu}\left(\partial_{\nu}\Omega_{\nu}^{3}\right)\right],
\end{gather}
\begin{gather}
	\Delta_{16}=\int d^{4}x\Omega_{\mu}^{\alpha}\Omega_{\mu}^{\alpha}\left(\partial_{\nu}\Omega_{\nu}^{3}\right),
\end{gather}
\begin{gather}
	\Delta_{17}=\int d^{4}x\Omega_{\mu}^{\alpha}\left(\partial_{\nu}\Omega_{\mu}^{\alpha}\right)\Omega_{\nu}^{3},
\end{gather}
\begin{gather}
	\Delta_{18}=\int d^{4}x\Omega_{\mu}^{\alpha}\Omega_{\nu}^{\alpha}\partial_{\mu}\Omega_{\nu}^{3},
\end{gather}
\begin{gather}
	\Delta_{19}=\int d^{4}x\left[\Omega_{\mu}^{\alpha}\left(\partial_{\mu}\Omega_{\nu}^{\alpha}\right)\Omega_{\nu}^{3}-g'\varepsilon^{\alpha\beta}\Omega_{\mu}^{\alpha}\left(X_{\mu}-\partial_{\mu}\xi\right)\Omega_{\nu}^{\beta}\Omega_{\nu}^{3}-\varepsilon^{\alpha\beta}\Omega_{\mu}^{\alpha}\Omega_{\nu}^{\beta}\Omega_{\mu}^{3}\Omega_{\nu}\right],
\end{gather}
\begin{gather}
	\Delta_{20}=\int d^{4}x\varepsilon^{\alpha\beta}\Omega_{\mu}^{\alpha}\Omega_{\nu}^{\beta}\left(\partial_{\mu}\Omega_{\nu}^{3}\right),
\end{gather}
\begin{gather}
	\Delta_{21}=\int d^{4}x\left[\varepsilon^{\alpha\beta}\Omega_{\mu}^{\alpha}\left(\partial_{\mu}\Omega_{\nu}^{\beta}\right)\Omega_{\nu}^{3}+g'\Omega_{\mu}^{\alpha}\left(X_{\mu}-\partial_{\mu}\xi\right)\Omega_{\nu}^{\alpha}\Omega_{\nu}^{3}-\Omega_{\mu}^{\alpha}\Omega_{\nu}^{\alpha}\Omega_{\mu}^{3}\Omega_{\nu}\right],
\end{gather}
\begin{gather}
	\Delta_{22}=\int d^{4}x\Omega_{\mu}^{3}\Omega_{\mu}^{3}\left(\partial_{\nu}\Omega_{\nu}^{3}\right),
\end{gather}
\begin{gather}
	\Delta_{23}=\int d^{4}x\Omega_{\mu}^{\alpha}\Omega_{\mu}^{\alpha}\Omega_{\nu}^{\beta}\Omega_{\nu}^{\beta},
\end{gather}
\begin{gather}
	\Delta_{24}=\int d^{4}x\Omega_{\mu}^{\alpha}\Omega_{\mu}^{\beta}\Omega_{\nu}^{\alpha}\Omega_{\nu}^{\beta},
\end{gather}
\begin{gather}
	\Delta_{25}=\int d^{4}x\varepsilon^{\alpha\beta}\varepsilon^{\gamma\delta}\Omega_{\mu}^{\alpha}\Omega_{\nu}^{\beta}\Omega_{\mu}^{\gamma}\Omega_{\nu}^{\delta},
\end{gather}
\begin{gather}
	\Delta_{26}=\int d^{4}x\Omega_{\mu}^{\alpha}\Omega_{\mu}^{\alpha}\Omega_{\nu}^{3}\Omega_{\nu}^{3},
\end{gather}
\begin{gather}
	\Delta_{27}=\int d^{4}x\Omega_{\mu}^{\alpha}\Omega_{\nu}^{\alpha}\Omega_{\mu}^{3}\Omega_{\nu}^{3},
\end{gather}
\begin{gather}
	\Delta_{28}=\int d^{4}x\Omega_{\mu}^{3}\Omega_{\mu}^{3}\Omega_{\nu}^{3}\Omega_{\nu}^{3}.
\end{gather}

\newpage

\section*{Appendix D: List of bare counterterms $\left(\hat{\Delta}_{i}\right)_{0}$ }

In this appendix, we present the set of bare counterterms $\left\{ \left(\hat{\Delta}_{i}\right)_{0} \right\} $ necessary to compute finite correlation functions. We have the following list:

\begin{gather}
	\left(\hat{\Delta}_{1}\right)_{\textrm{0}}=\int d^{4}x\left[-\frac{1}{2g'{}_{0}^{2}}\Omega_{0\mu}\partial^{2}\Omega_{0\mu}-\frac{1}{2g'{}_{0}^{2}}\left(\partial_{\mu}\Omega_{0\mu}\right)\left(\partial_{\nu}\Omega_{0\nu}\right)\right],
\end{gather}
\begin{gather}
	\left(\hat{\Delta}_{2}\right)_{\textrm{0}}=\int d^{4}x\left[\vartheta_{0}^{2}O_{0}-\frac{1}{2\lambda_{0}^{2}}\hat{J}_{0}^{2}-\frac{g'{}_{0}^{2}}{8\lambda_{0}^{2}}\left(X_{0\mu}-\partial_{\mu}\xi_{0}\right)\left(X_{0\nu}-\partial_{\nu}\xi_{0}\right)\Omega_{0\mu}^{3}\Omega_{0\nu}^{3}\right.\nonumber \\
	-\frac{g'}{4\lambda^{2}}\Omega_{0\mu}^{3}\left(X_{0\mu}-\partial_{\mu}\xi_{0}\right)\hat{J}_{0}-\frac{g'_{0}}{4\lambda_{0}^{2}}\Omega_{0\mu}^{3}\left(X_{0\mu}-\partial_{\mu}\xi_{0}\right)\hat{J}_{0}\nonumber \\
	+\frac{1}{4\lambda_{0}}\left(\Omega_{0\mu}\Omega_{0\mu}\left(O_{0}+\frac{1}{2\lambda_{0}}\hat{J}_{0}\right)+\frac{g'_{0}}{4\lambda_{0}}\Omega_{0\mu}^{3}\left(X_{0\mu}-\partial_{\mu}\xi_{0}\right)\Omega_{0\nu}\Omega_{0\nu}\right)\nonumber \\
	+\frac{1}{2\lambda_{0}}\left(\Omega_{0\mu}^{3}\Omega_{0\mu}\left(O_{0}+\frac{1}{2\lambda_{0}}\hat{J}_{0}\right)+\frac{g'_{0}}{4\lambda_{0}}\Omega_{0\mu}^{3}\left(X_{0\mu}-\partial_{\mu}\xi_{0}\right)\Omega_{0\nu}\Omega_{0\nu}^{3}\right)\nonumber \\
	-\frac{1}{4\lambda_{0}}\left(\Omega_{0\mu}^{3}\Omega_{0\mu}\left(\vartheta_{0}^{2}-\frac{1}{\lambda_{0}}\hat{J}_{0}\right)-\frac{g'_{0}}{2\lambda_{0}}\Omega_{0\mu}^{3}\left(X_{0\mu}-\partial_{\mu}\xi_{0}\right)\Omega_{0\nu}\Omega_{0\nu}^{3}\right)\nonumber \\
	-\frac{1}{8\lambda_{0}}\left(\Omega_{0\mu}\Omega_{0\mu}\left(\vartheta_{0}^{2}-\frac{1}{\lambda_{0}}\hat{J}_{0}\right)-\frac{g'_{0}}{2\lambda_{0}}\Omega_{0\mu}^{3}\left(X_{0\mu}-\partial_{\mu}\xi_{0}\right)\Omega_{0\nu}\Omega_{0\nu}\right)\nonumber \\
	\left.-\frac{1}{32\lambda_{0}^{2}}\left(\Omega_{0\mu}\Omega_{0\mu}\Omega_{0\nu}\Omega_{0\nu}+4\Omega_{0\mu}\Omega_{0\mu}\Omega_{0\nu}\Omega_{0\nu}^{3}+4\Omega_{0\mu}\Omega_{0\nu}\Omega_{0\nu}^{3}\Omega_{0\mu}^{3}\right)\right],
\end{gather}
\begin{gather}
	\left(\hat{\Delta}_{3}\right)_{\textrm{0}}=\int d^{4}x\left[\frac{1}{4\lambda_{0}^{2}}\hat{J}_{0}^{2}+\frac{g'{}_{0}^{2}}{16\lambda_{0}^{2}}\left(X_{0\mu}-\partial_{\mu}\xi_{0}\right)\left(X_{0\nu}-\partial_{\nu}\xi_{0}\right)\Omega_{0\mu}^{3}\Omega_{0\nu}^{3}+\frac{g'_{0}}{2\lambda_{0}}\Omega_{0\mu}^{3}\left(X_{0\mu}-\partial_{\mu}\xi_{0}\right)\frac{1}{2\lambda_{0}}\hat{J}_{0}\right.\nonumber \\
	-\frac{1}{2\lambda_{0}}\left(\Omega_{0\mu}^{3}\Omega_{0\mu}\left(O_{0}+\frac{1}{2\lambda_{0}}\hat{J}_{0}\right)+\frac{g'_{0}}{4\lambda_{0}}\Omega_{0\mu}^{3}\left(X_{0\mu}-\partial_{\mu}\xi_{0}\right)\Omega_{0\nu}\Omega_{0\nu}^{3}\right)\nonumber \\
	-\frac{1}{4\lambda_{0}}\left(\Omega_{0\mu}\Omega_{0\mu}\left(O_{0}+\frac{1}{2\lambda_{0}}\hat{J}_{0}\right)+\frac{g'_{0}}{4\lambda_{0}}\Omega_{0\mu}^{3}\left(X_{0\mu}-\partial_{\mu}\xi_{0}\right)\Omega_{0\nu}\Omega_{0\nu}\right)\nonumber \\
	\left.+\frac{1}{64\lambda_{0}^{2}}\left(\Omega_{0\mu}\Omega_{0\mu}\Omega_{0\nu}\Omega_{0\nu}+4\Omega_{0\mu}\Omega_{0\mu}\Omega_{0\nu}\Omega_{0\nu}^{3}+4\Omega_{0\mu}\Omega_{0\nu}\Omega_{0\nu}^{3}\Omega_{0\mu}^{3}\right)\right],
\end{gather}
\begin{gather}
	\left(\hat{\Delta}_{4}\right)_{0}=\int d^{4}x\left[\vartheta_{0}^{4}+\frac{1}{\lambda_{0}^{2}}\hat{J}_{0}^{2}+\frac{g'{}_{0}^{2}}{4\lambda_{0}^{2}}\left(X_{0\mu}-\partial_{\mu}\xi_{0}\right)\left(X_{0\nu}-\partial_{\nu}\xi_{0}\right)\Omega_{0\mu}^{3}\Omega_{0\nu}^{3}+\frac{g'_{0}}{\lambda_{0}^{2}}\Omega_{0\mu}^{3}\left(X_{0\mu}-\partial_{\mu}\xi_{0}\right)\hat{J}_{0}\right.\nonumber \\
	+\frac{1}{\lambda_{0}}\left(\Omega_{0\mu}^{3}\Omega_{0\mu}\left(\vartheta_{0}^{2}-\frac{1}{\lambda_{0}}\hat{J}_{0}\right)-\frac{g'_{0}}{2\lambda_{0}}\Omega_{0\mu}^{3}\left(X_{0\mu}-\partial_{\mu}\xi_{0}\right)\Omega_{0\nu}\Omega_{0\nu}^{3}\right)\nonumber \\
	+\frac{1}{2\lambda_{0}}\left(\Omega_{0\mu}\Omega_{0\mu}\left(\vartheta_{0}^{2}-\frac{1}{\lambda_{0}}\hat{J}_{0}\right)-\frac{g'_{0}}{2\lambda_{0}}\Omega_{0\mu}^{3}\left(X_{0\mu}-\partial_{\mu}\xi_{0}\right)\Omega_{0\nu}\Omega_{0\nu}\right)\nonumber \\
	\left.+\frac{1}{16\lambda_{0}^{2}}\left(\Omega_{0\mu}\Omega_{0\mu}\Omega_{0\nu}\Omega_{0\nu}+4\Omega_{0\mu}\Omega_{0\mu}\Omega_{0\nu}\Omega_{0\nu}^{3}+4\Omega_{0\mu}\Omega_{0\nu}\Omega_{0\nu}^{3}\Omega_{0\mu}^{3}\right)\right],
\end{gather}
\begin{gather}
	\left(\hat{\Delta}_{5}\right)_{0}=\int d^{4}x\left[\frac{1}{2}\left(\Omega_{0\mu}^{3}\Omega_{0\mu}\left(O_{0}+\frac{1}{2\lambda_{0}}\hat{J}_{0}\right)+\frac{g'_{0}}{4\lambda_{0}}\Omega_{0\mu}^{3}\left(X_{0\mu}-\partial_{\mu}\xi_{0}\right)\Omega_{0\nu}\Omega_{0\nu}^{3}\right)\right.\nonumber \\
	\left.+\frac{1}{4}\Omega_{0\mu}^{3}\Omega_{0\mu}\left(\vartheta_{0}^{2}-\frac{1}{\lambda_{0}}\hat{J}_{0}\right)-\frac{g'_{0}}{2\lambda_{0}}\Omega_{0\mu}^{3}\left(X_{0\mu}-\partial_{\mu}\xi_{0}\right)\Omega_{0\nu}\Omega_{0\nu}^{3}\right],
\end{gather}
\begin{gather}
	\left(\hat{\Delta}_{6}\right)_{0}=\int d^{4}x\Omega_{0\mu}^{3}\partial^{2}\Omega_{0\mu}^{3},
\end{gather}
\begin{gather}
	\left(\hat{\Delta}_{7}\right)_{0}=\int d^{4}x\left(\partial_{\mu}\Omega_{0\mu}^{3}\right)\left(\partial_{\nu}\Omega_{0\nu}^{3}\right),
\end{gather}
\begin{gather}
	\left(\hat{\Delta}_{8}\right)_{0}=\int d^{4}x\left[\Omega_{0\mu}^{3}\partial^{2}\Omega_{0\mu}+\left(\partial_{\mu}\Omega_{0\mu}^{3}\right)\left(\partial_{\nu}\Omega_{0\nu}\right)\right],
\end{gather}
\begin{gather}
	\left(\hat{\Delta}_{9}\right)_{0}=\int d^{4}x\left[\Omega_{0\mu}^{\alpha}\Omega_{0\mu}^{\alpha}\left(O_{0}+\frac{1}{2\lambda_{0}}\hat{J}_{0}\right)+\frac{g'_{0}}{4\lambda_{0}}\Omega_{0\mu}^{3}\left(X_{0\mu}-\partial_{\mu}\xi_{0}\right)\Omega_{0\nu}^{\alpha}\Omega_{0\nu}^{\alpha}\right.\nonumber \\
	\left.-\frac{1}{4\lambda_{0}}\left(\Omega_{0\mu}^{\alpha}\Omega_{0\mu}^{\alpha}\Omega_{0\nu}^{3}\Omega_{0\nu}+\frac{1}{2}\Omega_{0\mu}^{\alpha}\Omega_{0\mu}^{\alpha}\Omega_{0\nu}\Omega_{0\nu}\right)\right],
\end{gather}
\begin{gather}
	\left(\hat{\Delta}_{10}\right)_{0}=\int d^{4}x\left[\Omega_{0\mu}^{3}\Omega_{0\mu}^{3}\left(O_{0}+\frac{1}{2\lambda_{0}}\hat{J}_{0}\right)+\frac{g'_{0}}{4\lambda_{0}}\Omega_{0\mu}^{3}\left(X_{0\mu}-\partial_{\mu}\xi_{0}\right)\Omega_{0\nu}^{3}\Omega_{0\nu}^{3}\right.\nonumber \\
	\left.-\frac{1}{8\lambda_{0}}\left(\Omega_{0\mu}\Omega_{0\mu}\Omega_{0\nu}^{3}\Omega_{0\nu}^{3}+2\Omega_{0\mu}\Omega_{0\mu}^{3}\Omega_{0\nu}^{3}\Omega_{0\nu}^{3}\right)\right],
\end{gather}
\begin{gather}
	\left(\hat{\Delta}_{11}\right)_{0}=\int d^{4}x\left[\Omega_{0\mu}^{\alpha}\Omega_{0\mu}^{\alpha}\left(\vartheta_{0}^{2}-\frac{1}{\lambda_{0}}\hat{J}_{0}\right)-\frac{g'_{0}}{2\lambda_{0}}\Omega_{0\mu}^{3}\left(X_{0\mu}-\partial_{\mu}\xi_{0}\right)\Omega_{0\nu}^{\alpha}\Omega_{0\nu}^{\alpha}\right.\nonumber \\
	\left.+\frac{1}{2\lambda_{0}}\left(\Omega_{0\mu}^{\alpha}\Omega_{0\mu}^{\alpha}\Omega_{0\nu}^{3}\Omega_{0\nu}+\frac{1}{2}\Omega_{0\mu}^{\alpha}\Omega_{0\mu}^{\alpha}\Omega_{0\nu}\Omega_{0\nu}\right)\right],
\end{gather}
\begin{gather}
	\left(\hat{\Delta}_{12}\right)_{0}=\int d^{4}x\left[\Omega_{0\mu}^{3}\Omega_{0\mu}^{3}\left(\vartheta_{0}^{2}-\frac{1}{\lambda_{0}}\hat{J}_{0}\right)-\frac{g'_{0}}{2\lambda_{0}}\Omega_{0\mu}^{3}\left(X_{0\mu}-\partial_{\mu}\xi_{0}\right)\Omega_{0\nu}^{3}\Omega_{0\nu}^{3}\right.\nonumber \\
	\left.+\frac{1}{4\lambda_{0}}\left(\Omega_{0\mu}\Omega_{0\mu}\Omega_{0\nu}^{3}\Omega_{0\nu}^{3}+2\Omega_{0\mu}\Omega_{0\mu}^{3}\Omega_{0\nu}^{3}\Omega_{0\nu}^{3}\right)\right],
\end{gather}
\begin{gather}
	\left(\hat{\Delta}_{13}\right)_{0}=\int d^{4}x\left[\left(\partial_{\mu}\Omega_{0\mu}^{3}\right)\hat{J}_{0}+\frac{g'_{0}}{2}\left(X_{0\mu}-\partial_{\mu}\xi_{0}\right)\Omega_{0\mu}^{3}\partial_{\nu}\Omega_{0\nu}^{3}+\frac{1}{2}\Omega_{0\mu}^{3}\Omega_{0\nu}^{3}\left(\partial_{\mu}\Omega_{0\nu}\right)\right.\nonumber \\
	\left.+\frac{1}{2}\Omega_{0\mu}^{3}\left(\partial_{\mu}\Omega_{0\nu}^{3}\right)\Omega_{0\nu}-\frac{1}{4}\Omega_{0\mu}\Omega_{0\mu}\left(\partial_{\nu}\Omega_{0\nu}^{3}\right)\right],
\end{gather}
\begin{gather}
	\left(\hat{\Delta}_{14}\right)_{0}=\int d^{4}x\Omega_{0\mu}^{\alpha}\Omega_{0\mu}^{\alpha}\left(\partial_{\nu}\Omega_{0\nu}^{3}\right),
\end{gather}
\begin{gather}
	\left(\hat{\Delta}_{15}\right)_{0}=\int d^{4}x\Omega_{0\mu}^{\alpha}\left(\partial_{\nu}\Omega_{0\mu}^{\alpha}\right)\Omega_{0\nu}^{3},
\end{gather}
\begin{gather}
	\left(\hat{\Delta}_{16}\right)_{0}=\int d^{4}x\Omega_{0\mu}^{\alpha}\Omega_{0\nu}^{\alpha}\partial_{\mu}\Omega_{0\nu}^{3},
\end{gather}
\begin{gather}
	\left(\hat{\Delta}_{17}\right)_{0}=\int d^{4}x\left[\Omega_{0\mu}^{\alpha}\left(\partial_{\mu}\Omega_{0\nu}^{\alpha}\right)\Omega_{0\nu}^{3}-g'_{0}\varepsilon^{\alpha\beta}\Omega_{0\mu}^{\alpha}\left(X_{0\mu}-\partial_{\mu}\xi_{0}\right)\Omega_{0\nu}^{\beta}\Omega_{0\nu}^{3}-\varepsilon^{\alpha\beta}\Omega_{0\mu}^{\alpha}\Omega_{0\nu}^{\beta}\Omega_{0\mu}^{3}\Omega_{0\nu}\right],
\end{gather}
\begin{gather}
	\left(\hat{\Delta}_{18}\right)_{0}=\int d^{4}x\varepsilon^{\alpha\beta}\Omega_{0\mu}^{\alpha}\Omega_{0\nu}^{\beta}\left(\partial_{\mu}\Omega_{0\nu}^{3}\right),
\end{gather}
\begin{gather}
	\left(\hat{\Delta}_{19}\right)_{0}=\int d^{4}x\left[\varepsilon^{\alpha\beta}\Omega_{0\mu}^{\alpha}\left(\partial_{\mu}\Omega_{0\nu}^{\beta}\right)\Omega_{0\nu}^{3}+g'_{0}\Omega_{0\mu}^{\alpha}\left(X_{0\mu}-\partial_{\mu}\xi_{0}\right)\Omega_{0\nu}^{\alpha}\Omega_{0\nu}^{3}-\Omega_{0\mu}^{\alpha}\Omega_{0\nu}^{\alpha}\Omega_{0\mu}^{3}\Omega_{0\nu}\right],
\end{gather}
\begin{gather}
	\left(\hat{\Delta}_{20}\right)_{0}=\int d^{4}x\Omega_{0\mu}^{3}\Omega_{0\mu}^{3}\left(\partial_{\nu}\Omega_{0\nu}^{3}\right),
\end{gather}
\begin{gather}
	\left(\hat{\Delta}_{21}\right)_{0}=\int d^{4}x\Omega_{0\mu}^{\alpha}\Omega_{0\mu}^{\alpha}\Omega_{0\nu}^{\beta}\Omega_{0\nu}^{\beta},
\end{gather}
\begin{gather}
	\left(\hat{\Delta}_{22}\right)_{0}=\int d^{4}x\Omega_{0\mu}^{\alpha}\Omega_{0\mu}^{\beta}\Omega_{0\nu}^{\alpha}\Omega_{0\nu}^{\beta},
\end{gather}
\begin{gather}
	\left(\hat{\Delta}_{23}\right)_{0}=\int d^{4}x\varepsilon^{\alpha\beta}\varepsilon^{\gamma\delta}\Omega_{0\mu}^{\alpha}\Omega_{0\nu}^{\beta}\Omega_{0\mu}^{\gamma}\Omega_{0\nu}^{\delta},
\end{gather}
\begin{gather}
	\left(\hat{\Delta}_{24}\right)_{0}=\int d^{4}x\Omega_{0\mu}^{\alpha}\Omega_{0\mu}^{\alpha}\Omega_{0\nu}^{3}\Omega_{0\nu}^{3},
\end{gather}
\begin{gather}
	\left(\hat{\Delta}_{25}\right)_{0}=\int d^{4}x\Omega_{0\mu}^{\alpha}\Omega_{0\nu}^{\alpha}\Omega_{0\mu}^{3}\Omega_{0\nu}^{3},
\end{gather}
\begin{gather}
	\left(\hat{\Delta}_{26}\right)_{0}=\int d^{4}x\Omega_{0\mu}^{3}\Omega_{0\mu}^{3}\Omega_{0\nu}^{3}\Omega_{0\nu}^{3},
\end{gather}
\begin{eqnarray}
	\left(\hat{\Delta}_{27}\right)_{0} & = & \int d^{4}x\left(K_{0\mu}^{a}\zeta_{0\mu}^{a}+ib_{0\mu}^{a}\partial_{\mu}\Upsilon_{0\mu}^{a}-i\overline{c}_{0}^{a}\partial_{\mu}\zeta_{0\mu}^{a}\right).
\end{eqnarray}

\section*{Appendix E: Algrebraic analysis of the two-point function of $O_{\mu}$}

In this appendix we want to derive a series of important relations to the correlation
functions presented in the end of Subsection \ref{subsec:Non-integrated--equation-and}. Specially, we want to show that no propagating modes can
be associated to the longitudinal part of $\left\langle O_{\mu}\left(x\right)O_{\nu}\left(y\right)\right\rangle $.
The Ward identity for the tree-level action $\Sigma$, Eq. (\ref{eq:X_ward_identity}),
implies that
\begin{eqnarray}
	\frac{\delta\Gamma}{\delta X_{\mu}}+g'\frac{\delta\Gamma}{\delta\Omega_{\mu}}+\frac{g'}{2}\Omega_{\mu}\frac{\delta\Gamma}{\delta J} & = & -\partial_{\nu}X_{\nu\mu}+i\partial_{\mu}\theta-i\partial_{\mu}b-\frac{g'\vartheta^{2}}{4}\Omega_{\mu}+\Omega_{\mu}^{5}\vartheta^{2}+\partial^{2}\Omega_{\mu}^{6}-\partial_{\mu}\partial_{\nu}\Omega_{\nu}^{6}.
\end{eqnarray}
The effective action can be replaced by the generating functional of
the connected functions $W$ through the Legendre transformation
\begin{eqnarray}
	\Gamma\left[\phi\right] & = & W\left[J_{\phi}\right]-\int d^{4}xJ_{\phi}^{i}\phi^{i},
\end{eqnarray}
where $J_{\phi}^{i}$ denotes the source of the field $\phi^{i}$.
It follows that
\begin{eqnarray}
	-j_{\mu}+g'\frac{\delta W}{\delta\Omega_{\mu}}+\frac{g'}{2}\Omega_{\mu}\frac{\delta W}{\delta J} & = & -\partial_{\nu}\partial_{\nu}\frac{\delta W}{\delta j_{\mu}}+\partial_{\mu}\partial_{\nu}\frac{\delta W}{\delta j_{\nu}}+i\partial_{\mu}\frac{\delta W}{\delta J_{\theta}}-i\partial_{\mu}\frac{\delta W}{\delta J_{b}}\nonumber \\
	&  & -\frac{g'\vartheta^{2}}{4}\Omega_{\mu}+\Omega_{\mu}^{5}\vartheta^{2}+\partial^{2}\Omega_{\mu}^{6}-\partial_{\mu}\partial_{\nu}\Omega_{\nu}^{6},\label{eq:X_conneccted}
\end{eqnarray}
where $j_{\mu}$, $J_{\theta}$ and $J_{b}$ are the sources of $X_{\mu}$,
$\theta$ and $b$, respectively. By differentiating (\ref{eq:X_conneccted})
with respect to $j_{\nu}$ at a different point and setting all external
sources to zero, we obtain
\begin{eqnarray}
	-\delta_{\mu\nu}-g'\left\langle O_{\mu}\left(x\right)X_{\nu}\left(y\right)\right\rangle  & = & \left(\partial^{2}\delta_{\mu\lambda}-\partial_{\mu}\partial_{\lambda}\right)^{x}\left\langle X_{\lambda}\left(x\right)X_{\nu}\left(y\right)\right\rangle -i\partial_{\mu}^{x}\left\langle \theta\left(x\right)X_{\nu}\left(y\right)\right\rangle +i\partial_{\mu}^{x}\left\langle b\left(x\right)X_{\nu}\left(y\right)\right\rangle ,
\end{eqnarray}
where
\begin{eqnarray}
	\left\langle O_{\mu}\left(x\right)X_{\nu}\left(y\right)\right\rangle  & = & -\left.\frac{\delta^{2}W}{\delta\Omega_{\mu}\left(x\right)j_{\nu}\left(y\right)}\right|_{\textrm{sources=0}},\nonumber \\
	\left\langle X_{\lambda}\left(x\right)X_{\nu}\left(y\right)\right\rangle  & = & -\left.\frac{\delta^{2}W}{\delta j_{\lambda}\left(x\right)j_{\nu}\left(y\right)}\right|_{\textrm{sources=0}},\nonumber \\
	\left\langle \theta\left(x\right)X_{\nu}\left(y\right)\right\rangle  & = & -\left.\frac{\delta^{2}W}{\delta J_{\theta}\left(x\right)j_{\nu}\left(y\right)}\right|_{\textrm{sources=0}},\nonumber \\
	\left\langle b\left(x\right)X_{\nu}\left(y\right)\right\rangle  & = & -\left.\frac{\delta^{2}W}{\delta J_{b}\left(x\right)j_{\nu}\left(y\right)}\right|_{\textrm{sources=0}}.
\end{eqnarray}
In the Appendix A, we show that $\left\langle \theta\left(x\right)X_{\nu}\left(y\right)\right\rangle =0$
and
\begin{eqnarray}
	\left\langle b\left(x\right)X_{\mu}\left(y\right)\right\rangle  & = & -\int\frac{d^{4}p}{\left(2\pi\right)^{4}}e^{-ip\cdot\left(x-y\right)}\frac{p_{\mu}}{p^{2}},\label{eq:exact_W_b-1-1}
\end{eqnarray}
then
\begin{eqnarray}
	i\partial_{\mu}^{x}\left\langle b\left(x\right)X_{\nu}\left(y\right)\right\rangle  & = & -\int\frac{d^{4}p}{\left(2\pi\right)^{4}}e^{-ip\cdot\left(x-y\right)}\frac{p_{\mu}p_{\nu}}{p^{2}}\nonumber \\
	& = & -\left(\frac{\partial_{\mu}\partial_{\nu}}{\partial^{2}}\right)^{x}\delta\left(x-y\right).
\end{eqnarray}
Therefore, we obtain the following relationship between $\left\langle O_{\mu}\left(x\right)X_{\nu}\left(y\right)\right\rangle $
and $\left\langle X_{\lambda}\left(x\right)X_{\nu}\left(y\right)\right\rangle $:
\begin{eqnarray}
	g'\left\langle O_{\mu}\left(x\right)X_{\nu}\left(y\right)\right\rangle  & = & -\left(\partial^{2}\delta_{\mu\lambda}-\partial_{\mu}\partial_{\lambda}\right)^{x}\left\langle X_{\lambda}\left(x\right)X_{\nu}\left(y\right)\right\rangle -\left(\delta_{\mu\nu}-\frac{\partial_{\mu}\partial_{\nu}}{\partial^{2}}\right)^{x}\delta^{4}\left(x-y\right)\nonumber \\
	& = & -\left(\partial^{2}\right)^{x}\left\langle X_{\mu}\left(x\right)X_{\nu}\left(y\right)\right\rangle -\left(\delta_{\mu\nu}-\frac{\partial_{\mu}\partial_{\nu}}{\partial^{2}}\right)^{x}\delta^{4}\left(x-y\right),\label{eq:first_relation}
\end{eqnarray}
which establishes that $\left\langle O_{\mu}\left(x\right)X_{\nu}\left(y\right)\right\rangle $
is essentially given by $\left\langle X_{\mu}\left(x\right)X_{\nu}\left(y\right)\right\rangle $,
while also being transverse. By differentiating (\ref{eq:X_conneccted})
with respect to $\Omega_{\nu}$ at a different point and setting all
external sources to zero, we obtain a second relation, namely,
\begin{eqnarray}
	g'\left\langle O_{\mu}\left(x\right)O_{\nu}\left(y\right)\right\rangle  & = & -\left(\partial^{2}\delta_{\mu\lambda}-\partial_{\mu}\partial_{\lambda}\right)^{x}\left\langle X_{\lambda}\left(x\right)O_{\nu}\left(y\right)\right\rangle +\frac{g'}{2}\delta_{\mu\nu}\delta^{4}\left(x-y\right)\left(\left\langle O\left(x\right)\right\rangle +\frac{\vartheta^{2}}{2}\right),\label{eq:second_relation}
\end{eqnarray}
where
\begin{eqnarray}
	\left\langle O_{\mu}\left(x\right)O_{\nu}\left(y\right)\right\rangle  & = & -\left.\frac{\delta^{2}W}{\delta\Omega_{\mu}\left(x\right)\Omega_{\nu}\left(y\right)}\right|_{\textrm{sources=0}},\nonumber \\
	\left\langle O\left(x\right)\right\rangle  & = & \left.\frac{\delta W}{\delta J\left(x\right)}\right|_{\textrm{sources=0}}.
\end{eqnarray}
Finally, combining (\ref{eq:first_relation}) and (\ref{eq:second_relation}),
we derive the following expression for $\left\langle O_{\mu}\left(x\right)O_{\nu}\left(y\right)\right\rangle $:
\begin{eqnarray}
	g'{}^{2}\left\langle O_{\mu}\left(x\right)O_{\nu}\left(y\right)\right\rangle  & = & \left(\partial^{4}\right)^{x}\left\langle X_{\mu}\left(x\right)X_{\nu}\left(y\right)\right\rangle +\left(\partial^{2}\delta_{\mu\nu}-\partial_{\mu}\partial_{\nu}\right)^{x}\delta^{4}\left(x-y\right)\nonumber \\
	&  & +\frac{g'{}^{2}}{2}\delta_{\mu\nu}\delta^{4}\left(x-y\right)\left(\left\langle O\left(x\right)\right\rangle +\frac{\vartheta^{2}}{2}\right). \label{eq:OOvector}
\end{eqnarray}
As we anticipated, the transverse part of the two-point function of
$O_{\mu}$ is determined by $\left\langle X_{\mu}\left(x\right)X_{\nu}\left(y\right)\right\rangle $.
In addition to that, the longitudinal part is given by
\begin{gather}
	\frac{\partial_\mu \partial_\nu}{\partial^2}\left\langle O_{\mu}\left(x\right)O_{\nu}\left(y\right)\right\rangle =\frac{1}{2}\left(\left\langle O\left(x\right)\right\rangle +\frac{\vartheta^{2}}{2}\right). \label{eq:OO_transv}
\end{gather}


\end{document}